\documentclass[aps,twocolumn,nofootinbib]{revtex4}
\usepackage{graphicx}
\usepackage{amsmath}
\usepackage{amssymb}
\usepackage{bm}
\usepackage{mathtools}
\usepackage{hyperref}
\usepackage{xcolor}
\usepackage{xfrac}
\usepackage{bm}
\usepackage[T1]{fontenc}
\usepackage{upquote}

\DeclareMathOperator{\Ima}{im}

\DeclareMathOperator{\diag}{diag}

\newcommand{\rn}{{\rm n}}

\begin{document} 

\title{\bf Intrinsic and extrinsic thermodynamics for stochastic
population processes with multi-level large-deviation structure} 

% Intrinsic and extrinsic thermodynamics for discrete stochastic processes with multi-level large-deviation structure

% The intrinsic and extrinsic thermodynamics of chemical reaction networks

\author{Eric Smith}

\affiliation{Department of Biology, Georgia Institute of
Technology, 310 Ferst Drive NW, Atlanta, GA 30332, USA}

\affiliation{Earth-Life Science Institute, Tokyo Institute of
Technology, 2-12-1-IE-1 Ookayama, Meguro-ku, Tokyo 152-8550, Japan}

\affiliation{Santa Fe Institute, 1399 Hyde Park Road, Santa Fe, NM
87501, USA}

\affiliation{Ronin Institute, 127 Haddon Place, Montclair, NJ 07043,
USA} 

\date{\today}
\begin{abstract}

A set of core features is set forth as the essence of a thermodynamic
description, which derive from large-deviation properties in systems
with hierarchies of timescales, but which are \emph{not} dependent
upon conservation laws or microscopic reversibility in the substrate
hosting the process.  The most fundamental elements are the concept of
a macrostate in relation to the large-deviation entropy, and the
decomposition of contributions to irreversibility among interacting
subsystems, which is the origin of the dependence on a concept
of heat in both classical and stochastic thermodynamics.  A natural
decomposition is shown to exist, into a relative entropy and a
housekeeping entropy rate, which define respectively the
\textit{intensive} thermodynamics of a system and an
\textit{extensive} thermodynamic vector embedding the system in its
context.  Both intensive and extensive components are functions of
Hartley information of the momentary system stationary state, which is
information \emph{about} the joint effect of system processes on its
contribution to irreversibility.  Results are derived for stochastic
Chemical Reaction Networks, including a Legendre duality for the
housekeeping entropy rate to thermodynamically characterize
fully-irreversible processes on an equal footing with those at the
opposite limit of detailed-balance.  The work is meant to encourage
development of inherent thermodynamic descriptions for rule-based
systems and the living state, which are not conceived as reductive
explanations to heat flows.

\end{abstract}

\maketitle

\section{Introduction}

The statistical derivations underlying most thermodynamic phenomena
are understood to be widely applicable, and are mostly developed in
general terms.  Yet where thermodynamics is offered as an ontology to 
understand new patterns and causes in nature -- the
\textit{thermodynamics of
computation}~\cite{Szilard:MD:29,Landauer:IHGCP:61,Bennett:TC:82,Wolpert:stoch_thermo_comp:19}
or \textit{stochastic
thermodynamics}~\cite{Seifert:stoch_thermo_rev:12},\footnote{
Stochastic thermodynamics is the modern realization of a program to
create a non-equilibrium thermodynamics that began with
Onsager~\cite{Onsager:RRIP1:31,Onsager:RRIP2:31} and took much of its
modern form under Prigogine and
coworkers~\cite{Glansdorff:structure:71,Prigogine:MT:98}.  Since the
beginning its core method has been to derive rules or constraints for
non-stationary thermal dynamics from dissipation of free energies
defined by Gibbs equilibria, from any combination of thermal baths or
asymptotic reservoirs.  \\
\mbox{\quad} 
The parallel and contemporaneous development of thermodynamics of
computation can be viewed as a quasistatic analysis with discrete
changes in the boundary conditions on a thermal bath corresponding to
logical events in algorithms.  Early stochastic thermodynamics
combined elements of both traditions, with the quantities termed
``non-equilibrium entropies'' corresponding to the information
entropies over computer states, distinct from quasistatic entropies
associated with heat in a locally-equilibrium environment, with
boundary conditions altered by the explicitly-modeled stochastic state
transitions.\\
\mbox{\quad} 
More recently stochastic thermodynamics incorporated time-reversal
methods originally developed for measures in dynamical
systems~\cite{Searles:fluct_thm:99,Evans:fluct_thm:02,Gallavotti:dyn_ens_NESM:95,Gallavotti:dyn_ens_SS:95}
leading to a variety of fluctuation
theorems~\cite{Jarzynski:fluctuations:08,Chetrite:fluct_diff:08,Esposito:fluct_theorems:10}
and nonequilibrium work
relations~\cite{Crooks:NE_work_relns:99,Crooks:path_ens_aves:00,Kurchan:NEWRs:07},
still however relating path probability ratios either to dissipation
of heat or to differences in equilibrium free energies.}
or where these methods are taken to define a foundation for the  
\textit{statistical physics of reproduction or
adaptation}~\cite{England:statphys_selfrepl:13,Perunov:adaptation:15}
-- these problems are framed in terms of two
properties particular to the domain of mechanics: the conservation of
energy and microscopic reversibility.  Other applications of the same
mathematics, with designations such as 
\textit{intensity of
choice}~\cite{Luce:choice:59,McFadden:quantal_choice:76}
\textit{tipping points}, or 
\textit{early warning signs}~\cite{Lenton:climate_tip_ew:12}, are
recognized as analogies to thermodynamics, on the understanding that
they only become the ``thermodynamics of'' something when they derive
its causes from energy conservation connected to the entropies of
heat.

This accepted detachment of mathematics from phenomenology, with
thermodynamic phenomena interpreted in their historical terms and
mathematics kept interpretation-free, contrasts with the way
statistical mechanics was allowed to expand the ontological categories
of physics at the end of the last century.  The kinetic theory of
heat~\cite{Joule:sci_papers:44,Clausius:mech_the_heat:65,Boltzmann:second_law:86}
was not established as an analogy to gambling\footnote{The
large-deviation rate function underlies the solution of the gambler's
ruin problem of Pascal, Fermat, and
Bernoulli~\cite{Hald:prob_pre_1750:90}.} used to describe patterns in
the fluid caloric.\footnote{However, as late as 1957
Jaynes~\cite{Jaynes:ITSM_I:57,Jaynes:ITSM_II:57} needed to assert that
the information entropy of Shannon referred to the same quantity as
the physical entropy of Boltzmann, and was not merely the identical
mathematical function.  Even the adoption of ``entropy production'' to
refer to changes in the state-variable entropy by irreversible
transformations -- along the course of which the state-variable
entropy is not even defined -- is a retreat to a substance syntax; I
would prefer the less euphonious but categorically better expression
``loss of large-deviation accessibility''.}  Thermodynamics instead
took the experienced phenomena involving heat, and where there had
formerly been only names and metaphors to refer to them, it brought
into existence concepts capturing their essential nature, not
dissolving their reality as phenomena~\cite{Rota:lect_notes:08}, but
endowing it with a semantics.  The distinction is between
formalization in the service of reduction to remain within an existing
ontology, and formalization as the foundation for discovery of new
conceptual primitives.  Through generalizations and extensions that
could not have been imagined in the late 19th
century~\cite{GellMann:RG:54,Wilson:RG:74,Weinberg:phenom_Lagr:79,Polchinski:RGEL:84}
(revisited in Sec.~\ref{sec:discussion}), essentially thermodynamic
insights went on to do the same for our fundamental theory of objects
and interactions, the nature of the vacuum and the hierarchy of
matter, and the presence of stable macro-worlds at all.

This paper is written with the view that the essence of a
thermodynamic description is not found in its connection to
conservation laws, microscopic reversibility, or the equilibrium state
relations they entail, despite the central role those play in the
fields
mentioned~\cite{Landauer:IHGCP:61,England:statphys_selfrepl:13,Perunov:adaptation:15,Seifert:stoch_thermo_rev:12}.
At the same time, it grants an argument that has been maintained
across a half-century of enormous growth in both statistical methods
and applications~\cite{Fermi:TD:56,Bertini:macro_NEThermo:09}: that
thermodynamics should not be conflated with its statistical methods.
The focus will therefore be on the patterns and relations that make a
phenomenon essentially thermodynamic, which statistical mechanics made
it possible to articulate as concepts.  The paper proposes a sequence
of these and exhibits constructions of them unconnected to energy
conservation or microscopic reversibility.

Essential concepts are of three kinds: 1) the nature and origin of
macrostates; 2) the roles of entropy in relation to irreversibility
and fluctuation; and 3) the natural apportionment of irreversibility
between a system and its environment, which defines an
\textit{intrinsic thermodynamics} for the system and an
\textit{extrinsic thermodynamics} that embeds the system in its
context, in analogy to the way differential geometry defines an
intrinsic curvature for a manifold distinct from extrinsic curvatures
that may embed the manifold in another manifold of higher
dimension.\footnote{This analogy is not a reference to natural
\textit{information
geometries}~\cite{Amari:inf_geom:01,Ay:info_geom:17} that can also be
constructed, though perhaps those geometries would provide an
additional layer of semantics to the constructions here.}

All of these concepts originate in the large-deviation properties of
stochastic processes with multi-level timescale structure.  
They will be demonstrated here using a simple class of stochastic
population processes, further specified as Chemical Reaction Network
models as more complex relations need to be presented.  

Emphasis will be laid on the different roles of entropy as a
functional on general distributions versus entropy as a state function
on macrostates, and on the Lyapunov role of entropy in the 2nd
law~\cite{Boltzmann:second_law:86,Fermi:TD:56} versus its
large-deviation role in fluctuations~\cite{Ellis:ELDSM:85}, through
which the state-function entropy is most generally
definable~\cite{Touchette:large_dev:09}.  
Both Shannon's information entropy~\cite{Shannon:MTC:49} and the older
Hartley function~\cite{Hartley:information:28}, will appear here as
they do in stochastic thermodynamics
generally~\cite{Seifert:stoch_thermo_rev:12}.  The different meanings
these functions carry, which sound contradictory when described and
can be difficult to compare in derivations with different aims, become
clear as each appears in the course of a single calculation.  Most
important, they have unambiguous roles in relation to either large
deviations or system decomposition without need of a reference to
heat.

\subsubsection*{Main results and order of the derivation}

Sec.~\ref{sec:multi_level_sys} explains what is meant by multi-level
systems with respect to a robust separation of timescales, and
introduces a family of models constructed recursively from nested
population processes.  Macrostates are related to microstates
\emph{within} levels, and if timescale separations arise that create
new levels, they come from properties of a subset of long-lived,
metastable macrostates.  Such states within a level, and transitions
between them that are much shorter than their characteristic
lifetimes, map by coarse-graining to the elementary states and events
at the next level.

The \textit{environment} of a system that arises in some level of a
hierarchical model is understood to include both the thermalized
substrate at the level below, and other subsystems with explicit
dynamics within the same level.  Sec.~\ref{sec:sys_env_decomp}
introduces the problem of system/environment partitioning of changes
in entropy, and states\footnote{The technical requirements are either
evident or known from fluctuation
theorems~\cite{Speck:HS_heat_FT:05,Harris:fluct_thms:07}; variant
proofs are given in later sections.  Their significance for system
decomposition is the result of interest here.} the first main claim:
the natural partition employs a relative entropy~\cite{Cover:EIT:91}
within the system and a housekeeping entropy
rate~\cite{Hatano:NESS_Langevin:01} to embed the system in the
environment.  It results in two independently non-negative entropy
changes (plus a third entirely within the environment that is usually
ignored), which define the \textit{intensive thermodynamic} of the
focal system and the \textit{extensive}, or
\textit{embedding thermodynamics} of the system into the
$\mbox{system} \otimes
\mbox{environment}$ whole.

Sec.~\ref{sec:Ham_Jacob} expresses the relation of macrostates to
microstates in terms of the large-deviations concept of
\textit{separation of scale from
structure}~\cite{Touchette:large_dev:09}.  Large-deviations scaling,
defined as the convergence of distributions for aggregate statistics
toward exponential families, creates a formal concept of a macroworld
having definite structure, yet separated by an indefinite or even
infinite range of scale from the specifications of micro-worlds.

Large-deviations for population processes are handled with the
Hamilton-Jacobi theory~\cite{Bertini:macro_NEThermo:09} following from
the time dependence of the cumulant-generating function (CGF), and its
Legendre duality~\cite{Amari:inf_geom:01} to a large-deviation
function called the \textit{effective action}~\cite{Smith:LDP_SEA:11}.  
Macrostates are identified with the distributions that can be assigned
large-deviation probabilities from a system's stationary distribution.
Freedom to study any CGF makes this definition, although concrete,
flexible enough to require choosing what Gell-Mann and
Lloyd~\cite{GellMann:EC:96,GellMann:eff_complx:04} term ``a judge''.
The resulting definition, however, does not depend on whether the
system has any underlying mechanics or conservation laws, explicit or
implicit.

To make contact with multi-level dynamics and problems of interest in
stochastic
thermodynamics~\cite{Seifert:stoch_thermo_rev:12,Polettini:open_CNs_I:14,Polettini:stoch_macro_thermo:16},
while retaining a definite notation, Sec.~\ref{sec:CRN_form} assumes
generators of the form used for \textit{Chemical Reaction Networks}
(CRNs)~\cite{Horn:mass_action:72,Feinberg:notes:79,Krishnamurthy:CRN_moments:17,Smith:CRN_moments:17}.
The finite-to-infinite mapping that relates macrostates to microstates
has counterparts for CRNs in maps from the generator matrix to the
transition matrix, and from mass-action macroscopic currents to
probability currents between microstates.

This section formally distinguishes the Lyapunov~\cite{Fermi:TD:56}
and large-deviation~\cite{Ellis:ELDSM:85} roles of entropy, and shows
how the definition of macrostates from Sec.~\ref{sec:Ham_Jacob} first
introduces scale-dependence in the \textit{state function entropy}
that was not present in the Lyapunov function for arbitrary
multi-scale models of Sec.~\ref{sec:multi_level_sys}.  In the
Hamilton-Jacobi representation, Lyapunov and large-deviation entropy
changes occur within different manifolds and have different
interpretations.  These differences are subordinate to the more
fundamental difference between the entropy state function and entropy
as a functional on arbitrary distributions over microstates.  A
properly formulated 2nd law, which is never violated, is computed
from both deterministic and fluctuation macrostate-entropies.  The
roles of Hartley informations~\cite{Hartley:information:28} for
stationary states in stochastic
thermodynamics~\cite{Speck:HS_heat_FT:05,Seifert:FDT:10,Seifert:stoch_thermo_rev:12}
enter naturally as macrostate relative entropies.

Sec.~\ref{sec:cycle_decomp} derives the proofs of monotonicity of the
intrinsic and extrinsic entropy changes from
Sec.~\ref{sec:multi_level_sys}, using a cycle decomposition of
currents in the stationary distribution.\footnote{The decomposition is
related to that used by Schnakenberg~\cite{Schnakenberg:ME_graphs:76}
to compute dissipation in the stationary state, but more cycles are
required in order to compute dynamical quantities.}  Whether only
cycles or more complex hyperflows~\cite{Andersen:generic_strat:14} are
required in a basis for macroscopic currents distinguishes complexity
classes for CRNs.  Because cycles are \emph{always} a sufficient basis
in the microstate space, the breakdown of a structural equivalence
between micro- and macro-states for complex CRNs occurs in just the
terms responsible for the complex relations between state-function and
global entropies derived in Sec.~\ref{sec:CRN_form}.

Sec.~\ref{sec:examples} illustrates two uses of the
intensive/extensive decomposition of entropy changes.  It studies a
simple model of polymerization and hydrolysis under competing
spontaneous and driven reactions in an environment that can be in
various states of disequilibrium.  First a simple linear model of the
kind treated by Schnakenberg~\cite{Schnakenberg:ME_graphs:76} is
considered, in a limit where one elementary reaction becomes strictly
irreversible.  Although the housekeeping entropy rate becomes
uninformative because it is referenced to a diverging chemical
potential, this is a harmless divergence analogous to a scale
divergence of an extensive potential.  A Legendre dual to the entropy
rate remains regular, reflecting the existence of a
thermodynamics-of-events about which energy conservation, although
present on the path to the limit, asymptotically is not a source of
any relevant constraints.

A second example introduces autocatalysis into the coupling to the
disequilibrium environment, so that the system can become bistable.
The components of entropy rate usually attributed to the
``environment'' omit information about the interaction of reactions
responsible for the bistability, and attribute too much of the loss of
large-deviation accessibility to the environment.  Housekeeping
entropy rate gives the correct accounting, recognizing that part
of the system's irreversibility depends on the measure for
\emph{system} relative entropy created by the interaction.

Sec.\ref{sec:discussion} offers an alternative characterization of
thermodynamic descriptions when conservation and reversibility are not
central.  In 20th century physics the problem of the nature and source
of macroworlds has taken on a clear formulation, and provides an
alternative conceptual center for thermodynamics to relations between
work and heat.  System decomposition and the conditional independence
structures within the loss of large-deviation accessibility replace
adiabatic transformation and heat flow as the central abstractions to
describe irreversibility, and define the entropy interpretation of the
Hartley information.  Such a shift in view will be needed if path
ensembles are to be put on an equal footing with state ensembles to
create a fully non-equilibrium thermodynamics.  The alternative
formulation is meant to support the development of a native
thermodynamics of stochastic rule-based systems and a richer
phenomenology of living states.

\section{Multi-level systems}
\label{sec:multi_level_sys}

To provide a concrete class of examples for the large-deviation
relations that can arise in multi-scale systems, we consider systems
with natural \textit{levels}, such that within each level a given
system may be represented by a discrete population process.  The
population process in turn admits leading-exponential approximations
of its large-deviation behavior in the form of a continuous dynamical
system.  These dynamical systems are variously termed
\textit{momentum-space WKB approximations}~\cite{Assaf:mom_WKB:17},
\textit{Hamilton-Jacobi
representations}~\cite{Bertini:macro_NEThermo:09}, or \textit{eikonal
expansions}~\cite{Freidlin:RPDS:98}.  They exist for many other
classes of stochastic process besides the one assumed
here~\cite{Martin:MSR:73,Kamenev:DP:02}, and may be obtained either
directly from the G{\"{a}}rtner-Ellis theorem~\cite{Ellis:ELDSM:85}
for cumulant-generating functions (the approach taken here), by direct
WKB asymptotics, or through saddle-point methods in 2-field functional
integrals.

A connection between levels is made by supposing that the
large-deviation behavior possesses one or more fixed points of the
dynamical system.  These are coarse-grained to become the elementary
states in a similar discrete population process one level up in the
scaling hierarchy.  Nonlinearities in the dynamical system that
produce multiple isolated, metastable fixed points are of particular
interest, as transitions between these occur on timescales that are
exponentially stretched, in the large-deviation scale factor, relative
to the relaxation times.  The resulting robust criterion for
separation of timescales will be the basis for the thermodynamic
limits that distinguish levels and justify the coarse-graining.

\subsubsection{Micro to macro, within and between levels}

\begin{figure}[ht]
\begin{center} 
  \includegraphics[scale=0.45]{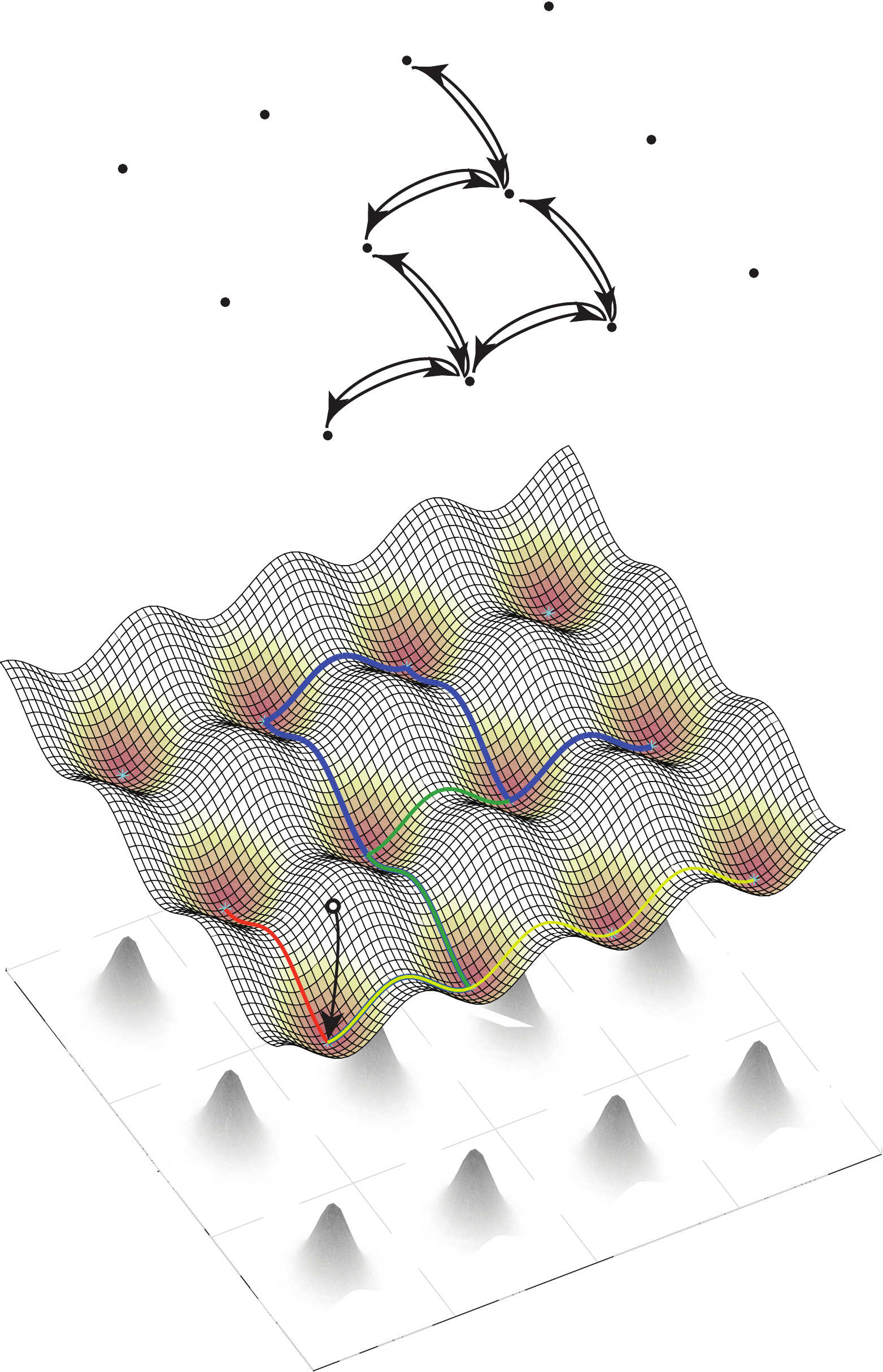} 
\caption{ 
  A multiscale population process with a lattice of fixed points.
  Middle layer is the grid of microstates and transitions in the
  mesoscale.  Surface $- \log \underline{\rho}$ for a stationary
  distribution is indicated by the color gradient.  Fixed points
  (asterisks) are at well bottoms.  Modes of $\underline{\rho}$
  concentrated near fixed points are indicated in greyscale beneath
  the surface.  A classical deterministic trajectory (black, with
  arrow) starts at the open circle and relaxes to the fixed point for
  that basin.  First-passage trajectories (colors) then mediate well
  switching by rare large deviations.  Fixed points of trajectory
  equations in the mesoscale are coarse-grained to become elementary
  microstates (black dots) in the macroscale (top layer), and first
  passages become elementary transition events (arrows).
  \label{fig:eggcrate_instantons_AI} 
}
\end{center}
\end{figure}

Models of this kind may be embedded recursively in scale through any
number of levels.  Here we will focus on three adjacent levels,
diagrammed in Fig.~\ref{fig:eggcrate_instantons_AI}, and on
the separation of timescales within a level, and the coarse-grainings
that define the maps between levels.  The middle level, termed the
\textit{mesoscale}, will be represented explicitly as a stochastic
process, and all results that come from large-deviations scaling will
be derived within this level.  The \textit{microscale} one level
below, and the \textit{macroscale} one level above, are described only
as needed to define the coarse-graining maps of variables between
adjacent levels.  Important properties such as bidirectionality of
escapes from metastable fixed point in the stationary distribution,
which the large-deviation analysis in the mesoscale supplies as
properties of elementary transitions in the macroscale, will be
assumed self-consistently as inputs from the microscale to the
mesoscale.

Within the description at a single level, the discrete population
states will be termed \textit{microstates}, as is standard in (both
classical and stochastic) thermodynamics.  Microstates are different
in kind from \textit{macrostates}, which correspond to a sub-class of
distributions (defined below), and from \textit{fixed points} of the
dynamical system.  There is no unique recipe for coarse-graining
descriptions to reduce dimensionality in a multi-scale system, because
the diversity of stochastic processes is vast.  Here, to obtain a
manageable terminology and class of models, we limit to cases in which
the dynamical-system fixed points at one level can be put in
correspondence with the microstates at the next higher level.
Table~\ref{tab:3scale_LD_mapping} shows the terms that arise within,
and correspondences between, levels.

\begin{table}[th]
\begin{tabular}[c]{|l|l|l|}
\hline 
  $\mbox{macro\textbf{scale}}$ & 
  $\mbox{meso\textbf{scale}}$ & 
  $\mbox{micro\textbf{scale}}$ \\
\hline
& 
  \parbox[t]{1.35in}{\raggedright
    micro\textbf{state} 
  } & 
  \parbox[t]{1.05in}{\raggedright
    H-J fixed point 
  } \\
  \rule{0pt}{4ex}
& 
  \parbox[t]{1.35in}{\raggedright
    thermalization of the microscale
  } & 
  \parbox[t]{1.05in}{\raggedright
   H-J relaxation trajectories 
  } \\
  \rule{0pt}{4ex}
& 
  \parbox[t]{1.35in}{\raggedright
    elementary transitions between microstates  
  } & 
  \parbox[t]{1.05in}{\raggedright
    H-J first-passages 
  } \\
  \rule{0pt}{4ex}
& 
  \parbox[t]{1.35in}{\raggedright
    arbitrary distribution on microstates  
  } & 
  \parbox[t]{1.05in}{\raggedright
    coarse-grained distribution on fixed points
  } \\
  \rule{0pt}{4ex}
& 
  \parbox[t]{1.45in}{\raggedright
    macro\textbf{state} (distribution) $\leftrightarrow$ 
    H-J variables
  } & \\
  \rule{0pt}{4ex}
  \parbox[t]{0.75in}{\raggedright
    micro\textbf{state} 
  } & 
  \parbox[t]{1.05in}{\raggedright
    H-J fixed point 
  } & 
\\ 
  \parbox[t]{0.75in}{\raggedright
    \ldots  
  } & 
  \parbox[t]{1.05in}{\raggedright
    \ldots 
  } & 
\\ 
\hline
\end{tabular}
\caption{ 
  Structures from micro to macro within and between levels.  H-J
  refers to the Hamilton-Jacobi dynamical system that arises in the
  large-deviation approximation.  Note that the distinction between
  micro\textbf{state} and macro\textbf{state} is a distinction of
  kind, which occurs within a level, whereas micro\textbf{scale},
  meso\textbf{scale}, and macro\textbf{scale} are designations of
  relative levels with respect to large-deviation timescale
  separations.  Terms in the same row are representations of the same
  quantity in different levels, related through coarse-graining.
}
\label{tab:3scale_LD_mapping}
\end{table}

\subsection{Models based on population processes}

The following elements furnish a description of a system: 

\paragraph{The multiscale distribution:}

The central object of study is any probability distribution $\rho$,
defined down to the smallest scale in the model, and the natural
coarse-grainings of $\rho$ produced by the dynamics.  To simplify
notation, we will write $\rho$ for the general distribution, across
all levels, and let the indexing of $\rho$ indicate which level is
being used in a given computation.

\paragraph{Fast relaxation to fixed points at the microscale:}

The counterpart, in our analysis of the mesoscale, to
Prigogine's~\cite{Glansdorff:structure:71,Prigogine:MT:98} assumption
of \textit{local equilibrium} in a bath, is fast relaxation of the
distribution on the microscale, to a distribution with modes around
the fixed points.  In the mesoscale these fixed points become the
elementary population states indexed $\rn$, and the coarse-grained
probability distribution is denoted ${\rho}_{\rn}$.  If a population
consists of individuals of \textit{types} indexed $p \in \left\{ 1 ,
\ldots , P \right\}$, then $\rn \equiv \left[ {\rn}_p \right]$ is a
vector in which the non-negative integer coefficient ${\rn}_p$ counts
the number of individuals of type $p$.

\paragraph{Elementary transitions in the mesoscale:}

First-passages occurring in the (implicit) large-deviations theory,
between fixed points ${\rn}^{\prime}$ and $\rn$ at the microscale,
appear in the mesoscale as elementary transitions ${\rn}^{\prime}
\rightarrow \rn$.  In Fig.~\ref{fig:eggcrate_instantons_AI}, the
elementary states are grid points and elementary transitions occur
along lines in the grid in the middle layer. We assume as input to the
mesoscale the usual condition of \textit{weak
reversibility}~\cite{Polettini:open_CNs_I:14}, meaning that if an
elementary transition ${\rn}^{\prime} \rightarrow \rn$ occurs with
nonzero rate, then the transition $\rn \rightarrow {\rn}^{\prime}$
also occurs with nonzero rate.  Weak reversibility is a property we
will derive for first passages within the mesoscale, motivating its
adoption at the lower level.

\paragraph{Thermalization in the microscale:} 

The elementary transitions ${\rn}^{\prime} \leftrightarrow \rn$ are
separated by typical intervals exponentially longer in some scale
factor than the typical time in which a single transition completes.
(In the large-deviation theory, they are \textit{instantons}.)  That
timescale separation defines \textit{thermalization} at the
microscale, and makes microscale fluctuations conditionally
independent of each other, given the index $\rn$ of the basin in which
they occur.  Thermalization also decouples components ${\rn}_p$ in the
mesoscale at different $p$ except through the allowed elementary state
transitions.

\paragraph{A System/Environment partition within the mesoscale:}

Generally, in addition to considering the thermalized microscale a
part of the ``environment'' in which mesoscale stochastic events take
place, we will choose some partition of the type-indices $p$ to
distinguish one subset, called the \textit{system} ($s$), from one or
more other subsets that also form part of the \textit{environment}
($e$).  Unlike the thermalized microscale, the environment-part in the
mesoscale is slow and explicitly stochastic, like the system.  The
vector $\rn$ indexes a tensor product space $s \otimes e$, so we write
$\rn \equiv \left( {\rn}_s , {\rn}_{e} \right)$.  

\paragraph{Marginal and conditional distributions in system and
environment:} 

On $s \otimes e$, ${\rho}_{\rn}$ is a joint distribution.  A
\textit{marginal} distribution ${\rho}^s$ for the system is defined by
${\rho}^s_{{\rn}_s} \equiv \sum_{\rn \mid {\rn}_s} {\rho}_{\rn}$,
where $\rn \mid {\rn}_s$ fixes the $s$ component of $\rn$ in the sum.
From the joint and the marginal a conditional distribution ${\rho}^{e
\mid s}$ at each ${\rn}_s$ is given by ${\rho}_{\rn = \left( {\rn}_s ,
{\rn}_e \right)} \equiv {\rho}^s_{{\rn}_s} {\rho}^{e \mid s}_{\rn}$.
The components of ${\rho}^{e \mid s}$ can fill the role often given to
chemostats in open-system models of CRNs.  Here we keep them as
explicit distributions, potentially having dynamics that can respond
to changes in $s$.  

\paragraph{Notations involving pairs of indices:}

Several different sums over the pairs of indices associated with state
transitions appear in the following derivations.  To make equations
easier to read, the following notations are used throughout:
\begin{trivlist}

\item $\left< \rn , {\rn}^{\prime} \right>$ is an unordered pair of
indices.

\item $\sum_{\rn} \sum_{{\rn}^{\prime} \neq \rn}$ counts every pair
in both orders.  \\ $\sum_{\left< \rn , {\rn}^{\prime} \right>}$
counts every unordered pair once.

\item Therefore for any function $f \! \left( \rn , {\rn}^{\prime}
\right)$, $\sum_{\rn} \sum_{{\rn}^{\prime} \neq \rn} f \! \left( \rn ,
{\rn}^{\prime} \right) = \sum_{\left< \rn , {\rn}^{\prime} \right>}
\left[ f \! \left( \rn , {\rn}^{\prime} \right) + f \! \left(
{\rn}^{\prime} , \rn \right) \right]$.  

\item $\sum_{\rn \mid {\rn}_s}$ is a sum on the ${\rn}_e$ component of
$\rn = \left( {\rn}_s , {\rn}_e \right)$.  

\item $\sum_{\left< \rn , {\rn}^{\prime} \right> \mid {\rn}_s}$ counts
all unordered pairs $\left< \rn , {\rn}^{\prime} \right>$ with common
$s$-component ${\rn}_s$.

\end{trivlist}

\subsection{Stochastic description within the mesoscale}

The coarse-grained distribution ${\rho}_{\rn}$ at the mesoscale
evolves in time under a master equation 
\begin{align}
  {\dot{\rho}}_{\rn} = 
  \sum_{{\rn}^{\prime}}
  {\mathbb{T}}_{\rn {\rn}^{\prime}}
  {\rho}_{{\rn}^{\prime}}
& = 
  \sum_{{\rn}^{\prime} \neq \rn}
  \left( 
    w_{\rn {\rn}^{\prime}} {\rho}_{{\rn}^{\prime}} - 
    w_{{\rn}^{\prime} \rn} {\rho}_{\rn}
  \right) . 
\label{eq:rhodot_genform}
\end{align}
Here and below, $(\dot{\mbox{ }})$ indicates the time derivative.  The
generator $\mathbb{T}$ is a stochastic matrix on the left, which we
write $1^T \mathbb{T} = 0$, where $1^T$ is the row-vector on index
$\rn$ corresponding to the uniform (unnormalized) measure.  $w_{\rn
{\rn}^{\prime}}$ (following a standard
notation~\cite{Polettini:open_CNs_I:14}) is the component of
$\mathbb{T}$ giving the transition rate from state ${\rn}^{\prime}$ to
state $\rn$.

For all of what follows, it will be necessary to restrict to systems
that possess a stationary, normalizable distribution denoted
$\underline{\rho}$, satisfying $\mathbb{T} \underline{\rho} = 0$.  The
stationary distribution will take the place of conservation laws as
the basis for the definition of macrostates.  Moreover, if
$\underline{\rho}$ is everywhere continuous, and the number of
distinct events generating transitions in the mesoscale (in a sense
made precise below) is finite, first passages between fixed points
corresponding to modes of $\underline{\rho}$ will occur at rates
satisfying a condition of detailed balance.  The joint, marginal, and
conditional stationary distributions are denoted 
${\underline{\rho}}_{\rn = \left( {\rn}_s , {\rn}_e \right)} \equiv
{\underline{\rho}}^s_{{\rn}_s} {\underline{\rho}}^{e \mid s}_{\rn}$.

\paragraph{The marginal stochastic process on $s$:}

The system-marginal distribution ${\rho}^s$ evolves under a master
equation ${\dot{\rho}}^s = {\mathbb{T}}^s \! \left( {\rho}^{e \mid s}
\right) {\rho}^s$, for which the transition matrix ${\mathbb{T}}^s$
has components that are functions of the instantaneous environmental
distribution, given by  
\begin{align}
  w^s_{{\rn}_s {\rn}^{\prime}_s} \! 
  \left( {\rho}^{e \mid s} \right) 
& \equiv 
  \sum_{\rn \mid {\rn}_s} 
  \sum_{{\rn}^{\prime} \mid {\rn}^{\prime}_s} 
  w_{\rn {\rn}^{\prime}} 
  {\rho}^{e \mid s}_{{\rn}^{\prime}} . 
\label{eq:ws_def}
\end{align}

\paragraph{Time-independent overall stationary distribution as a
reference:} 

Now we make an assumption that is crucial to being able to define a
Lyapunov function for the whole multi-level system distribution
$\rho$: namely, that the parameters in the mesoscale transition matrix
$\mathbb{T}$ and hence the stationary distribution $\underline{\rho}$
are time-independent.  All dynamics is made explicit as dynamics of
distributions ${\rho}^s$ and ${\rho}^{e \mid s}$, but the
explicitly-written distributions are the \emph{only} source of
dynamics: once the microscale has thermalized, there are no further
un-written sources of time-dependence in the system.

\paragraph{Detailed balance propagating up from the microscale:}

Finally we assume, propagating up from the microscale, a condition of
detailed balance that we will prove as a property of first-passages in
the mesoscale, and then apply recursively: 
\begin{align}
  w_{\rn {\rn}^{\prime}} {\underline{\rho}}_{{\rn}^{\prime}}
& = 
  w_{{\rn}^{\prime} \rn} {\underline{\rho}}_{\rn} . 
\label{eq:DB_mesoscale}
\end{align}
Note that condition~(\ref{eq:DB_mesoscale}) is \emph{not} an
assumption of microscopic reversibility in whatever faster stochastic
process is operating below in the microscale.  To understand why,
using constructions that will be carried out explicitly within the
mesoscale, see that even with rate constants satisfying
Eq.~(\ref{eq:DB_mesoscale}), the system-marginal transition
rates~(\ref{eq:ws_def}) need not satisfy a condition of detailed
balance.  Indeed we will want to be able to work in limits for the
environment's conditional distributions ${\rho}^{e \mid s}$ in which
some transitions can be made completely irreversible: that is
$w^s_{{\rn}_s {\rn}^{\prime}_s} \neq 0$ but $w^s_{{\rn}^{\prime}_s
{\rn}_s} = 0$.  Even from such irreversible dynamics among
microstates, first-passage rates with detailed balance in the
stationary distribution will result, and it is that property that is
assumed in Eq.~(\ref{eq:DB_mesoscale}).

From these assumptions on $\mathbb{T}$ and $\underline{\rho}$, it
follows that the relative entropy of any distribution $\rho$ from the
stationary distribution $\underline{\rho}$ is non-decreasing,
\begin{align}
  - \dot{D} \! \left( \rho \parallel \underline{\rho} \right)
& = 
  \sum_{\rn} 
  \sum_{{\rn}^{\prime} \neq \rn}
  \log
  \left(
    \frac{
      {\left( \rho / \underline{\rho} \right)}_{{\rn}^{\prime}}
    }{
      {\left( \rho / \underline{\rho} \right)}_{\rn}
    }
  \right) 
  w_{\rn {\rn}^{\prime}} {\rho}_{{\rn}^{\prime}}
\nonumber \\ 
& = 
  \sum_{\rn} 
  \sum_{{\rn}^{\prime} \neq \rn}
  \log
  \left(
    \frac{
      w_{\rn {\rn}^{\prime}} {\rho}_{{\rn}^{\prime}}
    }{
      w_{{\rn}^{\prime} \rn} {\rho}_{\rn}
    }
  \right) 
  w_{\rn {\rn}^{\prime}} {\rho}_{{\rn}^{\prime}}
\nonumber \\ 
& = 
  \sum_{\left< \rn , {\rn}^{\prime} \right>}
  \log
  \left(
    \frac{
      w_{\rn {\rn}^{\prime}} {\rho}_{{\rn}^{\prime}}
    }{
      w_{{\rn}^{\prime} \rn} {\rho}_{\rn}
    }
  \right) 
  \left( 
    w_{\rn {\rn}^{\prime}} {\rho}_{{\rn}^{\prime}} - 
    w_{{\rn}^{\prime} \rn} {\rho}_{\rn}
  \right) \ge 0 . 
\label{eq:Ddot_tot_genform}
\end{align}
The result is elementary for systems with detailed balance, because
each term in the third line of Eq.~(\ref{eq:Ddot_tot_genform}) is
individually non-negative.  We use \textit{relative entropy} to refer
to minus the Kullback-Leibler divergence of $\rho$ from
$\underline{\rho}$~\cite{Cover:EIT:91}, to follow the usual sign
convention for a non-decreasing entropy.

\subsection{System-environment decompositions of the entropy change}
\label{sec:sys_env_decomp}

The exchange of heat for work is central to classical thermodynamics
because energy conservation is a constraint on joint configurations
across sub-systems, either multiple thermal systems in contact or a
mechanical subsystem having only deterministic variables, some of
which set the boundary conditions on thermal subsystems that also host
fluctuations.  Only the state-function entropy, however, is a
``function'' of energy in any sense, so the only notion of a limiting
partition of irreversible effects between subsystems derivable from
energy conservation is the one defined by adiabatic transformations
passing through sequences of macrostates.

In more general cases, with or without conservation laws, the boundary
conditions on a system are imposed only through the elements of the
marginal transition matrix ${\mathbb{T}}^s$.  The problem remains, of
understanding how one subsystem can limit entropy change in another
through a boundary, but it is no longer organized with reference to
adiabatic transformations. 

We wish to understand what constitutes a thermodynamically natural
decomposition of the mesoscale process into a system and an
environment.  A widely-adopted
decomposition~\cite{Seifert:stoch_thermo_rev:12,Polettini:open_CNs_I:14,Polettini:stoch_macro_thermo:16}
for systems with energy conservation\footnote{The decomposition is the
same one used to define an energy cost of
computation~\cite{Landauer:IHGCP:61,Bennett:TC:82} by constructing
logical states through the analogues to heat
engines~\cite{Szilard:MD:29,Smith:NS_thermo_II:08}.}  separates a
Shannon entropy of ${\rho}_{\rn}^s$ from heat generation associated
with terms $-\log w^s_{{\rn}_s {\rn}^{\prime}_s}$ by the local
equilibrium assumption for the bath.  We begin by writing down this
information/heat decomposition, and arguing that it is not the natural
partition with respect to irreversibility.

\paragraph{Entropy relative to the stationary state rather than
Shannon entropy:}

The following will differ from the usual construction in replacing
Shannon entropy with a suitable relative entropy, without changing the
essence of the decomposition.  Two arguments can be given for
preferring the relative entropy: It would be clear, for a system with
a continuous state space in which ${\rho}_{\rn}$ would become a
density, that the logarithm of a dimensional quantity is undefined.
Hence some reference measure is always implicitly assumed.  A uniform
measure is not a coordinate-invariant concept, and a measure that is
uniform in one coordinate system makes those coordinates part of the
system specification.  Since discrete processes are often used as
approximations to continuum limits, the same concerns apply.  The more
general lesson is that a logarithmic entropy unit is always given
meaning with respect to some measure.  Only for systems such as symbol
strings, for which a combinatorial measure on integers is the natural
measure, is Shannon entropy the corresponding natural entropy.  For
other cases, such as CRNs, the natural entropy is relative entropy
referenced to the Gibbs equilibrium~\cite{Smith:NS_thermo_I:08}, and
its change gives the dissipation of chemical work.  For the processes
described here, the counterpart to Shannon entropy that solves these
consistency requirements, but does not yet address the question of
naturalness, is the relative entropy referenced to the steady state
marginal ${\underline{\rho}}^s$.  Its time derivative is given by
\begin{align}
  - \dot{D} \! \left( {\rho}^s \parallel {\underline{\rho}}^s \right)
& = 
  \sum_{\left< {\rn}_s , {\rn}^{\prime}_s \right>}
  \log
  \left(
    \frac{
      {\left( \rho / \underline{\rho} \right)}^s_{{\rn}^{\prime}_s}
    }{
      {\left( \rho / \underline{\rho} \right)}^s_{{\rn}_s}
    }
  \right) 
  \left( 
    w^s_{{\rn}_s {\rn}^{\prime}_s} {\rho}^s_{{\rn}^{\prime}_s} - 
    w^s_{{\rn}^{\prime}_s {\rn}_s} {\rho}^s_{{\rn}_s}
  \right) . 
\label{eq:Ddot_tot_genform_s}
\end{align}
The quantity~(\ref{eq:Ddot_tot_genform_s}) need not be either positive
or negative in general. 

A second term that separates out of the change in total relative
entropy~(\ref{eq:Ddot_tot_genform}) comes from changes in
environmental states through events that do not result in net change
of the system state.\footnote{Note $w_{\rn {\rn}^{\prime}}$ may depend
on the system index-component ${\rn}_s$ shared by both $\rn$ and
${\rn}^{\prime}$, so these rates can depend on system state.
Catalysis acts through such dependencies.}  The relative entropy of
the conditional distribution ${\rho}^{e \mid s}$ at a particular index
${\rn}_s$ from its stationary reference ${\underline{\rho}}^{e \mid
s}$ has time derivative
\begin{align}
\lefteqn{
  - {\dot{D}}^{e \mid s}_{{\rn}_s} \! 
  \left( 
    {\rho}^{e \mid s} \parallel {\underline{\rho}}^{e \mid s}
  \right) = 
} & 
\nonumber \\ 
& 
  \sum_{\left< {\rn}_s , {\rn}^{\prime}_s \right> \mid {\rn}_s}
  \log
  \left(
    \frac{
      {\left( \rho / \underline{\rho} \right)}^{e \mid s}_{{\rn}^{\prime}}
    }{
      {\left( \rho / \underline{\rho} \right)}^{e \mid s}_{\rn}
    }
  \right) 
  \left( 
    w_{\rn {\rn}^{\prime}} {\rho}^{e \mid s}_{{\rn}^{\prime}} - 
    w_{{\rn}^{\prime} \rn} {\rho}^{e \mid s}_{\rn}
  \right) 
\nonumber \\ 
& 
  \sum_{\left< {\rn}_s , {\rn}^{\prime}_s \right> \mid {\rn}_s}
  \log
  \left(
    \frac{
      w_{\rn {\rn}^{\prime}} {\rho}^{e \mid s}_{{\rn}^{\prime}} 
    }{
      w_{{\rn}^{\prime} \rn} {\rho}^{e \mid s}_{\rn}
    }
  \right) 
  \left( 
    w_{\rn {\rn}^{\prime}} {\rho}^{e \mid s}_{{\rn}^{\prime}} - 
    w_{{\rn}^{\prime} \rn} {\rho}^{e \mid s}_{\rn}
  \right) . 
\label{eq:Ddot_emids_genform_s}
\end{align}
Unlike the change of system relative
entropy~(\ref{eq:Ddot_tot_genform_s}), 
Eq.~(\ref{eq:Ddot_emids_genform_s}) is non-negative term-by-term, in
the same way as Eq.~(\ref{eq:Ddot_tot_genform}).  

The remaining terms to complete the entropy
change~(\ref{eq:Ddot_tot_genform}) come from joint transformations in
system and environment indices ${\rn}_s$ and ${\rn}_e$, and in usual
treatments have the interpretation of \textit{dissipated
heats}.\footnote{When more than one environment transition couples to
the same system transition, there can be reasons to further partition
these terms; an example is given in Sec.~\ref{sec:Apol_hydr}.}  They
are functions of the pair of indices $\left( {\rn}^{\prime}_s ,
{\rn}_s \right)$.  An average change in relative entropy of the
environment, over all processes that couple to a given system
state-change, is
\begin{align}
  {\sigma}_{{\rn}_s {\rn}^{\prime}_s} \! 
  \left( {\rho}^{e \mid s} \right) 
& \equiv 
  \frac{1}{w^s_{{\rn}_s {\rn}^{\prime}_s}}
  \sum_{\rn \mid {\rn}_s}
  \sum_{{\rn}^{\prime} \mid {\rn}^{\prime}_s}
  \log
  \left(
    \frac{
      {\left( \rho / \underline{\rho} \right)}^{e \mid s}_{{\rn}^{\prime}}
    }{
      {\left( \rho / \underline{\rho} \right)}^{e \mid s}_{\rn}
    }
  \right) 
  w_{\rn {\rn}^{\prime}} 
  {\rho}^{e \mid s}_{{\rn}^{\prime}} . 
\label{eq:sigdot_def}
\end{align}

(Note that if we had wished to use the un-referenced Shannon entropy
$- \sum_{{\rn}_s} {\rho}^s_{{\rn}_s} \log {\rho}^s_{{\rn}_s}$ in place
of the relative entropy~(\ref{eq:Ddot_tot_genform_s}) -- for instance,
in an application to digital computing -- we could shift the measures
${\underline{\rho}}^s$ to the dissipation term to produce what is
normally considered the ``environmental'' heat dissipation, given by
\begin{align}
  {\sigma}^{\rm env}_{{\rn}_s {\rn}^{\prime}_s} 
  w^s_{{\rn}_s {\rn}^{\prime}_s}
& \equiv 
  \left[ 
    {\sigma}_{{\rn}_s {\rn}^{\prime}_s} - 
    \log 
    \left( 
      \frac{
        {\underline{\rho}}^s_{{\rn}^{\prime}_s}
      }{
        {\underline{\rho}}^s_{{\rn}_s}
      } 
    \right)    
  \right] 
  w^s_{{\rn}_s {\rn}^{\prime}_s}
\nonumber \\
& = 
  \sum_{\rn \mid {\rn}_s}
  \sum_{{\rn}^{\prime} \mid {\rn}^{\prime}_s}
  \log
  \left(
    \frac{
      w_{\rn {\rn}^{\prime}} 
      {\rho}^{e \mid s}_{{\rn}^{\prime}} 
    }{
      w_{{\rn}^{\prime} \rn} 
      {\rho}^{e \mid s}_{{\rn}} 
    }
  \right) 
  w_{\rn {\rn}^{\prime}} 
  {\rho}^{e \mid s}_{{\rn}^{\prime}} . 
\label{eq:sigdot_env_def}
\end{align}
The quantity~(\ref{eq:sigdot_env_def}) is regarded as a property of
the environment (both slow variables and the thermal bath) because it
is a function only of the transition rates $w_{{\rn}^{\prime} \rn}$
and of the marginal distributions ${\rho}^{e \mid s}$.)

\subsubsection{The information/heat decomposition of total
relative-entropy change}

Eq.~(\ref{eq:Ddot_tot_genform}) is decomposed in terms of the
quantities in
equations~(\ref{eq:Ddot_tot_genform_s}--\ref{eq:sigdot_def}) as 
\begin{align}
  - \dot{D} \! \left( \rho \parallel \underline{\rho} \right)
& = 
  - \dot{D} \! \left( {\rho}^s \parallel {\underline{\rho}}^s \right)
\nonumber \\ 
& \mbox{} + 
  \sum_{{\rn}_s}
  \sum_{{\rn}^{\prime}_s \neq {\rn}_s}
  {\sigma}_{{\rn}_s {\rn}^{\prime}_s} \! 
  w^s_{{\rn}_s {\rn}^{\prime}_s} {\rho}^s_{{\rn}^{\prime}_s} 
\nonumber \\ 
& \mbox{} - 
  \sum_{{\rn}_s}
  {\rho}^s_{{\rn}_s} 
  {\dot{D}}^{e \mid s}_{{\rn}_s} \! 
  \left( 
    {\rho}^{e \mid s} \parallel {\underline{\rho}}^{e \mid s}
  \right) . 
\label{eq:Ddot_tot_decomp_raw}
\end{align}
The total is non-negative and the third line is independently
non-negative, as already mentioned.  The sum of the first two lines is
also non-negative, a result that can be proved as a
\textit{fluctuation theorem} for what is normally called ``total
entropy change''~\cite{Esposito:fluct_theorems:10}.\footnote{More
detailed proofs for a decomposition of the same sum will be given
below.}

Here we encounter the first property that makes a decomposition of a
thermal system ``natural''.  The term in ${\dot{D}}^{e \mid
s}_{{\rn}_s} \! \left( {\rho}^{e \mid s} \parallel
{\underline{\rho}}^{e \mid s} \right)$ is not generally considered,
and it \emph{does not need to be considered}, because thermal
relaxation at the microscale makes transitions in ${\rn}_e$ at fixed
${\rn}_s$ conditionally independent of transitions that change
${\rn}_s$.  Total relative entropy changes as a sum of two
independently non-decreasing contributions.

The first and second lines in Eq.~(\ref{eq:Ddot_tot_decomp_raw}) are
not likewise independently non-negative.  Negative values of the
second or first line, respectively, describe phenomena such as
randomization-driven endothermic reactions, or heat-driven information
generators.  To the extent that they do not use thermalization in the
microscale to make the system and environment conditionally
independent, as the third term in Eq.~(\ref{eq:Ddot_tot_decomp_raw})
is independent, we say they do not provide a natural
system/environment decomposition.

\subsubsection{Relative entropy referencing the system steady state at
instantaneous parameters}

Remarkably, a natural division does exist, based on the
\textit{housekeeping heat} introduced by Hatano and
Sasa~\cite{Hatano:NESS_Langevin:01}.  The decomposition uses the
solution ${\bar{\rho}}^s$ to ${\mathbb{T}}^s {\bar{\rho}}^s = 0$,
which would be the stationary marginal distribution for the system $s$
at the instantaneous value of ${\rho}^{e \mid s}$.  As for the
whole-system stationary distribution $\underline{\rho}$, we restrict
to cases in which ${\bar{\rho}}^s$ exists and is
normalizable.\footnote{This can be a significant further restriction
when we wish to study limits of sequences of environments to model
chemostats.  For systems with unbounded state spaces, such as arise in
polymerization models, it is quite natural for a chemostat-driven
system to possess no normalizable steady-state distributions.  In such
cases other methods of analysis must be
used~\cite{Esposito:copol_eff:10}.}

Treating ${\bar{\rho}}^s$ as fixed and considering only the dynamics
of ${\rho}^s$ with ${\bar{\rho}}^s$ as a reference, we may consider
a time derivative $\dot{D} \! \left( {\rho}^s \parallel {\bar{\rho}}^s
\right)$ in place of Eq.~(\ref{eq:Ddot_tot_genform_s}).  Note that the
rates $w^s_{{\rn}_s {\rn}^{\prime}_s}$ in the transition matrix
${\mathbb{T}}^s$ no longer need satisfy any simplified balance
condition in relation to ${\bar{\rho}}^s$, such as detailed balance.
Non-negativity of $- \dot{D} \! \left( {\rho}^s \parallel
{\bar{\rho}}^s \right)$ was proved by
Schnakenberg~\cite{Schnakenberg:ME_graphs:76} by an argument that
applies to discrete population processes with normalized stationary
distributions of the kind assumed here.  We will derive a slightly
more detailed decomposition proving this result for the case of CRNs,
in a later section.

The dissipation term that complements the change in $\dot{D} \!
\left( {\rho}^s \parallel {\bar{\rho}}^s \right)$ is obtained by
shifting the ``environmental'' entropy
change~(\ref{eq:sigdot_env_def}) by $\log {\bar{\rho}}^s$ to obtain
the housekeeping heat\footnote{These terms are used because this is
how they are known.  No energy interpretation is assumed here, so a
better term would be ``housekeeping entropy change''.}
\begin{align}
  {\sigma}^{\rm HK}_{{\rn}_s {\rn}^{\prime}_s} 
  w^s_{{\rn}_s {\rn}^{\prime}_s}
& \equiv 
  \left[ 
    {\sigma}^{\rm env}_{{\rn}_s {\rn}^{\prime}_s} + 
    \log 
    \left( 
      \frac{
        {\bar{\rho}}^s_{{\rn}^{\prime}_s}
      }{
        {\bar{\rho}}^s_{{\rn}_s}
      } 
    \right)    
  \right] 
  w^s_{{\rn}_s {\rn}^{\prime}_s}
\nonumber \\
& = 
  \sum_{\rn \mid {\rn}_s}
  \sum_{{\rn}^{\prime} \mid {\rn}^{\prime}_s}
  \log
  \left(
    \frac{
      {\bar{\rho}}^s_{{\rn}^{\prime}_s} 
      {\rho}^{e \mid s}_{{\rn}^{\prime}} / 
      \underline{\rho}_{{\rn}^{\prime}}
    }{
      {\bar{\rho}}^s_{{\rn}_s} 
      {\rho}^{e \mid s}_{{\rn}} / 
      \underline{\rho}_{\rn}
    }
  \right) 
  w_{\rn {\rn}^{\prime}} 
  {\rho}^{e \mid s}_{{\rn}^{\prime}} . 
\label{eq:sigdot_HK_def}
\end{align}
Non-negativity of Eq.~(\ref{eq:sigdot_HK_def}) is implied by a
fluctuation theorem~\cite{Speck:HS_heat_FT:05}, and a time-local proof
with interesting further structure for CRNs will be given below.

The total change in relative entropy~(\ref{eq:Ddot_tot_genform}) is
then the sum  
\begin{align}
  - \dot{D} \! \left( \rho \parallel \underline{\rho} \right)
& = 
  - \dot{D} \! \left( {\rho}^s \parallel {\bar{\rho}}^s \right)
\nonumber \\ 
& \mbox{} + 
  \sum_{{\rn}_s}
  \sum_{{\rn}^{\prime}_s \neq {\rn}_s}
  {\sigma}^{\rm HK}_{{\rn}_s {\rn}^{\prime}_s} \! 
  w^s_{{\rn}_s {\rn}^{\prime}_s} {\rho}^s_{{\rn}^{\prime}_s} 
\nonumber \\ 
& \mbox{} - 
  \sum_{{\rn}_s}
  {\rho}^s_{{\rn}_s} 
  {\dot{D}}^{e \mid s}_{{\rn}_s} \! 
  \left( 
    {\rho}^{e \mid s} \parallel {\underline{\rho}}^{e \mid s}
  \right) . 
\label{eq:Ddot_tot_decomp_HK}
\end{align}
Each line in Eq.~(\ref{eq:Ddot_tot_decomp_HK}) is now independently
non-negative.  The first measures a gain of entropy within the system
$s$, conditionally independent of changes in the environment given the
marginal transition matrix ${\mathbb{T}}^s$.  The third measures a
gain of entropy in the environment $e$ independent of any changes in
the system at all.  The second, housekeeping entropy rate, measures a
change of entropy in the environment that is conditionally independent
of changes of entropy within the system, given ${\mathbb{T}}^s$ as
represented in ${\bar{\rho}}^s$.  Any of the terms may be changed,
holding the conditioning data ${\bar{\rho}}^s$ or ${\rn}_s$, or
omitted, without changing the limits for the others.  They respect the
conditional independence created by thermalization in the microscale,
and by that criterion constitute a natural decomposition of the
system.

\subsubsection{Intrinsic and extrinsic thermodynamics}

We take ${\mathbb{T}}^s$ and $D \! \left( {\rho}^s \parallel
{\bar{\rho}}^s \right)$ as a specification of the \textit{intrinsic
thermodynamics} of the system $s$, analogous to the role of intrinsic
curvature of a manifold in differential geometry.  The vector (indexed
by pairs of system indices) of housekeeping entropy differentials,
$\left\{ {\sigma}^{\rm HK}_{{\rn}_s {\rn}^{\prime}_s} \right\}$,
correspondingly defines the way $s$ is \textit{thermodynamically
embedded} in the mesoscale system, analogous to the role of components
of an embedding curvature for a submanifold within larger manifold.

\subsubsection{System Hartley information as a temporal connection}

The natural decomposition~(\ref{eq:Ddot_tot_decomp_HK}) differs from
the information/heat decomposition~(\ref{eq:Ddot_tot_decomp_raw}) in
what is regarded as inherent to the system versus the environment.
The attribution of Shannon entropy as a ``system'' property follows
from the fact that it involves only ${\rho}^s_{{\rn}_s}$, and its
change counts only actual transitions ${\rn}_s^{\prime} \rightarrow
{\rn}_s$ with rate $w^s_{{\rn}_s {\rn}_s^{\prime}}$.  Likewise, the
``environment'' heat~(\ref{eq:sigdot_env_def}) is a function only of
the actual distribution ${\rho}^{e \mid s}$ and the realized
currents.  

Those events that occur \emph{within} $s$ or $e$, however, fail to
capture the additional features of ${\mathbb{T}}^s$: that specific
transitions are coupled \emph{as} a system, and that they have the
dependence on ${\rho}^{e \mid s}$ of Eq.~(\ref{eq:ws_def}).  The
stationary distribution ${\bar{\rho}}^s$ is the function of
${\mathbb{T}}^s$ reflecting its status as a system. 

The differences between the three entropy-change
terms~(\ref{eq:sigdot_def},\ref{eq:sigdot_env_def},\ref{eq:sigdot_HK_def})
are differences of the \textit{Hartley
informations}~\cite{Hartley:information:28}, respectively for
${\underline{\rho}}^s$ or ${\bar{\rho}}^s$.  In the natural
decomposition~(\ref{eq:Ddot_tot_decomp_HK}), they are not acted upon
by the time derivative, but rather define the tangent plane of zero
change for $\log \rho$ terms that are acted upon by the transition
matrix, and resemble connection coefficients specifying parallel
transport in differential geometry.

\section{Hamilton-Jacobi theory for large deviations}
\label{sec:Ham_Jacob}

The central concepts in thermodynamics are that of the macrostate, and
of the entropy as a state function from which properties of
macrostates and constraints on their transformations are derived.  In
classical thermodynamics~\cite{Fermi:TD:56,Kittel:TP:80}, macrostates
are introduced in association with average values of conserved
quantities (\textit{e.g.}~energy, particle numbers), because
conservation laws naturally generate conditional independence between
subsystems in contact, given a simple boundary condition (the
partition of the conserved quantity between them).

Here we wish to separate the construction that \emph{defines} a
macrostate from the properties that make one or another class of
macrostates dynamically \emph{robust} in a given system (conservation
laws are important for the latter).  The defining construction can be
quite general, but it must in all cases create a dimensional reduction
by an indefinite (or in the limit infinite) factor, from the
dimensionality of the microstate space that is definite but
arbitrarily large, to the dimensionality of a space of macrostate
variables that is fixed and independent of the dimensionality of
microstates.  Only dimensional reductions of this kind are compatible
with the large-deviation definition of macroworlds as worlds in which
structure can be characterized asymptotically separate from
scale~\cite{Touchette:large_dev:09}.  Robustness can then be
characterized separately within the large-deviation analysis in terms
of closure approximations or spectra of relaxation times for various
classes of macrostates.

Dimensional reduction by an indefinite degree is achieved by
associating macrostates with particular classes of
\textit{distributions} over microstates: namely, those distributions
produced in exponential families to define generating functions.  The
coordinates in the tilting weights that define the family are
independent of the dimension of the microstate space for a given
family and become the intensive state variables
(see~\cite{Amari:inf_geom:01}).  They are related by Legendre
transform to deviations that are the dual extensive state variables.
Relative entropies such as $- D \! \left( \rho \parallel
\underline{\rho} \right)$, defined as functionals on arbitrary
distributions, and dual under Legendre transform to suitable
cumulant-generating functions, become \textit{state functions} when
restricted to the distributions for macrostates.  The extensive state
variables are their arguments and the intensive state variables their
gradients.  Legendre duality leads to a system of Hamiltonian
equations~\cite{Bertini:macro_NEThermo:09} for time evolution of
macrostate variables, and from these the large-deviation scaling
behavior, timescale structure, and moment closure or other properties
of the chosen system of macrostates are derived.

In the treatment of Sec.~\ref{sec:multi_level_sys}, the relative
entropy increased deterministically without reference to any
particular level of system timescales, or the size-scale factors
associated with various levels.  As such it fulfilled the Lyapunov
role of (minus) the entropy, but not the large-deviation role that is
the other defining characteristic of
entropy~\cite{Ellis:ELDSM:85,Touchette:large_dev:09}.  The selection
of a subclass of distributions as macrostates introduces
level-dependence, scale-dependence, and the large-deviation role of
entropy, and lets us construct the relation between the Lyapunov and
large-deviation roles of entropy for macroworlds, which are generally
distinct.  As a quantity capable of fluctuations, the macrostate
entropy can decrease along subsets of classical trajectories; these
fluctuations are the objects of study in stochastic thermodynamics.
The Hamiltonian dynamical system is a particularly clarifying
representation for the way it separates relaxation and fluctuation
trajectories for macrostates into distinct sub-manifolds.  In
distinguishing the unconditional versus conditional nature of the two
kinds of histories, it shows how these macro-fluctuations are not
``violations'' of the 2nd law, but rather a partitioning of the
elementary events through which the only properly formulated 2nd law
is realized.\footnote{See a brief discussion making essentially this
point in Sec.~1.2 of~\cite{Seifert:stoch_thermo_rev:12}.  Seifert
refers to the 2nd law as characterizing ``mean entropy production'',
in keeping with other interpretations of entropies such as the Hartley
information in terms of heat.  The characterization adopted here is
more categorical: the Hartley function and its mean, Shannon
information, are not quantities with the same interpretation;
likewise, the entropy change~(\ref{eq:Ddot_tot_decomp_HK}) is the only
entropy relative to the boundary conditions in $\mathbb{T}$ that is
the object of a well-formulated 2nd law.  The ``entropy productions''
resulting from the Prigogine local-equilibrium assumption are
conditional entropies for macrostates defined through various
large-deviation functions, shown explicitly below.}

\subsection{Generating functions, Liouville equation, and the
Hamilton-Jacobi construction for saddle points}

\subsubsection{Relation of the Liouville operator to the
cumulant-generating function}

The $P$-dimensional Laplace transform of a distribution ${\rho}_{\rn}$
on discrete population states gives the moment-generating function
(MGF) for
the species-number counts ${\left\{ {\rn}_p \right\} }_{p \in 1 ,
\ldots , P}$.  For time-dependent problems, it is convenient to work
in the formal Doi operator algebra for generating functions, in which
a vector of raising operators $a^{\dagger} \equiv \left[ a^{\dagger}_p
\right]$ are the arguments of the MGF, and a conjugate vector of
lowering operators $a \equiv \left[ a_p \right]$ are the formal
counterparts to $\partial / \partial a^{\dagger}$.  MGFs are written
as vectors 
$\left| \rho \right) \equiv \sum_{\rn} {\rho}_{\rn}
{\left( a^{\dagger} \right)}^{\rn} \left| 0 \right) \equiv 
\sum_{\rn} {\rho}_{\rn} \left| \rn \right)$ in a Hilbert space built
upon a ground state $\left| 0 \right)$, and the commutation relations
of the raising and lowering operators acting in the Hilbert space are
$\left[ a_p , a^{\dagger}_q \right] = {\delta}_{pq}$.

The basis vectors corresponding to specific population states are
denoted $\left| \rn \right)$.  They are eigenvectors of the number
operators $a^{\dagger}_p a_p$ with eigenvalues ${\rn}_p$:
\begin{align}
  a^{\dagger}_p a_p  
  \left| \rn \right) 
& = 
  {\rn}_p  \left| \rn \right) . 
\label{eq:number_state_def}
\end{align}
Through by-now-standard constructions~\cite{Smith:LDP_SEA:11}, the
master equation $\dot{\rho} = \mathbb{T} \rho$ is converted to a
Liouville equation for time evolution of the MGF,
\begin{align}
  \frac{d}{dt}
  \left| \rho \right) 
& = 
  - \mathcal{L} \! \left( a^{\dagger} , a \right)
  \left| \rho \right) , 
\label{eq:MGF_time_evol}
\end{align}
in which the \textit{Liouville operator} $\mathcal{L} \! \left(
a^{\dagger} , a \right)$ is derived from the elements of the matrix $-
\mathbb{T}$.  

To define exponential families and a cumulant-generating function
(CGF), it is convenient to work with the Laplace transform with an
argument that is a vector $z \equiv \left[ z_p \right]$ of complex
coefficients.  The corresponding CGF, $-\Gamma \! 
\left( \log z \right)$,\footnote{We adopt the sign for $\Gamma$
corresponding to the free energy in thermodynamics.  Other standard
notations, such as $\psi$ for the CGF~\cite{Amari:inf_geom:01}, are
unavailable because they collide with notations used in the
representation of CRNs below.} for which the natural argument is $\log
z$, is constructed in the Doi Hilbert space as the inner product with
a variant on the Glauber norm,
\begin{align}
  e^{- \Gamma \left( \log z \right)} 
& = 
  \left( 0 \right| 
  e^{z a}
  \left| \rho \right) . 
\label{eq:MGF_eval_z}
\end{align}
$z$ will be called the \textit{tilt} of the exponential family,
corresponding to its usage in importance
sampling~\cite{Siegmund:IS_seq_tests:76}.

An important quantity will be the vector of expectations of the number
operators in the tilted distribution
\begin{align}
  {n_{\rho}}_p  \! \left( z \right) 
& \equiv 
  \frac{
    \left( 0 \right| 
    e^{z a}
    a^{\dagger}_p a_p 
    \left| \rho \right)
  }{
    \left( 0 \right| 
    e^{z a}
    \left| \rho \right) 
  } . 
\label{eq:mean_n_z}
\end{align}
Time evolution of the CGF follows from Eq.~(\ref{eq:MGF_time_evol}),
as 
\begin{align}
  \frac{\partial}{\partial t}
  e^{- \Gamma \left( \log z \right)} 
& = 
  - \left( 0 \right| 
  e^{z a}
  \mathcal{L} \! \left( a^{\dagger} , a \right)
  \left| \rho \right) . 
\label{eq:MGF_time_evol_z}
\end{align}
Under the saddle-point or leading-exponential approximation that
defines the large-deviation limit (developed further below), the
expectation of $\mathcal{L}$ in Eq.~(\ref{eq:MGF_time_evol_z}) is
replaced by the same function at classical arguments
\begin{align}
  - \frac{\partial}{\partial t}
  \Gamma \! \left( \log z \right)
& = 
  - \mathcal{L} \! 
  \left( 
    z , n_{\rho}  \! \left( z \right) / z 
  \right) . 
\label{eq:CGF_time_evol_DP}
\end{align}

\paragraph{From coherent-state to number-potential coordinates:}

With a suitable ordering of operators in $\mathcal{L}$ (with all
lowering operators to the right of any raising operator), $z$ is the
exact value assigned to $a^{\dagger}$ in the
expectation~(\ref{eq:MGF_time_evol_z}), and only the value $n_{\rho}
\! \left( z \right) / z$ for $a$ depends on saddle-point
approximations.  In special cases, where $\left| \rho \right)$ are
eigenstates of $a$ known as \textit{coherent states}, the assignment
to $a$ is also exact.  Therefore the arguments of $\mathcal{L}$ in
Eq.~(\ref{eq:CGF_time_evol_DP}) are called \textit{coherent-state
coordinates}.  

However, $\log z$, which we henceforth denote by $\theta \equiv \left[
{\theta}_p \right]$, is the affine coordinate system in which the CGF
is locally convex,\footnote{$\theta$ also provides the affine
coordinate system in the exponential family, which defines
contravariant coordinates in information
geometry~\cite{Amari:inf_geom:01}.} and it will be preferable to work
in coordinates $\left( \theta , n_{\rho} \right)$, which we term
\textit{number-potential} coordinates, because for applications in
chemistry $\theta$ has the dimensions of a chemical potential.  We
abbreviate exponentials and other functions acting component-wise on
vectors as $z \equiv e^{\theta}$, and simply assign 
$n_{\rho}  \! \left( \theta \right) \equiv n_{\rho}  \! \left( z
\right)$.  Likewise, $\mathcal{L}$ is defined in either coordinate
system by mapping its arguments: 
${\mathcal{L}}^{\rm (N-P)} \! \left( \theta , n \right) \equiv 
{\mathcal{L}}^{\rm (CS)} \! \left( z , n \! \left( z \right) / z
\right)$.

\subsubsection{Legendre transform of the CGF}

To establish notation and methods, consider first distributions
${\rho}_{\rn}$ that are convex with an interior maximum in $\rn$.
Then the gradient of the CGF 
\begin{align}
  - \frac{
    \partial \Gamma
  }{
    \partial \theta 
  } = 
  n_{\rho} \! \left( \theta \right)
\label{eq:argmax_z}
\end{align}
gives the mean~(\ref{eq:mean_n_z}) in the tilted distribution.  

The \textit{stochastic effective action} is the Legendre transform of
$- \Gamma$, defined as
\begin{align}
  S_{\rm eff} \! \left( n \right)
& = 
  \max_{\theta} 
  \left\{ 
    \theta n + 
    \Gamma \! \left( \theta \right) 
  \right\} . 
\label{eq:rho_as_LF_ofGamma}
\end{align}
For distributions over discrete states, to leading exponential
order, $S_{\rm eff}$ is a continuously-indexed approximation
to minus the log-probability: 
$S_{\rm eff} \! \left( n \right) \sim - { \left. \log {\rho}_{\rn}
\right| }_{\rn \approx n}$.  Its gradient recovers the tilt coordinate
$\theta$, 
\begin{align}
  \frac{\partial S_{\rm eff}}{\partial n} = 
  {\theta}_{\rho} \! \left( n \right) , 
\label{eq:argmax_form}
\end{align}
and the CGF is obtained by inverse Legendre transform.  
\begin{align}
  - \Gamma \! \left( \theta \right)
& = 
  \max_n 
  \left\{ 
    \theta n - 
    S_{\rm eff} \! \left( n \right)
  \right\} . 
\label{eq:Gamma_as_LF_ofrho}
\end{align}

The time evolution of $S_{\rm eff}$ can be obtained by taking a total
time derivative of Eq.~(\ref{eq:rho_as_LF_ofGamma}) along any
trajectory, and using Eq.~(\ref{eq:argmax_z}) to cancel the term in
$\dot{\theta}$.  The partial derivative that remains, evaluated using
Eq.~(\ref{eq:CGF_time_evol_DP}), gives 
\begin{align}
  {
    \left. 
      \frac{\partial S_{\rm eff}}{\partial t}
    \right| 
  }_{
    n
  } = 
  \mathcal{L} \! 
  \left( 
    {\theta}_{\rho} \! \left( n \right) , n
  \right) . 
\label{eq:HJ_equation_native}
\end{align}
Eq.~(\ref{eq:HJ_equation_native}) is of Hamilton-Jacobi form, with $-
\mathcal{L}$ filling the role of the Hamiltonian.  (The reason for
this sign correspondence, which affects nothing in the derivation,
will become clear below.)

\paragraph{Multiple modes, Legendre-Fenchel transform, and
locally-defined extrema}

Systems with interesting multi-level structure do not have $S_{\rm
eff}$ or $\rho$ globally convex, but rather only locally convex.  For
these the Legendre-Fenchel transform takes the place of the Legendre
transform in Eq.~(\ref{eq:rho_as_LF_ofGamma}), and if constructed with
a single coordinate $\theta$, $- \Gamma$ may have discontinuous
derivative.  

For these one begins, rather than with the CGF, with $S_{\rm eff}$,
evaluated as a line integral of Eq.~(\ref{eq:argmax_form}) in basins
around the stationary points $n_{\rho} \! \left( 0 \right)$ at $\theta
= 0$.  Each such basin defines invertible pair of functions
${\theta}_{\rho} \! \left( n \mid n_{\rho} \! \left( 0 \right)
\right)$ and $n_{\rho} \! \left( \theta \mid n_{\rho} \! \left( 0
\right) \right)$.  We will not be concerned with the large-deviation
construction for general distributions in this paper, which is better
carried out using a path integral.  We return in a later section to
the special case of multiple metastable fixed points in the stationary
distribution, and the modes associated with the stationary
distribution $\underline{\rho}$, and provide a more complete
treatment.

\subsubsection{Hamiltonian equations of motion and the action}

Partial derivatives with respect to $t$ and $n$ commute, so the
relation~(\ref{eq:argmax_form}), used to evaluate $\partial / \partial
n$ of Eq.~(\ref{eq:HJ_equation_native}), gives the relation
\begin{align}
  \frac{
    \partial {\theta}_{\rho} \! \left( n \right)
  }{
    \partial t
  }
& = 
  \frac{\partial \mathcal{L}}{\partial n} . 
\label{eq:eta_EOMz_from_HJ}
\end{align}
The dual construction for the time dependence of $n$ from
Eq.~(\ref{eq:argmax_z}) gives 
\begin{align}
  \frac{
    \partial n_{\rho} \! \left( \theta \right)
  }{
    \partial t
  }
& = 
  - \frac{\partial \mathcal{L}}{\partial \theta} . 
\label{eq:eta_EOMn_from_HJ}
\end{align}

The evolution
equations~(\ref{eq:eta_EOMz_from_HJ},\ref{eq:eta_EOMn_from_HJ})
describe stationary trajectories of an extended-time Lagrange-Hamilton
action functional which may be written in either coherent-state or
number-potential coordinates, as\footnote{The same action functionals
are arrived at somewhat more indirectly via 2-field functional
integral constructions such as the Doi-Peliti method.}
\begin{align}
  S 
& = 
  \int dt 
  \left\{ 
    - \left( d_t z \right) \left( n / z \right) + 
    {\mathcal{L}}^{\rm (CS)} \! \left( z , n/z \right)    
  \right\}
\nonumber \\ 
& = 
  \int dt 
  \left\{ 
    - \left( d_t \theta \right) n + 
    {\mathcal{L}}^{\rm (N-P)} \! \left( \theta , n \right)    
  \right\} . 
\label{eq:S_oflogzn_from_HJ}
\end{align}
From the form of the first term in either line, it is clear that the
two sets of coordinates relate to each other through a canonical
transformation~\cite{Goldstein:ClassMech:01}.

\paragraph{Circulation-free vector field of $\theta$ for the
stationary distribution} 

In order for $S_{\rm eff}$ to be a continuum approximation to $- \log
\rho$, if $\rho$ exists and is smooth everywhere, the vector field
$\theta$ obtained from stationary trajectories of the
action~(\ref{eq:S_oflogzn_from_HJ}) must have zero circulation in
order to be a well-defined gradient through
Eq.~(\ref{eq:argmax_form}).  To check that this is the case, consider
the increment of $\theta$ under a small interval $dt$ under
Eq.~(\ref{eq:eta_EOMz_from_HJ}): 
\begin{align}
  d{\theta}_p 
& \equiv 
  dt 
  \frac{\partial \mathcal{L}}{\partial n_p} . 
\label{eq:offset_logz_HJ}
\end{align}
The gradient in $n$ of $\theta$ therefore increments in time as 
\begin{align}
  \frac{
    \partial \left( {\theta}_p + d{\theta}_p \right)
  }{
    \partial n_q 
  }
& = 
  \frac{
    \partial {\theta}_p 
  }{
    \partial n_q 
  } + 
  dt 
  \frac{\partial {\mathcal{L}}^2}{\partial n_p \partial n_q} . 
\label{eq:grad_log_z}
\end{align}
Contraction of Eq.~(\ref{eq:grad_log_z}) with the antisymmetric symbol
in $p$ and $q$ vanishes
\begin{align}
  \frac{d}{dt}
  {\epsilon}_{pq}
  \frac{\partial {\theta}_p}{\partial n_q} 
& = 
  0 ,
\label{eq:time_der_no_curl}
\end{align}
so the circulation of $\theta$ is the same everywhere as at the fixed
points.  

From Eq.~(\ref{eq:argmax_form}), it is required to be the case that 
$\partial {\theta}_p / \partial n_q = {\partial}^2 S_{\rm eff} /
\partial n_p \partial n_q \equiv g^{-1}_{pq}$,
the inverse of the Fisher metric~\cite{Amari:inf_geom:01}, 
symmetric by construction and thus giving ${\epsilon}_{pq} g^{-1}_{pq}
= 0$.  The only difference between the fixed point and any other point
is that for distant points we are relying on Hamiltonian trajectories
to evaluate $\theta$, whereas the the fixed point the Fisher metric
may be calculated by means not relying on the large-deviation
saddle-point approximation.  Therefore Eq.~(\ref{eq:time_der_no_curl})
may be read as a check that symmetry of the Fisher metric is preserved
by Hamiltonian trajectories directly from the symmetric partial
derivative of $\mathcal{L}$ in Eq.~(\ref{eq:grad_log_z}).

\subsection{The stationary distribution and macrostates}
\label{sec:def_macrostates}

Up to this point only the Lyapunov role~(\ref{eq:Ddot_tot_genform}) of
the relative entropy has been developed.  While the increase of $- D
\! \left( \rho \parallel \underline{\rho} \right)$ has the appearance
of the classical 2nd law, we can understand from three observations
that this relative entropy is not the desired generalization of the
entropy state function of classical thermodynamics to express the
phenomenology of multi-level systems:
\begin{enumerate}

\item The relative entropy $- D \! \left( \rho \parallel
\underline{\rho} \right)$ is a functional on arbitrary distributions,
like the Shannon entropy that is a special case.  It identifies no
concept of macrostate, and has no dependence on state variables.

\item In a multilevel system that may have arbitrarily fine-grained
descriptions, there is no upper limit to $D \! \left( \rho \parallel
\underline{\rho} \right)$, and no appearance of the system scale at
any particular level, which characterizes state-function entropies.

\item Eq.~(\ref{eq:Ddot_tot_genform}) describes a deterministic
increase of relative entropy; the large-deviation role of entropy
as a log-probability for macrostate fluctuations~\cite{Ellis:ELDSM:85}
does not appear.  

\end{enumerate}

The step that has not yet been taken in our construction is, of
course, the identification of a macrostate concept.  Here we depart
from the usual development based on conservation laws, and follow
Gell-Mann and Lloyd~\cite{GellMann:EC:96,GellMann:eff_complx:04} in
claiming that the concept of macrostate is not inherent in features of
a system's dynamics, but requires one to explicitly choose a procedure
for aggregation or coarse-graining -- what they call a ``judge'' -- as
part of the commitment to which phenomenology is being described.  

We will put forth the definition of macrostates as \emph{the tilted
distributions arising in generating functions for the stationary
distribution $\underline{\rho}$}.  In the case of generating functions
for number, these are the distributions appearing in
Eq.~(\ref{eq:MGF_eval_z}), which we will denote by ${\rho}^{\left(
\bar{n} \right)}$.  They are the least-improbable distributions with a
given non-stationary mean to arise through aggregate microscopic
fluctuations, and therefore dominate the construction of the
large-deviation probability.

The \textit{extensive state variable} associated with this definition
of macrostate is the tilted mean from Eq.~(\ref{eq:argmax_z}), which
we will denote $\bar{n} = n_{\underline{\rho}} \! \left( \theta
\right)$.  If $\underline{\rho}$ and ${\rho}^{\left( \bar{n}
\right)}$ are sharply peaked -- the limit in which the large-deviation
approximation is informative -- the relative entropy of the macrostate
is dominated at the saddle point of ${\rho}^{\left( \bar{n} \right)}$,
where $\log {\rho}^{\left( \bar{n} \right)} \sim 0$, and thus 
\begin{align}
  D \! 
  \left( 
    {\rho}^{\left( \bar{n} \right)}
  \parallel 
    \underline{\rho} 
  \right)
& \sim 
  {
    \left. 
      - \log {\underline{\rho}}_{\rn}
    \right| 
  }_{\rn \sim \bar{n}} = 
  {\underline{S}}_{\rm eff} \! \left( \bar{n} \right) . 
\label{eq:DKL_ms_from_ss}
\end{align}
The general relative entropy functional, applied to the macrostate,
becomes the \textit{entropy state function} ${\underline{S}}_{\rm
eff}$, which takes as its argument the extensive state variable
$\bar{n}$.  Moreover, because probability under $\underline{\rho}$ is
concentrated on configurations with the scale that characterizes the
system, tilted means $\bar{n}$ that are not suppressed by very large
exponential probabilities will have comparable scale.  If $\bar{n}$ is
also the scale factor in the large-deviations function (a property
that may or may not hold, depending on the system studied), then $
{\underline{S}}_{\rm eff} \sim \bar{n}$ in scale, and the entropy
state function now has the characteristic scale of the mesoscale level
of the process description.

The three classes of distributions that enter a thermodynamic
description are summarized in Table~\ref{tab:dist_notations}.  

\begin{table}[th]
\begin{tabular}[c]{|l|l|l|}
\hline 
  notation & 
  definition & 
  comment \\
\hline
  % \rule{0pt}{2ex}
  $\underline{\rho}$ , 
  ${\underline{\rho}}^s$ , 
  ${\underline{\rho}}^{e \mid s}$ & 
  \parbox[t]{1.35in}{\raggedright
    whole-system, $s$-marginal, $\left( e \mid s \right)$-conditional
    distributions in the global stationary state  
  } &
  \parbox[t]{1.35in}{\raggedright
  } \\
  \rule{0pt}{4ex}
  ${\bar{\rho}}^s$ & 
  \parbox[t]{1.35in}{\raggedright
    marginal system steady-state distribution
  } &
  \parbox[t]{1.35in}{\raggedright
    function of instantaneous environment conditional distribution 
    ${\rho}^{e \mid s}$ 
  } \\
  \rule{0pt}{4ex}
  ${\rho}^{\left( \bar{n} \right)}$ & 
  \parbox[t]{1.35in}{\raggedright
    macrostate tilted to saddle-point value $\bar{n}$
  } &
  \parbox[t]{1.35in}{\raggedright
    defined relative to global stationary distributions
    $\underline{\rho}$; \\
    may be defined for whole-system, $s$, or $e \mid s$
  } \\
\hline
\end{tabular}
\caption{ 
  Notation conventions adopted for three classes of distributions
  arising in large-deviation systems with multi-scale relaxation
  structure.  
}
\label{tab:dist_notations}
\end{table}

\subsubsection{Coherent states, dimensional reduction, and the
$f$-divergence} 
\label{sec:coh_states}

A special case, which is illustrative for its simplicity and which
arises for an important sub-class of stochastic CRNs, is the case when
$\underline{\rho}$ is a \textit{coherent state}, an eigenvector of the
lowering operator $a$.  Coherent states are the generating functions
of product-form Poisson distributions, or cross-sections through such
products if the transitions in the population process satisfy
conservation laws.  They are known~\cite{Anderson:product_dist:10} to
be general solutions for CRN steady states satisfying a condition
termed \textit{complex balance}, and the fixed points associated with
such stationary distributions are also known to be unique and interior
(no zero-expectations for any ${\rn}_p$)~\cite{Feinberg:notes:79}.

Let $\underline{n}$ be the
eigenvalue of the stationary coherent state: 
$a \left| {\rho}^{\left( \underline{n} \right)} \right) =
\underline{n} \left| {\rho}^{\left( \underline{n} \right)} \right)$.
Then the mean in the tilted distribution lies in a simple exponential
family, $\bar{n} \equiv n_{\underline{\rho}} \! \left( \theta \right)
= e^{\theta} \underline{n}$ (component-wise), and the tilted
macrostate ${\rho}^{\left( \bar{n} \right)}$ is also a coherent state: 
$a \left| {\rho}^{\left( \bar{n} \right)} \right) = \bar{n} \left|
{\rho}^{\left( \bar{n} \right)} \right)$.  

The logarithm of a product-form Poisson distribution in Stirling's
approximation is given by
\begin{align}
  - \log {\rho}^{\left( \bar{n} \right)}_{\rn}
& \approx  
  \rn \cdot 
  \log \left( \frac{\rn}{\bar{n}} \right) - 
  \left( \rn - \bar{n} \right) \cdot 1 \equiv 
  D_f \! \left( \rn \parallel \bar{n} \right) . 
\label{eq:coh_state_log}
\end{align}
$D_f \! \left( \rn \parallel \bar{n} \right)$, known as the
$f$-divergence, is a generalization of the Kullback-Leibler divergence
to measures such as $n$ which need not have a conserved sum.  The
Lyapunov function from Eq.~(\ref{eq:Ddot_tot_genform}) reduces in the
same Stirling approximation to
\begin{align}
  D \! 
  \left(  
    {\rho}^{\left( \bar{n} \right)}
  \parallel  
    \underline{\rho} 
  \right) 
& \approx 
  D_f \! \left( \bar{n} \parallel \underline{n} \right) , 
\label{eq:coh_state_KLtof}
\end{align}
giving ${\underline{S}}_{\rm eff} \! \left( \bar{n} \right)$ in
Eq.~(\ref{eq:DKL_ms_from_ss}).  On coherent states, the
Kullback-Leibler divergence on distributions, which may be of
arbitrarily large dimension, reduces to the $f$-divergence on their
extensive state variables which have dimension $P$.

The coherent states play a much more general role than their role as
exact solutions for the restricted case of complex-balanced CRNs.  In
the Doi-Peliti 2-field functional integral
formalism~\cite{Doi:SecQuant:76,Doi:RDQFT:76,Peliti:PIBD:85,Peliti:AAZero:86}
for generating functionals over discrete-state stochastic processes,
the coherent states form an over-complete basis in the Peliti
representation of unity.  The saddle-point approximation on
trajectories, which yields the classical
actions~(\ref{eq:S_oflogzn_from_HJ}) and the resulting Hamilton-Jacobi
equations, approximates expectations in the exact distribution by
those in the nearest coherent-state basis element.  Observables in
macrostates are thus mapped to observables in coherent states,
although in cases when the coherent state is not an exact solution,
the saddle-point condition may be sensitive to which observable is
being evaluated.

\subsubsection{Multiple fixed points and instantons}
\label{sec:multi_FP_inst}

Systems with multiple metastable fixed points correspond to 
non-convex $\underline{\rho}$ and thus multiple modes.  For these,
monotone decrease of $\dot{D} \! \left( \rho \parallel
\underline{\rho} \right)$ in Eq.~(\ref{eq:Ddot_tot_genform}) does not
entail monotonicity of the $f$-divergence in
Eq.~(\ref{eq:coh_state_KLtof}).
In such systems, first passages between basins of attraction are
solutions to the Hamiltonian
equations~(\ref{eq:eta_EOMz_from_HJ},\ref{eq:eta_EOMn_from_HJ}) with
momentum coordinate $\theta \neq 0$.  Along these
${\underline{S}}_{\rm eff} \! \left( \bar{n} \right)$ increases, and
that increase is what is sometimes termed the ``violation of the
2nd law''.  

For unimodal $\underline{\rho}$, the $\theta \neq 0$ large-deviation
trajectories have a separate use and interpretation from the
relaxation trajectories at $\theta = 0$ that give the classical 2nd
law in Eq.~(\ref{eq:DKL_dot_ctm_mimicdisc_class}).  For multi-modal
$\underline{\rho}$ a special sub-class of $\theta \neq 0$
trajectories, those known as \textit{instantons} and responsible for
first-passages between
fixed-points~\cite{Cardy:Instantons:78,Coleman:AoS:85}, must be used
to refine the interpretation of classical relaxation trajectories.
That refinement relates the transient increases in the large-deviation
function to the deterministic 2nd law~(\ref{eq:Ddot_tot_genform})
that continues to apply. 

This section briefly introduces the Legendre duality that defines
first-passage probabilities in metastable systems, arriving at the
chain rule for entropy that separates the roles of classical and
instanton trajectories.  
Let $\underline{n}$ be a fixed point of the Hamiltonian equations for
$\underline{\rho}$, and denote by $n_{\underline{\rho}} \! \left(
\theta \mid \underline{n} \right)$ the values of classical state
variables $\bar{n}$ obtained along $\theta \neq 0$ trajectories from
$\underline{n}$.  Call these \textit{escape trajectories}.  The set
of all $\bar{n}$ is partitioned among basins of repulsion from fixed
points.  Saddle points and escape separatrices are limit points of
escapes from two or more basins.

Within one such basin, we may construct ${\underline{S}}_{\rm eff}$ as
a Legendre transform of a summand ${\Gamma}_{\underline{n}}$ in the
overall CGF, as
\begin{align}
  {\underline{S}}_{\rm eff} \! \left( \bar{n} \right)
& = 
  \max_{\theta}
  \left\{ 
    \theta \, 
    \bar{n} + 
    {\Gamma}_{\underline{n}} \! \left( \theta \right) 
  \right\} . 
\label{eq:Seff_as_LF_ofGamma_multi}
\end{align}
$\theta$ ranges only over the values that arise on escape trajectories
from $\underline{n}$, which generally are
bounded~\cite{Smith:CRN_CTM:20}, and within that range 
\begin{align}
  {
    \left. 
      - \frac{
        \partial {\Gamma}_{\underline{n}}
      }{
        \partial \theta
      }
    \right| 
  }_{
    \theta \left( \bar{n} \right)
  } = 
  \bar{n} . 
\label{eq:argmax_z_multi}
\end{align}

Next, let the function 
$\underline{n} \! \left( \bar{n} \right)$\footnote{Note that the maps
$\underline{n} \! \left( \bar{n} \right)$ and $n_{\underline{\rho}} \!
\left( \theta \mid \underline{n} \right)$ need not be reflexive.  That
is, we may have $n_{\underline{\rho}} \! \left( \theta \mid
\underline{n} \! \left( \bar{n} \right) \right) \neq \bar{n}$ for any
$\theta$, because escape and relaxation separatrices may differ.} 
denote the fixed point to which a trajectory with $\theta \equiv 0$
relaxes, starting from $\bar{n}$.  
From the large-deviation identification of $S_{\rm eff} \! \left( n
\right) \sim - { \left. \log {\rho}_{\rn} \right| }_{\rn \approx n}$, 
recognize that the large-deviation function for the stationary
distribution is given by 
${\underline{S}}_{\rm eff} \! \left( \underline{n} \right) = D \!
\left( 1_{\underline{n}} \parallel \underline{\rho} \right)$,
where $1_{\underline{n}}$ denotes the special macrostate for which the
stationary value is a fixed-point value $\underline{n}$.

Then the expression~(\ref{eq:DKL_ms_from_ss}) for the Kullback-Leibler
divergence appearing in Eq.~(\ref{eq:Ddot_tot_genform}) may be
approximated to leading exponential order as
\begin{align}
  D \! 
  \left( 
    {\rho}^{\left( \bar{n} \right)}
  \parallel 
    \underline{\rho} 
  \right) 
& \sim 
  {\underline{S}}_{\rm eff} \! 
  \left( \bar{n} \right) - 
  {\underline{S}}_{\rm eff} \! 
  \left( \underline{n} \! \left( \bar{n} \right) \right) + 
  D \! 
  \left( 
    1_{\underline{n} \left( \bar{n} \right)} 
  \parallel 
    \underline{\rho}
  \right) . 
\label{eq:DKL_ray_Seff_rays_chain}
\end{align}
Eq.~(\ref{eq:DKL_ray_Seff_rays_chain}) is the chain rule for relative
entropy.  ${\underline{S}}_{\rm eff} \! \left( \underline{n} \! \left(
\bar{n} \right) \right) - {\underline{S}}_{\rm eff} \! \left( \bar{n}
\right)$ is the conditional entropy of the macrostate $\bar{n}$
relative to the macrostate $1_{\underline{n} \left( \bar{n} \right)}$
to which it is connected by a $\theta = 0$ Hamiltonian trajectory.  $-
D \! \left( 1_{\underline{n} \left( \bar{n} \right)} \parallel
\underline{\rho} \right)$ is the unconditioned entropy of
$1_{\underline{n} \left( \bar{n} \right)}$ relative to
$\underline{\rho}$. 

Relaxation along $\theta = 0$ trajectories describes a classical
2nd law only for the conditional part of the relative entropy.
Deterministic relaxation of the unconditioned entropy is derived from
the refinement of the (as it turns out, only apparent) classical
trajectory with an instanton sum.  The general method is described in
depth in~\cite{Cardy:Instantons:78} and~\cite{Coleman:AoS:85}~Ch.7,
using path integrals that require too much digression to fit within
the scope of this paper.  The structure of the instanton sum, and in
particular the way it creates a new elementary stochastic process at
the macroscale for which $D \! \left( 1_{\underline{n} \left( \bar{n}
\right)} \parallel \underline{\rho} \right)$ is the Lyapunov function,
will be explained in Sec.~\ref{sec:inst_relent_cont}
following~\cite{Smith:LDP_SEA:11}, after the behavior of
${\underline{S}}_{\rm eff}$ along $\theta = 0$ and $\theta
\neq 0$ trajectories has been characterized.

\section{Population processes with CRN-form generators}
\label{sec:CRN_form}

The results up to this point apply to general processes with discrete
state spaces, normalizable stationary distributions, and some kind of
$\mbox{system} \otimes \mbox{environment}$ tensor-product structure on
states.  For the next steps we restrict to stochastic population
processes that can be written in a form equivalent to Chemical
Reaction
Networks~\cite{Horn:mass_action:72,Feinberg:notes:79,Gunawardena:CRN_for_bio:03}.
CRNs are an expressive enough class to include many non-mechanical
systems such as evolving Darwinian
populations~\cite{Smith:evo_games:15}, and to implement
algorithmically complex processes~\cite{Andersen:NP_autocat:12}.  Yet
they possess generators with a compact and simple
structure~\cite{Krishnamurthy:CRN_moments:17,Smith:CRN_moments:17}, in
which similarities of microstate and macrostate phenomena are simply
reflected in the formalism.

\subsection{Hypergraph generator of state-space transitions}

Population states $\rn$ give counts of individuals, grouped according
to their types $p$ which are termed \textit{species}.  Stochastic
Chemical Reaction Networks assume independent elementary events
grouped into types termed \textit{reactions}.  Each reaction event
removes a multiset\footnote{A multiset is a collection of distinct
individuals, which may contain more than one individual from the same
species.}  of individuals termed a \textit{complex} from the
population, and places some (generally different) multiset of
individuals back into it.  The map from species to complexes is called
the \textit{stoichiometry} of the CRN.  Reactions occur with
probabilities proportional per unit time to \textit{rates}.  The
simplest rate model, used here, multiplies a half-reaction rate
constant by a combinatorial factor for proportional sampling without
replacement from the population to form the complex, which in the
mean-field limit leads to \textit{mass-action kinetics}.  

Complexes will be indexed with subscripts $i , j , \ldots$, and an
ordered pair such as $ji$ labels the reaction removing $i$ and
creating $j$.  With these conventions the Liouville-operator
representation of the generator from Eq.~(\ref{eq:MGF_time_evol})
takes the
form~\cite{Krishnamurthy:CRN_moments:17,Smith:CRN_moments:17} 
\begin{align}
  -\mathcal{L} 
& = 
  \sum_{ji} k_{ji} 
  \left[ 
    {\psi}_j \! \left( a^{\dagger} \right) - 
    {\psi}_i \! \left( a^{\dagger} \right)  
  \right] 
  {\psi}_i \! \left( a \right)
\nonumber \\ 
& \equiv  
  {\psi}^T \! \left( a^{\dagger} \right) 
  \mathbb{A} \, 
  \psi \! \left( a \right) . 
\label{eq:gen_L_defform_Doi}
\end{align}
$\left\{ k_{ji} \right\}$ are  half-reaction rate constants, organized
in the second line into an adjacency matrix $\mathbb{A}$ on complexes.
$\psi \equiv \left[ {\psi}_i \right]$ is a column vector with
components ${\psi}_i \! \left( a \right) \equiv \prod_p a_p^{y_{ip}}
\equiv a^{Y_i}$ that, in the Doi algebra, produce the combinatorial
factors and state shifts reflecting proportional sampling without
replacement.  Here $Y \equiv \left[ y_{ip} \right]$ is the matrix of
stoichiometric coefficients, with entry $y_{ip}$ giving the number of
individuals of species $p$ that make up complex $i$.  Further
condensing notation, $Y_i \equiv \left[ y_{ip} \right]$ are column
vectors in index $p$, and $Y_p \equiv \left[ y_{ip} \right]$ are
row vectors on index $i$.  $a^{Y_i}$ is understood as the
component-wise product of powers $a_p^{y_{ip}}$ over the species index
$p$.  ${\psi}^T$ is a transpose (row) vector, and vector and matrix
products are used in the second line of
Eq.~(\ref{eq:gen_L_defform_Doi}).

The generator~(\ref{eq:gen_L_defform_Doi}) defines a
\textit{hypergraph}~\cite{Berge:hypergraphs:73} in which $\mathbb{A}$
is the adjacency matrix on an ordinary graph over complexes known as
the \textit{complex
graph}~\cite{Gunawardena:CRN_for_bio:03,Baez:QTRN_eq:14}.  The
stoichiometric vectors $\left\{ Y_i \right\}$, defining complexes as
multisets, make the edges in $\mathbb{A}$ directed hyper-edges
relative to the population states that are the Markov states for the
process.\footnote{Properly, the generator should be called a
``directed multi-hypergraph'' because the complexes are multisets
rather than sets.}  The concurrent removal or addition of complexes is
the source of both expressive power and analytic difficulty provided
by hypergraph-generators.

A master equation~(\ref{eq:rhodot_genform}) acting in the state space
rather than on the generating function may be written in terms of the
same operators, as
\begin{align}
  {\dot{\rho}}_{\rn} 
& = 
  {\psi}^T \! 
  \left( e^{- \partial / \partial \rn} \right)
  \mathbb{A}
  \diag 
  \left[ 
    \psi \! \left( e^{\partial / \partial \rn} \right) 
  \right] 
  \Psi \! \left( \rn \right) 
  {\rho}_{\rn} . 
\label{eq:rho_state_timeevol}
\end{align}
Here a formal shift operator $e^{\partial / \partial \rn} \equiv
\left[ e^{\partial / \partial {\rn}_p} \right]$ is used in place of an
explicit sum over shifted indices, and ${\psi}_i \! \left( e^{\pm
\partial / \partial \rn} \right) = e^{\pm Y_i^T \partial / \partial
\rn}$ creates shifts by the stoichiometric vector $Y_i$.  $\diag 
\left[ \psi \right]$ refers to the matrix with diagonal entries given
by the components ${\psi}_i$.

In the master equation the combinatorial factors must be given
explicitly.  These are written as a vector $\Psi \equiv \left[
{\Psi}_i \right]$ with components 
${\Psi}_i \! \left( \rn \right) \equiv \prod_p \left[ {\rn}_p ! /
\left( {\rn}_p - y_{ip} \right) ! \right] \equiv
{\rn}^{\underline{Y_i}}$ that are falling factorials from $\rn$,
denoted with the underscore as ${\rn}^{\underline{Y_i}}$.

The matrix elements of Eq.~(\ref{eq:rhodot_genform}) may be read off
from Eq.~(\ref{eq:rho_state_timeevol}) in terms of elements in the
hypergraph, as 
\begin{align}
  w^{\left( ji \right)}_{{\rn}^{\prime} \rn} 
& \equiv 
  k_{ji}
  {\Psi}_i \! \left( \rn \right) 
& \mbox{where} \qquad 
  {\rn}^{\prime} 
& = 
  Y_j - Y_i + \rn , 
\label {eq:state_trans_rates_per}
\end{align}
between all pairs $\left( {\rn}^{\prime} , \rn \right)$ separated by
the stoichiometric difference vector $Y_j - Y_i$.  If multiple
reactions produce transitions between the same pairs of states, the
aggregate rates become
\begin{align}
  w_{{\rn}^{\prime} \rn} 
& \equiv 
  \sum_{
    ji \; 
  \mid \; 
    Y_j - Y_i  = 
    {\rn}^{\prime}  - \rn
  }
  w^{\left( ji \right)}_{{\rn}^{\prime} \rn} . 
\label{eq:state_trans_rates}
\end{align}

In this way a finite adjacency matrix $\mathbb{A}$ on complexes may
generate an infinite-rank transition matrix $\mathbb{T}$, which is the
adjacency matrix for an ordinary graph over states.  We will see that
for CRNs, the hypergraph furnishes a representation for macrostates
similar to the representation given by the simple graph for
microstates.

From Eq.~(\ref{eq:state_trans_rates}), marginal transition rates
$w^s_{{\rn}_s {\rn}^{\prime}_s}$ for the system may be defined using
the average~(\ref{eq:ws_def}) over ${\rho}^{e \mid s}$.  Note that the
dependence of the activity products ${\Psi}_i$ on species $p$ within
the system remains that of a falling factorial, even if the average
over activities of species in the environment is complicated.  Denote
by ${\Psi}^s \! \left( {\rn}_s \right)$ and ${\psi}^s \! \left( e^{\pm
\partial / \partial {\rn}_s} \right)$ the restrictions of the activity
and shift functions to the species in the system $s$, and by $k^s_{ji}
\! \left( {\rho}^{e \mid s} \right)$ the rate constants after
computing the sum in Eq.~(\ref{eq:ws_def}) over the index for species
in the environment.

\subsubsection{Descaling of transition matrices for microstates}

Proofs of monotonic change, whether of total relative
entropy~(\ref{eq:rhodot_genform}) or for the components of entropy
change partitioned out in Sec.~\ref{sec:sys_env_decomp}, take a
particularly simple form for CRNs generated by finitely many
reactions.  They make use of finite cycle decompositions of the
current through any microstate or complex, which are derived from
descaled adjacency matrices.

The descaling that produces a transition matrix that annihilates the
uniform measure on both the left and the right is 
\begin{align}
  \hat{\mathbb{T}}
& \equiv 
  {\psi}^T \! 
  \left( e^{- \partial / \partial \rn} \right)
  \mathbb{A}
  \diag 
  \left[ 
    \psi \! \left( e^{\partial / \partial \rn}\right) 
  \right] 
  \Psi \! \left( \rn \right) 
  \diag \left[ {\underline{\rho}}_{\rn} \right] 
\label{eq:state_space_TM}
\end{align}
for the whole mesoscale, or 
\begin{align}
  {\hat{\mathbb{T}}}^s
& \equiv 
  {\psi}^{sT} \! 
  \left( e^{- \partial / \partial {\rn}_s} \right)
  {\mathbb{A}}^s
  \diag 
  \left[ 
    {\psi}^s \! \left( e^{\partial / \partial {\rn}_s}\right) 
  \right] 
  {\Psi}^s \! \left( {\rn}_s \right) 
  \diag \left[ {\bar{\rho}}^s_{{\rn}_s} \right] 
\label{eq:state_space_TM_s}
\end{align}
for the subsystem $s$, by definition of the stationary states.
${\mathbb{A}}^s$ in Eq.~(\ref{eq:state_space_TM_s}) is the adjacency
matrix on complexes that gives rate constants $w^s_{{\rn}_s
{\rn}^{\prime}_s}$ from Eq.~(\ref{eq:ws_def}).  These descalings are
familiar as the ones leading to the dual time-reversed generators in
the fluctuation theorem for housekeeping
heat~\cite{Speck:HS_heat_FT:05}. 

We return to use the descaled microstate transition
matrices~(\ref{eq:state_space_TM},\ref{eq:state_space_TM_s}) in
monotonicity proofs for general CRNs in Sec.~\ref{sec:cycle_decomp},
but before doing that, we use the descaling in the state space to
motivate an analogous and simpler descaling for macrostates at the
level of the hypergraph.  That descaling illustrates the cycle
decomposition on finitely many states, though it only yields the
$f$-Divergence of Eq.~(\ref{eq:coh_state_KLtof}) as a Lyapunov
function for the complex-balanced CRNs.

\subsubsection{Descaling of transition matrices for macrostates}

The \textit{coherent states}, which are the moment-generating
functions of Poisson distributions and their duals in the Doi Hilbert
space, are defined as 
\begin{align}
  \left( {\phi}^{\dagger} \right|
& \equiv 
  e^{- {\phi}^{\dagger} \phi}
  \left( 0 \right| 
  e^{{\phi}^{\dagger} a}
& 
  \left| \phi \right) 
& \equiv 
  e^{a^{\dagger} \phi}
  \left| 0 \right) , 
\label{eq:coh_states_def}
\end{align}
where $\phi \equiv \left[ {\phi}_p \right]$ is a vector of (generally
complex) numbers and ${\phi}^{\dagger}$ is its Hermitian conjugate.  
They are eigenstates respectively of the raising and lowering
operators with eigenvalues 
$\left( {\phi}^{\dagger} \right| a^{\dagger}_p = 
\left( {\phi}^{\dagger} \right| {\phi}^{\ast}_p$ and 
$a_p \left| \phi \right) = {\phi}_p \left| \phi \right)$. 

On the constant trajectories corresponding to a fixed point of the
Hamiltonian equations~(\ref{eq:eta_EOMz_from_HJ},
\ref{eq:eta_EOMn_from_HJ}), ${\phi}^{\dagger} \equiv 1$ and $\phi
\equiv \underline{\phi} = \underline{n}$, the fixed-point number.  We
may descale the coherent-state parameters $\phi$ and
${\phi}^{\dagger}$, which correspond to classical state variables, by
defining 
\begin{align}
  {\phi}_p
& \equiv 
  {\underline{\phi}}_p
  {\varphi}_p
& 
  {\phi}^{\ast}_p
  {\underline{\phi}}_p
& \equiv 
  {\varphi}^{\ast}_p . 
\label{eq:Baish_descale}
\end{align}
In vector notation, $\phi \equiv \diag \left[ \underline{\phi} \right]
\varphi$, ${\phi}^{\dagger} \diag \left[ \underline{\phi} \right]
\equiv {\varphi}^{\dagger}$.  The scaling~(\ref{eq:Baish_descale}) was
introduced by Baish~\cite{Baish:DP_duality:15} to study dualities of
correlation functions in Doi-Peliti functional integrals.

The compensating descaling of the adjacency matrix 
\begin{align}
  \hat{\mathbb{A}}
& \equiv 
  \mathbb{A}
  \diag \left[ \psi \! \left( \underline{\phi} \right) \right]
\label{eq:bbAhat_def}
\end{align}
may be compared with Eq.~(\ref{eq:state_space_TM}) for $\mathbb{T}$.
A similar descaling may be done for ${\mathbb{A}}^s$ using
${\bar{\rho}}^s$.  Henceforth we omit the duplicate notation, and
carry out proofs with respect to whatever is the stationary
distribution for a given adjacency matrix and hypergraph. 

The coherent-state action~(\ref{eq:S_oflogzn_from_HJ}) with Liouville
operator~(\ref{eq:gen_L_defform_Doi}) has a symmetric form in descaled
variables that is particularly useful for understanding the relaxation
and escape trajectories in the Hamilton-Jacobi system:
\begin{align}
  S 
& = 
  \int dt 
  \left\{ 
    {\phi}^{\dagger} 
    \diag \left[ \underline{\phi} \right] 
    d_t \varphi - 
    {\psi}^T \! \left( {\phi}^{\ast} \right) 
    \hat{\mathbb{A}} \, 
    \psi \! \left( \varphi \right) 
  \right\} . 
\label{eq:gen_S_twoforms}
\end{align}

\subsubsection{Equations of motion and the $\mathcal{L} = 0$ manifold}

The Hamiltonian equations derived by variation of the
action~(\ref{eq:gen_S_twoforms}) can be written 
\begin{align}
  \diag \left[ \underline{\phi} \right] 
  \dot{\varphi}
& = 
  \frac{\partial {\psi}^T}{\partial {\phi}^{\dagger}}
  \hat{\mathbb{A}}
  \psi \! \left( \varphi \right) 
&& \underset{{\phi}^{\dagger} = 1}{\rightarrow}
  Y \hat{\mathbb{A}}
  \psi \! \left( \varphi \right) 
\nonumber \\ 
  {\dot{\phi}}^{\dagger}
  \diag \left[ \underline{\phi} \right] 
& = 
  - {\psi}^T \! \left( {\phi}^{\ast} \right)
  \hat{\mathbb{A}}
  \frac{\partial {\psi}}{\partial \varphi}
&& \underset{\varphi = 1}{\rightarrow}
  - {\psi}^T \! \left( {\phi}^{\ast} \right)
  \hat{\mathbb{A}}
  Y^T . 
\label{eq:gen_EOM_twoforms}
\end{align}
The general solution is shown first in each line and then particular
limiting forms are shown.

The sub-manifold ${\phi}^{\dagger} \equiv 1$ contains all relaxation
trajectories for any CRN.  It is dynamically stable because
$\mathbb{A}$ is a stochastic matrix on the left, and thus ${\psi}^T \! 
\left( 1 \right) \hat{\mathbb{A}} = 0$ in the second line of
Eq.~(\ref{eq:gen_EOM_twoforms}).
The image of $Y \mathbb{A}$ is called the \textit{stoichiometric
subspace}, and its dimension is denoted $s \equiv \dim \left[ \Ima
\left( Y \mathbb{A} \right) \right]$. 

An important simplifying property of some CRNs is known as
\textit{complex balance} of the stationary
distributions.\footnote{Complex balance can be ensured at all
parameters if a topological character of the CRN known as
\textit{deficiency} equals zero, but may also be true for
nonzero-deficiency networks for suitably tuned parameters.  In this
work nothing requires us to distinguish these reasons for complex
balance.}  It is the condition that the fixed point $\psi \! \left(
\underline{n} \right) \in \ker \mathbb{A}$ and not only $\in \ker Y
\mathbb{A}$.   
Since $\underline{n} = \underline{\phi}$ corresponds to $\varphi = 1$
and thus $\psi \! \left( \varphi \right) = 1$, $1 \in \ker
\hat{\mathbb{A}}$ and $\dot{\varphi} = 0$ in the first line of
Eq.~(\ref{eq:gen_EOM_twoforms}) at any ${\phi}^{\dagger}$.  
Complex-balanced CRNs (with suitable conditions on $\mathbb{A}$ to
ensure ergodicity on the complex
network~\cite{Gunawardena:CRN_for_bio:03}) always possess unique
interior fixed points~\cite{Feinberg:notes:79,Feinberg:def_01:87} and
simple product-form stationary distributions at these fixed
points~\cite{Anderson:product_dist:10}. 

For non-complex-balanced stationary solutions, although escapes may
have $\varphi = 1$ as initial conditions, that value is not
dynamically maintained.  Recalling the
definition~(\ref{eq:number_state_def}) of the number operator, the
field $n_p = {\phi}^{\dagger}_p {\phi}_p$, so non-constant $\phi$ is
required for instantons to escape from stable fixed points and
terminate in saddle fixed points, in both of which limits
${\phi}^{\dagger} \rightarrow 1$.

All relaxations and also the escape trajectories from fixed points
(the instantons) share the property that $\mathcal{L} \equiv 0$ for a
CRN with time-independent parameters.\footnote{We can see that this
must be true because $-\mathcal{L}$ is a Hamiltonian conserved under
the equations of motion, and because instantons trace the values of
$n$ and $\theta$ in the stationary distribution $\underline{\rho}$,
which must then have $\mathcal{L} = 0$ to satisfy the Hamilton-Jacobi
equation~(\ref{eq:HJ_equation_native}).}  This submanifold separates
into two branches,  with ${\phi}^{\dagger} \equiv 1$  for relaxations
and ${\phi}^{\dagger} \neq 1$ for escapes.

\subsubsection{The Schl{\"{o}}gl cubic model to illustrate}

The features of CRNs with multiple metastable fixed
points are exhibited in a cubic 1-species model introduced by
Schl{\"{o}}gl~\cite{Schlogl:near_SS_thermo:71}, which has been
extensively
studied~\cite{Dykman:chem_paths:94,Krishnamurthy:CRN_moments:17,Smith:CRN_moments:17}
as an example of dynamical bistability.

The reaction schema in simplified form is 
\begin{align}
  \varnothing 
& \overset{k_d}{\underset{{\bar{k}}_d}{\rightleftharpoons}} 
  A
& 
  2A 
& \overset{k_c}{\underset{{\bar{k}}_c}{\rightleftharpoons}} 
  3A . 
\label{eq:1spec_autocat_canon}
\end{align}
We choose rate constants so that the mass-action equation for the
number of $A$ particles, given by $n$, is 
\begin{equation}
  \dot{n} = 
  \left( n_3 - n \right) 
  \left( n_2 - n \right) 
  \left( n_1 - n \right) . 
\label{eq:cubic_n_EOM}
\end{equation}
The three fixed points are 
$\underline{n} \in \left\{ n_1, n_2, n_3 \right\}$, of which $n_1$ and
$n_3$ are stable, and $n_2$ is a saddle.

\begin{figure}[ht]
\begin{center} 
  \includegraphics[scale=0.425]{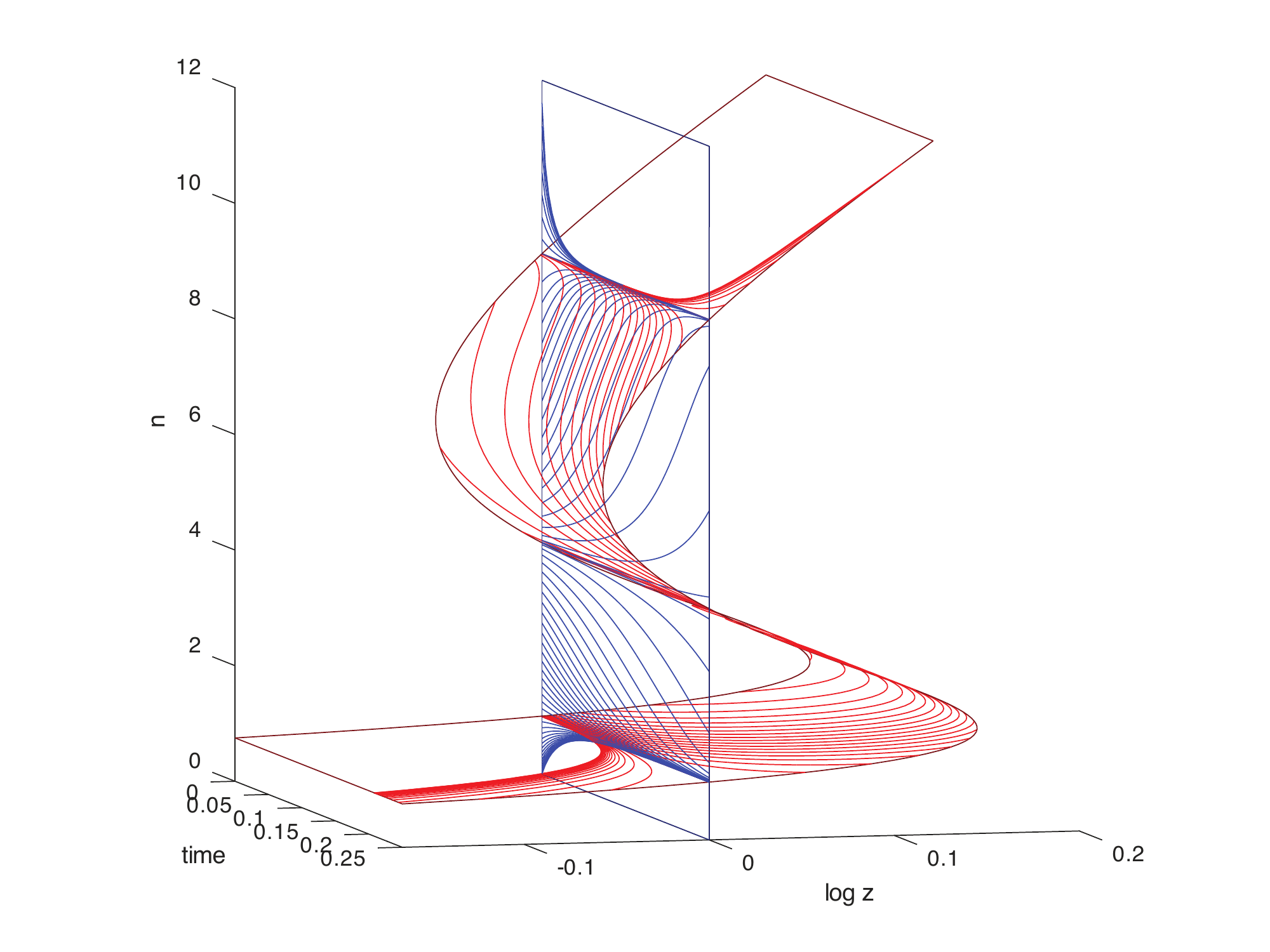} 
\caption{ 
  The two branches of the $\mathcal{L} = 0$ manifold, corresponding to
  the rate equation~(\ref{eq:cubic_n_EOM}).  ${\phi}^{\dagger} \equiv
  1$ (blue); ${\phi}^{\dagger} \neq 1$ (red).  Trajectories at common
  $n$ have the same value on the vertical axis.  Time is shown moving
  toward the front.
  \label{fig:L0man_1spec} 
}
\end{center}
\end{figure}

The relaxation and escape branches of the $\mathcal{L} = 0$ manifold
are shown in Fig.~\ref{fig:L0man_1spec}.  Because the Schl{\"{o}}gl
model is 1-dimensional, the condition $\mathcal{L} = 0$ fixes $\theta
\! \left( n \right)$ along escape trajectories.  Because the
stochastic model is a birth-death process, it is also exactly
solvable~\cite{vanKampen:Stoch_Proc:07}.  We will return to the
example in Sec.~\ref{sec:Schlogl_examp} to study properties of the
intensive and extensive thermodynamic potentials for it.

\subsection{Large-deviation and Lyapunov roles of the effective
action} 

The role of entropy in the understanding of Boltzmann and Gibbs was
that of a Lyapunov function~\cite{Fermi:TD:56,Kittel:TP:80},
accounting for unidirectionality from microscopically reversible
mechanics.  The much later understanding of entropy in relation to
fluctuations~\cite{Ellis:ELDSM:85,Touchette:large_dev:09} -- precisely
the opposite of deterministic evolution -- is that of a
large-deviation function.  

The chain rule in Eq.~(\ref{eq:DKL_ray_Seff_rays_chain}) of
Sec.~\ref{sec:multi_FP_inst} relates the conditional relative entropy
associated with quasi-deterministic relaxation to an additional
unconditioned entropy of metastable states that are stable fixed
points of the Hamiltonian dynamical system.  Here we complete the
description of the relation between the Lyapunov and large-deviation
roles of the macrostate entropy~(\ref{eq:DKL_ms_from_ss}), and show
how the sum over instantons rather than a single Hamiltonian
trajectory results in deterministic increase of the unconditioned
relative entropy $D \! \left( 1_{\underline{n}} \parallel
\underline{\rho} \right)$.  In the Hamilton-Jacobi representation this
means constructing relations between the ${\phi}^{\dagger} = 1$ and
the ${\phi}^{\dagger} \neq 1$ branches of the $\mathcal{L} = 0$
manifold.  For special cases, such as CRNs with detailed balance or
one-dimensional systems, the mapping is one of simple time reversal of
the $n$ fields in trajectories.  More generally, even for
complex-balanced CRNs where the Lyapunov and large-deviation functions
are the same, the relations between relaxation and escape trajectories
become more variable.

\subsubsection{Convexity proof of the Lyapunov property of macrostate
entropy in Hamilton-Jacobi variables}
\label{sec:ray_monoton_pf}

Begin with the dynamics of the relative entropy ${\underline{S}}_{\rm
eff} \! \left( \bar{n} \right)$ from Eq.~(\ref{eq:DKL_ms_from_ss}), as
the state variable $\bar{n}$ evolves along a relaxation solution to
the Hamiltonian equations of motion.  The properties of
${\underline{S}}_{\rm eff} \! \left( \bar{n} \right)$ follow from its
construction as a large-deviation function along escape trajectories.
From Eq.~(\ref{eq:argmax_form}), the time derivative of
${\underline{S}}_{\rm eff}$ along an escape trajectories is given by
\begin{align}
  \frac{
    {\underline{S}}_{\rm eff} \! \left( n \right) 
  }{
    \partial n 
  } 
  {\dot{n}}_{\rm esc} 
& = 
  {\theta}_{\underline{\rho}} \! \left( n \right) \cdot 
  {\dot{n}}_{\rm esc} \equiv 
  {
    \left. 
      {\dot{\underline{S}}}_{\rm eff}
    \right| 
  }_{\rm esc} . 
\label{eq:dotSeff_onpath}
\end{align}
Hamiltonian trajectories are least-improbable paths of fluctuation, so
escapes are conditionally dependent along trajectories.  The
conditional probability to extend a path, having reached any position
along the path, is always positive, giving ${
\left. {\dot{\underline{S}}}_{\rm eff} \right| }_{\rm esc} \ge 0$ in
Eq.~(\ref{eq:dotSeff_onpath}).

Next compute ${\dot{\underline{S}}}_{\rm eff}$ along a relaxation
trajectory, for simplicity considering $\mathbb{A}$ (an equivalent
construction exists for ${\mathbb{A}}^s$ in terms of
${\bar{\rho}}^s$).  The continuum limit of the relative entropy from
Eq.~(\ref{eq:Ddot_tot_genform}) replaces $\sum_{\rn}
\rightarrow \int dn$, and continuously-indexed $-\log {\rho}^{\left(
\bar{n} \right)}_n$ and $-\log {\underline{\rho}}_n$ are defined
through the large-deviation functions.  

Writing the CRN Liouville operator~(\ref{eq:gen_L_defform_Doi}) in
coherent-state arguments, the time dependence is evaluated as
\begin{align}
  \dot{D} \! 
  \left( 
    {\rho}^{\left( \bar{n} \right)}
  \parallel 
    \underline{\rho} 
  \right) 
& \rightarrow 
  \int d^s \! n \, 
  \log 
  \left( 
    \frac{{\rho}^{\left( \bar{n} \right)}_n}{{\underline{\rho}}_n} 
  \right)
  {\dot{\rho}}^{\left( \bar{n} \right)}_n 
\nonumber \\
& \sim 
  \int d^s \! n \, 
  \log 
  \left( 
    \frac{{\rho}^{\left( \bar{n} \right)}_n}{{\underline{\rho}}_n} 
  \right)
  \left[ 
    {\psi}^T \! \left( z \right)
    \mathbb{A}
    \psi \! \left( \frac{n}{z} \right) 
  \right] 
  {\rho}^{\left( \bar{n} \right)}_n
\nonumber \\
& = 
  \sum_{ji}
  k_{ji}
  \int d^s \! n \, 
  \log 
  \left( 
    \frac{{\rho}^{\left( \bar{n} \right)}_n}{{\underline{\rho}}_n} 
  \right)
  e^{\theta \left( Y_j - Y_i \right)}
  {\psi}_i \! \left( n \right)
  {\rho}^{\left( \bar{n} \right)}_n
\nonumber \\
& \approx 
  \sum_{ji}
  k_{ji}
  \int d^s \! n \, 
  {
    \left. 
      \frac{\partial}{\partial n_p}
      \log 
      \left( 
        \frac{{\rho}^{\left( \bar{n} \right)}_n}{{\underline{\rho}}_n} 
      \right)
    \right|
  }_{\bar{n}} 
  {\left( n - \bar{n} \right)}_p
\nonumber \\
& \mbox{} \times 
  {\left( n - \bar{n} \right)}_q
  {
    \left. 
      \frac{
        -{\partial}^2 \log {\rho}^{\left( \bar{n} \right)}_n
      }{
        \partial n_q \partial n_r
      } 
    \right|
  }_{\bar{n}} 
  {\left( Y_j - Y_i \right)}_r
  {\psi}_i \! \left( n \right)
  {\rho}^{\left( \bar{n} \right)}_n
\nonumber \\
& = 
  \sum_{ji}
  k_{ji}
  {
    \left. 
      \frac{-\partial \log {\underline{\rho}}_n}{\partial n_p}
    \right|
  }_{\bar{n}} 
  {\left( Y_j - Y_i \right)}_p
  {\psi}_i \! \left( \bar{n} \right)
\nonumber \\
& = 
  {
    \left. 
      \frac{-\partial \log {\underline{\rho}}_n}{\partial n}
    \right|
  }_{\bar{n}} 
  \cdot 
  Y \hat{\mathbb{A}}
  {\psi}_i \! \left( \bar{n} \right)
\nonumber \\
& = 
  {\theta}_{\underline{\rho}} \! \left( \bar{n} \right)
  \cdot 
  \dot{n}_{{\rho}^{\left( \bar{n} \right)}} \! \left( 0 \right)
\nonumber \\
& = 
  { 
    \left. 
      {\dot{\underline{S}}}_{\rm eff} \! 
      \left( \bar{n} \right) 
    \right| 
  }_{\rm rel} . 
\label{eq:DKL_dot_ctm_mimicdisc_class}
\end{align}
In Eq.~(\ref{eq:DKL_dot_ctm_mimicdisc_class}) $z$ abbreviates
$z_{{\rho}^{\left( \bar{n} \right)}} \! \left( n \right) \equiv \exp
{\theta}_{{\rho}^{\left( \bar{n} \right)}} \! \left( n \right)$ from
Eq.~(\ref{eq:HJ_equation_native}), for the $S_{\rm eff}$ that is the
continuum approximation of $-\log {\rho}^{\left( \bar{n} \right)}$.
The third through fifth lines expand $\mathbb{A}$ explicitly in terms
of rate constants $k_{ji}$ following Eq.~(\ref{eq:gen_L_defform_Doi}),
to collocate all terms in $\theta \equiv \log z$ in the Liouville
operator.  The fourth line expands $\log \left( {\rho}^{\left( \bar{n}
\right)} / \underline{\rho} \right)$ and $\theta$ to linear order in
$n - \bar{n}$ in neighborhoods of the saddle point $\bar{n}$ of
${\rho}^{\left( \bar{n} \right)}$.  The matrix $\left[ -{\partial}^2
\log \rho / \partial n_q \partial n_r \right]$ is the inverse of the
Fisher metric that is the variance $\left[ \left< {\left( n - \bar{n}
\right)}_p {\left( n - \bar{n} \right)}_q \right>
\right]$~\cite{Amari:inf_geom:01}, so the product of the two is just
the identity $\left[ {\delta}_{pr} \right]$.

In the penultimate line of Eq.~(\ref{eq:DKL_dot_ctm_mimicdisc_class}),
${\theta}_{\underline{\rho}} \! \left( \bar{n} \right)$ is the value
of $\partial {\underline{S}}_{\rm eff} / \partial n$ for the
\emph{escape} trajectory passing through $\bar{n}$, and now
$\dot{n}_{{\rho}^{\left( \bar{n} \right)}} \! \left( 0 \right)$ is the
velocity along the \emph{relaxation} trajectory rather than the
Hamilton-Jacobi escape solution at $\bar{n}$.  So the net effect of
the large-deviation approximation on relative entropy has been to
replace escape with relaxation velocity vectors at a fixed value of
$\theta$ Legendre dual to $\bar{n}$.

\noindent \textbf{Lemma:} 
${\theta}_{\underline{\rho}} \! \left( \bar{n} \right) \cdot
\dot{n}_{{\rho}^{\left( \bar{n} \right)}} \! \left( 0 \right) \le 0$.

\noindent \textbf{Proof:} 
The proof follows from four observations:
\begin{enumerate}

\item \textbf{$\mathcal{L} = 0$:} As noted, for both escapes and
relaxations, $\mathcal{L} \! \left( \theta , n \right) = 0$.

\item \textbf{Convexity:} Both the potential value
${\theta}_{\underline{\rho}}$ for the escape trajectory, and the
velocity $\dot{n}$ of the relaxation trajectory, are evaluated at the
same location $\bar{n} = n_{{\rho}^{\left( \bar{n} \right)}} \! \left(
0 \right)$.  The Liouville function $- \mathcal{L} = \sum_{ji} k_{ji}
\left( e^{\theta \left( Y_j - Y_i \right)} - 1 \right) {\psi}_i \!
\left( n \right)$, with all $k_{ji} > 0$, is convex on the
$s$-dimensional sub-manifold of fixed $n$.  $\mathcal{L}$ is bounded
above at fixed $n$, and in order for cycles to be possible, shift
vectors $\left( Y_j - Y_i \right)$ giving positive exponentials must
exist for all directions of $\theta$ in the stoichiometric subspace.
Therefore $\mathcal{L} \rightarrow - \infty$ at large $\left| \theta
\right|$ in every direction, and the region $\mathcal{L} > 0$ at fixed
$n$ is bounded.  The boundary $\mathcal{L} \! \left( \theta , n
\right) = 0$ at fixed $n$ is likewise convex with respect to $\theta$
as affine coordinates, and $\mathcal{L} > 0$ is its interior.

\item \textbf{Chord:} The vector $\left(\theta - 0 \right)$ is thus a
chord spanning the $\mathcal{L} = 0$ submanifold of co-dimension 1
within the $s$-dimensional manifold of fixed $n$.

\item \textbf{Outward-directedness:} The equation of motion $\dot{n} =
- \partial \mathcal{L} / \partial \theta$ gives $\dot{n} \! \left(
\theta , n \right)$ as the \emph{outward} normal function to the
surface $\mathcal{L} \! \left( \theta , n \right) = 0$.  The outward
normal at $\theta = 0$ is the classical relaxation trajectory.  Every
chord $\left(\theta - 0 \right)$ of the surface lies in its interior,
implying that ${\theta}_{\underline{\rho}} \! \left( n \right) \cdot
\dot{n} \! \left( 0 , n \right) < 0$ for any $n$, and thus 
${\theta}_{\underline{\rho}} \! \left( \bar{n} \right) \cdot
\dot{n}_{{\rho}^{\left( \bar{n} \right)}} \! \left( 0 \right) \le 0$.
\textcolor{red}{$\blacksquare$}

\end{enumerate}

The conditional part of the relative entropy, $-D \! \left(
{\rho}^{\left( \bar{n} \right)} \parallel \underline{\rho} \right)$,
is thus shown to be monotone increasing along relaxation trajectories,
which is the Lyapunov role for the entropy state function familiar
from classical thermodynamics.  That increase ends when $\bar{n}$
terminates in the trajectory fixed point $\underline{n} \! \left(
\bar{n} \right)$.

\subsubsection{Instantons and the loss of Large-deviation
accessibility from first passages}
\label{sec:inst_relent_cont}

The deterministic analysis of
Eq.~(\ref{eq:DKL_dot_ctm_mimicdisc_class}) is refined by the inclusion
of instanton trajectories through the following sequence of
observations, completing the discussion begun in
Sec.~\ref{sec:multi_FP_inst}.  Relevant trajectories are shown in
Fig.~\ref{fig:eggcrate_instantons_AI}.  

\begin{enumerate}

\item The 2nd law as formulated in Eq.~(\ref{eq:Ddot_tot_genform})
is approximated in the large-deviation limit not by a \emph{single}
Hamiltonian trajectory, but by the \emph{sum of all} Hamiltonian
trajectories, from an initial condition.  Along a single trajectory,
$- D \! \left( {\rho}^{\left( \bar{n} \right)} \parallel
\underline{\rho} \right)$ could increase or decrease.

\item $- D \! \left( {\rho}^{\left( \bar{n} \right)} \parallel
\underline{\rho} \right)$ \emph{increases} everywhere in the submanifold
$\theta = 0$ of the manifold $\mathcal{L} = 0$, by
Eq.~(\ref{eq:DKL_dot_ctm_mimicdisc_class}).  This is the classical
increase of (relative) entropy of Boltzmann and Gibbs.  $- D \! \left(
{\rho}^{\left( \bar{n} \right)} \parallel \underline{\rho} \right)$
\emph{decreases} everywhere in the submanifold
$\theta \neq 0$ of the manifold $\mathcal{L} = 0$, by
Eq.~(\ref{eq:dotSeff_onpath}).  This is the construction of the
log-probability for large deviations.  These escape paths, however,
simply lead to the evaluations of ${\underline{S}}_{\rm eff} \! \left(
n \right) \sim { \left. -\log {\underline{\rho}}_{\rn} \right| }_{\rn
\approx n}$, the stationary distribution.

\item If a CRN has a single fixed point $\underline{n}$, there is a
unique $\theta = 0$ trajectory from any starting $\bar{n}$ to it, and
$- D \! \left( {\rho}^{\left( \bar{n} \right)} \parallel
\underline{\rho} \right)$ increases deterministically by
Eq.~(\ref{eq:DKL_dot_ctm_mimicdisc_class}) along that single path.
The black trajectory with arrow converging exactly in a domain fixed
point is such a path in Fig.~\ref{fig:eggcrate_instantons_AI}.

\item If a CRN has multiple fixed points and instantons, all
trajectories are exponentially close to the exact $\theta = 0$
trajectory before they enter a small neighborhood around the terminus
$\underline{n} \! \left( \bar{n} \right)$ of the exact trajectory;
that is: they give the appearance of being the black deterministic
trajectory in Fig.~\ref{fig:eggcrate_instantons_AI}.
The trajectory sum is correspondingly close to the deterministic
relaxation that increases the conditional entropy
${\underline{S}}_{\rm eff} \! \left( \underline{n} \! \left( \bar{n}
\right) \right) - {\underline{S}}_{\rm eff} \! \left( \bar{n} \right)$
in Eq.~(\ref{eq:DKL_ray_Seff_rays_chain}). 

\item On longer times, however, the infinite sum of formally distinct
Hamiltonian trajectories disperses into a sum over series of
instantons making a random walk among fixed points, with an integral
for each passage over the possible times at which the escape occurs.
(See~\cite{Coleman:AoS:85}~Ch.7.)  Such a sum is shown as a tree of
colored first-passage trajectories in
Fig.~\ref{fig:eggcrate_instantons_AI}.  The ``cross-sectional'' sum at
a single observation time \emph{over} instantons distinguished by
their escaping times gives the same result as a longitudinal line
integral
\emph{along} a single instanton between the start time and the
observation.  That integral of $- \left( \partial
\log {\rho}^{\left( \bar{n} \right)}_n / \partial n \right) \dot {n}$
through a full passage (escape instanton + succeeding relaxation)
gives ${\underline{S}}_{\rm eff} \! \left( \underline{n} \right) -
{\underline{S}}_{\rm eff} \! \left( {\underline{n}}^{\prime} \right) =
\log \left( w^{\rm macro}_{{\underline{n}}^{\prime} \underline{n}} /
w^{\rm macro}_{\underline{n} {\underline{n}}^{\prime}} \right)$.  The
escape from fixed point $\underline{n}$ to a saddle between
$\underline{n}$ and a fixed point ${\underline{n}}^{\prime}$ in an
adjacent basin, which we denote $\bar{n} = {\underline{n}}^{\prime} \!
\ddagger \! \underline{n}$, is an integral over
Eq.~(\ref{eq:dotSeff_onpath}), while the relaxation from the saddle
$\bar{n}$ to the fixed point ${\underline{n}}^{\prime}$ is an integral
over Eq.~(\ref{eq:DKL_dot_ctm_mimicdisc_class}).  These are the
classical ``entropy fluctuations'' of stochastic thermodynamics.

\item The contribution to the probability of a trajectory from each
instanton comes only from the $\theta \neq 0$ sub-manifold, and is
given by $w^{\rm macro}_{{\underline{n}}^{\prime} \underline{n}} \sim
e^{- \left[ {\underline{S}}_{\rm eff} \! \left( \bar{n} \right) -
{\underline{S}}_{\rm eff} \! \left( \underline{n} \right) \right]}
\equiv e^{- \Delta {\underline{S}}_{\rm eff}}$, just the leaving rate
from the macrostate $1_{\underline{n}}$.  The result, upon
coarse-graining to the macroscale (see
Table~\ref{tab:3scale_LD_mapping} and the top diagram in
Fig.~\ref{fig:eggcrate_instantons_AI}) where first-passages become
instantaneous elementary events, is a new stochastic process on
discrete states corresponding to the mesoscale Hamiltonian fixed
points $\left\{ \underline{n}, {\underline{n}}^{\prime} , \ldots
\right\}$.  The coarse-grained counterpart to $D \! \left(
1_{\underline{n} \left( \bar{n} \right)} \parallel \underline{\rho}
\right)$ from Eq.~(\ref{eq:DKL_ray_Seff_rays_chain}) is the Lyapunov
function reduced by a transition matrix ${\mathbb{T}}^{\rm macro}$
with matrix elements $w^{\rm macro}_{{\underline{n}}^{\prime}
\underline{n}}$.  The coarse-graining and the reduction of $D \! \left(
1_{\underline{n} \left( \bar{n} \right)} \parallel \underline{\rho}
\right)$ are described in detail for a 2-basin example
in~\cite{Smith:LDP_SEA:11}. 

\item The properties of ${\mathbb{T}}^{\rm macro}$ are exactly those
we have assumed for $\mathbb{T}$ as inputs to the mesoscale,
completing the self-consistency of our recursively-defined multi-level
model universe.

\end{enumerate}

\section{Cycle decompositions and non-decrease of intrinsic and
extrinsic relative entropies}
\label{sec:cycle_decomp}

Schnakenberg~\cite{Schnakenberg:ME_graphs:76} introduced a method to
solve for the dissipation of a driven CRN in terms of steady cyclic
currents within the network, coupled to flows into or out of
environmental buffers modeled as
chemostats~\cite{Polettini:open_CNs_I:14}.  A similar decomposition
can be used here to derive the positive-semidefiniteness of the
relative entropy and housekeeping terms in
Eq.~(\ref{eq:Ddot_tot_decomp_HK}).  The form of the CRN generator
leads to a decomposition of entropy changes into sums of densities
convected around a basis of cycles by the stationary-state currents.
Positivity is proved by convexity arguments similar to those for the
standard fluctuation
theorems~\cite{Speck:HS_heat_FT:05,Harris:fluct_thms:07,Esposito:fluct_theorems:10},
and the cycle decomposition expresses the duality relations of those
theorems in terms of shifts forward or backward around cycles.

Sec.~\ref{sec:CRN_form} constructed the relation between the
hypergraph adjacency matrix $\mathbb{A}$ and stoichiometry $Y$, and
the microstate transition matrix $\mathbb{T}$.  Each reaction in the
hypergraph acts regularly as a \textit{rule} on indefinitely many
microstates, making CRNs an example of
\textit{rule-based systems}~\cite{Danos:rule_based_modeling:08}. 
This section extends that mapping to cycle decompositions.  For
finitely-generated CRNs, there is a finite basis of either cyclic
flows through complexes or stoichiometrically coupled cycles called
\textit{hyperflows}~\cite{Andersen:generic_strat:14}, with the
interpretation of mass-action currents.  From these, a (generally
infinite) basis of finite-length cycles can be constructed for flows
in the microstate space, which projects onto the basis in the
hypergraph.  We first use the finite basis of macrostate currents to
prove monotonicity results for $f$-divergences where those exist,
showing how the projection of cycles can result in a projection of the
Lyapunov property between levels.

The parallelism between microstates and macrostates only holds,
however, when the bases for currents at both levels contain only
\emph{cycles}.  More generally, when macrostate currents include
non-cyclic hyperflows, the analogy between the two levels is broken,
leading to loss of the $f$-divergence as a Lyapunov function on
macrostate variables and generally to much greater difficulty of
solving for the stationary distribution on microstates.  The
conditions for different classes of flows on the hypergraph are the
basis for a complexity classification of CRNs, and both the form of
the large-deviation function and the symmetry between its Lyapunov and
large-deviation roles differ across classes.

\subsection{Complexity classes and cycle decompositions of stationary
currents on the hypergraph}  

Remarkably, the properties of classical fixed points in the
Hamilton-Jacobi representation are sufficient to classify CRNs in a
hierarchy with three nested levels of complexity, according to the
mass-action currents at the fixed points.  Let $\underline{n}$ again
denote the fixed point.

\begin{enumerate}

\item The CRNs with \textbf{detailed balance} are those in which
$k_{ji} {\psi}_i \! \left( \underline{n} \right) = k_{ij} {\psi}_j \!
\left( \underline{n} \right)$, for all pairs of complexes $\left< j ,
i \right>$ connected by a reaction.  Under the
descaling~(\ref{eq:bbAhat_def}), this condition is that
${\hat{\mathbb{A}}}^T = \hat{\mathbb{A}}$. 

\item The CRNs with \textbf{complex balance} only require 
$\sum_i k_{ji} {\psi}_i \! \left( \underline{n} \right) = \sum_j
k_{ij} {\psi}_j \! \left( \underline{n} \right)$, for each complex
$j$, or under descaling, $\hat{\mathbb{A}} \psi \! \left( 1 \right) =
0$. 

\item The general case requires only $Y \mathbb{A} \psi \!
\left( \underline{n} \right)$, the condition of current balance at
each species $p$, or under  descaling, only $Y \hat{\mathbb{A}} \psi
\! \left( 1 \right) = 0$. 

\end{enumerate}
The classes are summarized in Table.~\ref{tab:CRN_eq_categories}, 

\begin{table}[th]
\begin{tabular}[c]{|l|l|}
\hline 
  case & complexity class \\
\hline
  $Y \hat{\mathbb{A}} 1 = 0$ & 
  general case \\ 
  $\hat{\mathbb{A}} 1 = 0$ & 
  \textit{complex-balanced} \\ 
  ${\hat{\mathbb{A}}}^T = \hat{\mathbb{A}}$ & 
  \textit{detailed-balanced} \\ 
\hline
\end{tabular}
\caption{ 
  Hierarchical categories of stationary distributions of CRNs.
}
\label{tab:CRN_eq_categories}
\end{table}

\subsubsection{Complex balance and relations of the Lyapunov and
large-deviation roles of $S_{\rm eff}$}

It is known that complex-balanced CRNs (with technical conditions to
ensure ergodicity on the complex network under
$\mathbb{A}$~\cite{Gunawardena:CRN_for_bio:03,Baez:QTRN:13}) possess
unique interior fixed points~\cite{Feinberg:notes:79}, and moreover
that their \emph{exact} steady states $\underline{\rho}$ are products
of Poisson distributions or sections through such
products~\cite{Anderson:product_dist:10} defined by conserved
quantities under the stoichiometry.  Their generating functions
are the coherent states from Sec.~\ref{sec:coh_states}.  It is
immediate that all classical states obtained by exponential tilts with
parameter $z = e^{\theta}$ are likewise coherent states with state
variables  
\begin{align}
  {\bar{n}}_p 
& \equiv 
  e^{{\theta}_p}
  {\underline{n}}_p , 
\label{eq:exp_fam_not}
\end{align}
and that the $f$-divergence of Eq.~(\ref{eq:coh_state_log}) is the
macrostate relative entropy function.

From the equations of motion~(\ref{eq:gen_EOM_twoforms}), within
either the ${\phi}^{\dagger} \equiv 1$ or ${\phi}^{\dagger} \neq 1$
sub-manifolds, the time derivatives of ${\underline{S}}_{\rm eff}$
along escape and relaxation trajectories,
Eq.~(\ref{eq:DKL_dot_ctm_mimicdisc_class}) and
Eq.~(\ref{eq:dotSeff_onpath}) respectively, take the forms
\begin{align}
  {
    \left. 
      {\dot{\underline{S}}}_{\rm eff}
    \right| 
  }_{\rm rel}
& = 
  \log \left( \frac{n}{\underline{n}} \right) \cdot 
  {\dot{n}}_{\rm rel} = 
  \log \left( \frac{n}{\underline{n}} \right) \cdot 
  Y \hat{\mathbb{A}} 
  \psi \! \left( \frac{n}{\underline{n}} \right)   
\nonumber \\ 
  {
    \left. 
      {\dot{\underline{S}}}_{\rm eff}
    \right| 
  }_{\rm esc}
& = 
  \log \left( \frac{n}{\underline{n}} \right) \cdot 
  {\dot{n}}_{\rm esc} = 
  - \log \left( \frac{n}{\underline{n}} \right) \cdot 
  Y {\hat{\mathbb{A}}}^T 
  \psi \! \left( \frac{n}{\underline{n}} \right) .  
\label{eq:dotSeff_rel_esc_form}
\end{align}
By Eq.~(\ref{eq:exp_fam_not}), ${\psi}_i$ appearing in
Eq.~(\ref{eq:dotSeff_rel_esc_form}) evaluate to 
\begin{align}
  {\psi}_i \! \left( \frac{n}{\underline{n}} \right) 
& = 
  e^{\theta Y_i} . 
\label{eq:complx_act_exp}
\end{align}

\paragraph{Finite cycle decomposition of the steady state current:}

By definition of complex balance, $\hat{\mathbb{A}} 1 = 0$, in
addition to the general condition that $1^T \hat{\mathbb{A}} = 0^T$.
Any such matrix acting over finitely many complexes may be written as
a sum of adjacency matrices for cyclic flows, which we index $\alpha$,
with positive coefficients ${\bar{\jmath}}_{\alpha}$ equal to the
currents around those cycles in the stationary state.  For the
subclass with detailed balance, a decomposition in cycles of length 2
is always possible.

Letting $\sum_{ji \mid \alpha}$ denote the sum over directed links
$ji$ in order around the cycle $\alpha$, the trajectory
derivatives~(\ref{eq:dotSeff_rel_esc_form}) may be decomposed as
\begin{align}
  {
    \left. 
      {\dot{\underline{S}}}_{\rm eff}
    \right| 
  }_{\rm rel} = 
  \theta \cdot 
  {\dot{n}}_{\rm rel} 
& = 
  - \sum_{\alpha} 
  {\bar{\jmath}}_{\alpha}
  \sum_{ji \mid \alpha}
  e^{\theta Y_i} \cdot 
  \log
  \left(
    \frac{
      e^{\theta Y_i}
    }{
      e^{\theta Y_j}
    }
  \right) 
\nonumber \\ 
  {
    \left. 
      {\dot{\underline{S}}}_{\rm eff}
    \right| 
  }_{\rm esc} = 
  \theta \cdot 
  {\dot{n}}_{\rm esc} 
& = 
  \sum_{\alpha} 
  {\bar{\jmath}}_{\alpha}
  \sum_{ji \mid \alpha}
  e^{\theta Y_j} \cdot 
  \log
  \left(
    \frac{
      e^{\theta Y_j}
    }{
      e^{\theta Y_i}
    }
  \right) . 
\label{eq:cycle_disses_nonorm}
\end{align}
($\cdot$ indicates the vector inner product over the species index
$p$.) 

Letting $\sum_{i \mid \alpha}$ denote the sum over all complexes in
the cycle $\alpha$, the complex activities~(\ref{eq:complx_act_exp})
may be normalized to vectors $p_{\alpha} \equiv \left[ p_{i \mid
\alpha} \right]$ with unit measure, as
\begin{equation}
  p_{i \mid \alpha} \equiv 
  \frac{
    e^{\theta Y_i}
  }{
    \sum_{j \mid \alpha} 
    e^{\theta Y_j}
  } . 
\label{eq:cycle_prob_def}
\end{equation}
Then the trajectory derivatives~(\ref{eq:cycle_disses_nonorm}),
themselves the time derivatives of the
$f$-divergence~(\ref{eq:DKL_ray_Seff_rays_chain}), may be written in
terms of positive semidefinite KL divergences of the measures
$p_{\alpha}$ from their own images advanced or retarded by one complex
around each cycle:
\begin{align}
  \theta \cdot 
  {
    \left. 
      \dot{n}
    \right| 
  }_{
    \rm rel
  } 
& = 
  - \sum_{\alpha} 
  \left( 
    {\bar{\jmath}}_{\alpha}
    \sum_{i \mid \alpha} 
    e^{\theta Y_i}
  \right) 
  \sum_{ji \mid \alpha}
  p_{i \mid \alpha} \cdot 
  \log
  \left(
    \frac{
      p_{i \mid \alpha} 
    }{
      p_{j \mid \alpha} 
    }
  \right) 
\nonumber \\ 
  \theta \cdot 
  {
    \left. 
      \dot{n}
    \right| 
  }_{
    \rm esc
  } 
& = 
  \sum_{\alpha} 
  \left( 
    {\bar{\jmath}}_{\alpha}
    \sum_{i \mid \alpha} 
    e^{\theta Y_i}
  \right) 
  \sum_{ji \mid \alpha}
  p_{j \mid \alpha} \cdot 
  \log
  \left(
    \frac{
      p_{j \mid \alpha} 
    }{
      p_{i \mid \alpha} 
    }
  \right) . 
\label{eq:cycle_disses}
\end{align}

Non-negativity of the KL divergences in every term of
Eq.~(\ref{eq:cycle_disses}) recovers the monotonicities of
${\underline{S}}_{\rm eff}$ along relaxation and escape trajectories
from Sec.~\ref{sec:ray_monoton_pf}.  The total time derivatives are
decomposed into terms over finitely many cycles, each independently
having the same sign, a stronger decomposition than can be obtained
from the convexity proofs for increase of relative
entropy~\cite{Schnakenberg:ME_graphs:76} for general CRNs.

\paragraph{Locally linear coordinates for classical entropy change in
complex-balanced CRNs}

\begin{figure}[ht]
\begin{center} 
  \includegraphics[scale=0.425]{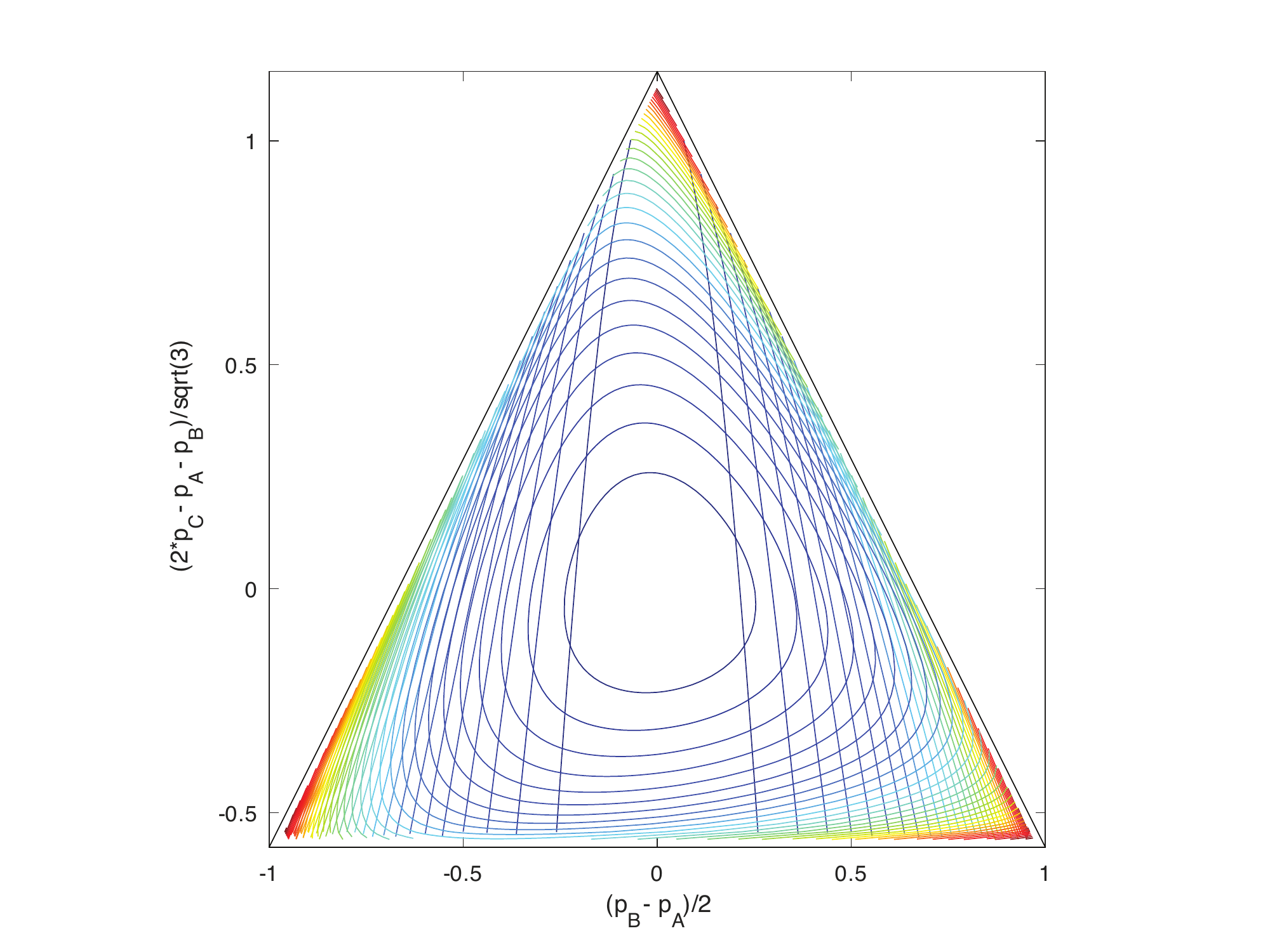} 
\caption{ 
  Simplex for a probability distribution $p = \left( p_A , p_B , p_C
  \right)$, with two coordinates $x_{AB}$ and $x_{ABC}$ of the form in
  Eq.~(\ref{eq:DKL_coords}) on a 2-cycle and a 3-cycle.
  \label{fig:DKL_coords_2simp} 
}
\end{center}
\end{figure}

Note that the cycle-KL divergences, multiplied by the density factors
$\sum_{i \mid \alpha} e^{\theta Y_i}$, define a coordinate system that
is locally invertible for the coordinates $\theta$ except at isolated
points:
\begin{align}
  x_{\alpha -}
& \equiv 
  \sum_{ji \mid \alpha}
  e^{\theta Y_i}
  \log
  \left(
    \frac{
      e^{\theta Y_i}
    }{
      e^{\theta Y_j}
    }
  \right) 
\nonumber \\ 
  x_{\alpha +}
& \equiv 
  \sum_{ji \mid \alpha}
  e^{\theta Y_j}
  \log
  \left(
    \frac{
      e^{\theta Y_j}
    }{
      e^{\theta Y_i}
    }
  \right) . 
\label{eq:DKL_coords}
\end{align}
The time derivatives of ${\underline{S}}_{\rm eff}$ from
Eq.~(\ref{eq:cycle_disses_nonorm}) may be written in these coordinates
as a linear system, 
\begin{align}
  {
    \left. 
      {\dot{\underline{S}}}_{\rm eff}
    \right| 
  }_{\rm rel} = 
  {
    \dot{D} \! \left( e^{\theta Y} \parallel 1 \right)
  }_{\rm rel} = 
  \theta \cdot 
  {\dot{n}}_{\rm rel} 
& = 
  - \sum_{\alpha} 
  {\bar{\jmath}}_{\alpha}
  x_{\alpha -}
\nonumber \\ 
  {
    \left. 
      {\dot{\underline{S}}}_{\rm eff}
    \right| 
  }_{\rm esc} = 
  {
    \dot{D} \! \left( e^{\theta Y} \parallel 1 \right)
  }_{\rm esc} = 
  \theta \cdot 
  {\dot{n}}_{\rm esc} 
& = 
  \sum_{\alpha} 
  {\bar{\jmath}}_{\alpha}
  x_{\alpha +} . 
\label{eq:cycle_disses_DKL_coords}
\end{align}
Fig.~\ref{fig:DKL_coords_2simp} shows examples of the
coordinates~(\ref{eq:DKL_coords}) for a 2-cycle and a 3-cycle in the
simplex of normalized activities ${\psi}_i / \sum_j {\psi}_j$ for
three species $i \in \left\{ A, B, C \right\}$.  

\subsubsection{Vorticity in the flowfield of stationary trajectories}

Recall (Table~\ref{tab:CRN_eq_categories}) that detailed balance means
${\hat{\mathbb{A}}}^T = \hat{\mathbb{A}}$.  In this case
Eq.~(\ref{eq:dotSeff_rel_esc_form}) implies equal and opposite change
of ${\underline{S}}_{\rm eff}$ along relaxations and escapes, because
the two trajectories are time-reverses of each other.  Detailed
balance thus generalizes the time-reversal symmetry of 1-dimensional
systems to any number of dimensions, and is a correspondingly
restrictive condition.  This is the assumption in classical
thermodynamics that identifies the Lyapunov and large-deviation roles
of the entropy state function.

Already for complex-balanced CRNs, exact time reversal is generally
broken.  The gradients of the tangent vectors to relaxation and escape
trajectories, with respect to the exponential-family coordinates
$\theta$ in Eq.~(\ref{eq:exp_fam_not}), evaluate to 
\begin{align}
  \frac{
    \partial 
    {
      \left(
        {\dot{n}}_{\rm rel} 
      \right) 
    }_p 
  }{
    \partial {\theta}_q
  } 
& = 
  Y_p 
  \hat{\mathbb{A}} 
  \diag \left[ \psi \! \left( e^{\theta} \right) \right]
  Y^T_q 
\nonumber \\ 
  \frac{
    \partial 
    {
      \left(
        {\dot{n}}_{\rm esc} 
      \right) 
    }_p 
  }{
    \partial {\theta}_q
  } 
& = 
  - Y_p 
  {\hat{\mathbb{A}}}^T 
  \diag \left[ \psi \! \left( e^{\theta} \right) \right]
  Y^T_q . 
\label{eq:ndot_by_logn_sum}
\end{align}
A measure of the time-asymmetry of the CRN is the vorticity, defined
as the antisymmetric part of the matrices~(\ref{eq:ndot_by_logn_sum}).
This vorticity equals the gradient of the \emph{sum} of tangent
vectors to relaxations and escapes at the same point, 
\begin{align}
  \frac{
    \partial 
    {
      \left(
        {\dot{n}}_{\rm esc} + 
        {\dot{n}}_{\rm rel} 
      \right) 
    }_p 
  }{
    \partial {\theta}_q
  } 
& = 
  Y_p 
  \left( 
    \hat{\mathbb{A}} - 
    {\hat{\mathbb{A}}}^T 
  \right) 
  \diag \left[ \psi \! \left( e^{\theta} \right) \right]
  Y^T_q . 
\label{eq:ndot_by_logn_sum_sum}
\end{align}
At the fixed point where $\theta = 0$, the vorticity is the
antisymmetric part of the matrix $Y A Y^T$.

\subsubsection{Hyperflow decomposition for non-complex-balanced CRNs}

For complex-balanced CRNs, $\hat{\mathbb{A}}$ describes flows on the
complex network, which is an ordinary directed graph.  For non-complex
balanced flows, $Y \hat{\mathbb{A}} 1 = 0$ but ${\mathbb{A}} 1 \neq
0$.  $\hat{\mathbb{A}}$ therefore cannot be written as a sum of
adjacency matrices for currents on cycles in the complex graph.
However, because net current still equals zero for every species, a
basis for the currents at any fixed point still exists in
\textit{balanced hyperflows}.  If the elements of $Y$ are all integer
values (as they are for chemistry or for biological population
processes), the basis elements are $\textit{balanced integer
hyperflows}$~\cite{Andersen:generic_strat:14}.  Each such flow is an
assignment of a set of non-negative integers to every current with
nonzero $k_{ji}$, such that the net currents at each species vanish.

We may continue to refer to basis hyperflows with index $\alpha$ by
extension of the complex-balanced case, and for each basis element
there is a non-negative current ${\bar{\jmath}}_{\alpha}$ which is the
coefficient of that flow in the stationary-state current solution.
Integer hyperflows cannot be used to express
${\dot{\underline{S}}}_{\rm eff}$ as sums of KL divergences with
uniform sign, in keeping with the variety of complications that arise
for ${\dot{\underline{S}}}_{\rm eff}$ in for non-complex-balanced
CRNs.

\subsection{Cycle decomposition in the microstate space, and
non-decrease of relative entropy components}
\label{sec:cycle_decomp_micro}

Next we apply a cycle decomposition similar to the one in the previous
section, but in the microstate space rather than on the hypergraph, to
prove non-negativity of the first two terms in
Eq.~(\ref{eq:Ddot_tot_decomp_HK}).  The crucial distinction between
the two levels is that balanced integer flows can always be mapped to
cycles in the state space.  For cycles in the hypergraph a natural map
is unique; for more general flows many mappings may be possible.
Relative entropies thus give Lyapunov functions more generally for
distributions over microstates than for macrostate variables.
However, unlike the hypergraph where basis decompositions are
\emph{finite}, in the state space they generally require solution for
infinitely many cycle currents and are difficult except in special
cases.

\subsubsection{The system-marginal relative entropy from
${\bar{\rho}}^s$} 

The system $s$ and environment $e$ must now be considered explicitly
because housekeeping heat is an embedding of $s$ in $s \otimes e$.  We
therefore work in system indices $\left\{ {\rn}_s \right\}$ and
descale with the steady state ${\bar{\rho}}^s$ at the instantaneous
${\rho}^{e \mid s}$.  In the following derivations ${\rho}^s$ and
${\rho}^{e \mid s}$ can be any normalizable distributions; macrostates
are no longer singled out.  The time-change of $D \! \left( {\rho}^s
\parallel {\bar{\rho}}^s \right)$, fixing ${\bar{\rho}}^s$ as before,
follows from Eq.~(\ref{eq:rhodot_genform}) and the
form~(\ref{eq:state_space_TM_s}), as
\begin{align}
\lefteqn{
  \dot{D} \! 
  \left( {\rho}^s \parallel {\bar{\rho}}^s \right) 
= 
  \sum_{{\rn}_s}
  \log 
  \left( \frac{{\rho}^s_{{\rn}_s}}{{\bar{\rho}}^s_{{\rn}_s}} \right)
  {\dot{\rho}}^s_{{\rn}_s} 
} & 
\nonumber \\
& = 
  \sum_{{\rn}_s}
  \log 
  \left( \frac{{\rho}^s_{{\rn}_s}}{{\bar{\rho}}^s_{{\rn}_s}} \right)
\nonumber \\ 
& \phantom{\sum_{{\rn}_s}} \times 
  {\psi}^{sT} \! 
  \left( e^{- \partial / \partial {\rn}_s} \right)
  {\mathbb{A}}^s
  \diag 
  \left[ 
    {\psi}^s \! \left( e^{\partial / \partial {\rn}_s}\right) 
  \right] 
  {\Psi}^s \! \left( {\rn}_s \right) 
  {\rho}^s_{{\rn}_s}
\nonumber \\
& = 
  \sum_{{\rn}^{\prime}_s {\rn}_s}
  \log 
  \left( 
    \frac{{\rho}^s_{{\rn}^{\prime}_s}}{{\bar{\rho}}^s_{{\rn}^{\prime}_s}}
  \right) \cdot 
  {\hat{\mathbb{T}}}^s_{{\rn}^{\prime}_s, {\rn}_s}
  \left( 
    \frac{{\rho}^s_{{\rn}_s}}{{\bar{\rho}}^s_{{\rn}_s}}
  \right) . 
\label{eq:time_der_KL_rho_rhobar}
\end{align}

From the definition~(\ref{eq:state_space_TM_s}), $1^T
{\hat{\mathbb{T}}}^s = 0$ and ${\hat{\mathbb{T}}}^s 1 = 0$, as was the
case for \emph{complex-balanced} ${\mathbb{A}}^s$.  Therefore
${\hat{\mathbb{T}}}^s$ can be written as a sum of adjacency matrices
for cycles, weighted by the currents around those cycles in the steady
state.  For a general CRN, it would not be assured that these cycles
were all of finite length or that there were only finitely many of
them passing through any given state ${\rn}_s$.  However, for a CRN in
which both the dimensions of $\mathbb{A}$ and $Y$ and their matrix
elements are finite, the integer hyperflow decomposition under
${\mathbb{A}}^s$ and $Y^s$ is in turn finite, and the basis flows can
be embedded in the microstate space to give a decomposition of
${\hat{\mathbb{T}}}^s$.

Let $\alpha$ index any basis of integer hyperflows spanning $\ker Y
\mathbb{A}$.  The cyclic flows spanning $\ker \mathbb{A}$ embed
without ambiguity in the state space, as images the cycles on the
complex network.  Each cycle $\alpha$ defines a sequence of state
shifts $\left\{ \rn \rightarrow \rn + Y_j - Y_i \right\}$ when the
transitions $\left\{ ji \mid \alpha \right\}$ in the cycle are
activated in order.

Non-complex-balanced integer flows also create cycles through states,
because by construction the sum of stoichiometric shifts in a balanced
flow is zero.  However, there may be a choice in the way a given flow
projects into the state space, depending on the \textit{realization},
defined as the order in which the reaction events in the flow are
activated.  Any order is acceptable, as long as each state in the
cycle can be reached by a reaction that can execute.\footnote{This
condition is called \textit{reachability}.  It can be violated if
there are boundary states with too few individuals to permit the input
complex to some reaction to be formed.  States may be grouped into
equivalence classes under reachability, as a means to study network
features such as autocatalysis.  Reachability is beyond the scope of
this paper.  For the large-population limits to which large-deviation
scaling applies, under finitely generated reactions, all states will
have arbitrarily many members and will be equivalent under
reachability, so all embeddings of a finite integer hyperflow will
always complete.}  We then extend the indexing $\alpha$ from the
images of the complex-balanced cycles to include all the integer
flows.

Once a cycle realization has been chosen for each integer hyperflow,
an embedding of these flows in the state space is defined as follows.
Pick an arbitrary starting complex in the cycle, and for each
${\rn}_s$, embed an image of the cycle with the starting complex
sampled at ${\rn}_s$. Then every state is passed through by every
cycle with each complex in the cycle sampling from that state exactly
once.  Every link has a finite number of cycles passing through it,
because the number of integer flows spanning $\ker Y \mathbb{A}$ and
the length of each flow are both finite.

A set of currents $\left\{ {\bar{\jmath}}_{\alpha , \rn} \right\}$
that sum to the steady-state currents over each transition is then
assigned to the cycles, indexed by the state $\rn$ where which cycle
$\alpha$ samples at its starting complex.  Solving for the values of
the $\left\{ {\bar{\jmath}}_{\alpha , \rn} \right\}$ is of course
equivalent to solving for $\bar{\rho}$, and is not generally a finite
problem.

With these notations in place, Eq.~(\ref{eq:time_der_KL_rho_rhobar})
becomes  
\begin{align}
  \dot{D} \! \left( {\rho}^s \parallel {\bar{\rho}}^s \right) 
& = 
  - \sum_{{\rn}_s}
  \sum_{\alpha}
  {\bar{\jmath}}^s_{\alpha , {\rn}_s}
  \sum_{{\rn}^{\prime}_s {\rn}_s \mid \alpha}
  \left( 
    \frac{{\rho}^s_{{\rn}_s}}{{\bar{\rho}}^s_{{\rn}_s}}
  \right) 
  \log 
  \left(
    \frac{
      {\rho}^s_{{\rn}_s} / {\bar{\rho}}^s_{{\rn}_s}
    }{
      {\rho}^s_{{\rn}^{\prime}_s} / {\bar{\rho}}^s_{{\rn}^{\prime}_s}
    }
  \right) . 
\label{eq:That_states_eval}
\end{align}
Eq.~(\ref{eq:That_states_eval}) is a non-positive sum of KL
divergences of probability ratios $\left( {\rho}^s_{{\rn}_s} /
{\bar{\rho}}^s_{{\rn}_s} \right)$ referenced to their own values
advanced around cycles, of the same form as
Eq.~(\ref{eq:cycle_disses_nonorm}) for complex activities $e^{\theta
Y_i}$ for macrostates on the hypergraph.  The state space divergence
involves an additional sum over the reference state ${\rn}_s$.
Because the sum over $\alpha$ is finite, the contribution at each
${\rn}_s$ is finite and proportional by finite factors to
${\bar{\rho}}^s_{{\rn}^{\prime}_s}$ within a finite neighborhood of
${\rn}_s$.  Therefore the sum~(\ref{eq:That_states_eval}) is finite.
This proves that the first line of Eq.~(\ref{eq:Ddot_tot_decomp_HK})
is non-decreasing. \textcolor{red}{$\blacksquare$}

\subsubsection{Non-negativity of the housekeeping entropy rate} 

The cycle decomposition on the state space may also be used to prove
non-negativity of the housekeeping entropy
rate~(\ref{eq:sigdot_HK_def}) introduced by Hatano and
Sasa~\cite{Hatano:NESS_Langevin:01}.  Here rather than work with
entropy changes over extended-time paths, we prove positivity moment
by moment from the master equation.

Unlike the relative entropy within the system, the housekeeping
entropy rate is not only a function of aggregated within-system
transition rates $w^s_{{\rn}_s {\rn}^{\prime}_s}$, but must be
disaggregated to all distinct system-environment interactions.  Begin
by expressing the sum of all transition currents both in terms of
elementary events and in the previous cycle decomposition:
\begin{align}
  J^s_{\rho}
& \equiv 
  \sum_{{\rn}_s}
  {\rho}^s_{{\rn}_s}
  \sum_{{\rn}^{\prime}_s \neq {\rn}_s}
  \sum_{\rn \mid {\rn}_s}
  \sum_{{\rn}^{\prime} \mid {\rn}^{\prime}_s}
  \sum_{
    ji \; 
  \mid \; 
    y_j - y_i  = 
    {\rn}^{\prime}  - \rn
  }
  w^{\left( ji \right)}_{{\rn}^{\prime} \rn} 
  {\rho}^{e \mid s}_{\rn}
\nonumber \\ 
& \equiv 
  \sum_{{\rn}_s}
  {\rho}^s_{{\rn}_s}
  \sum_{{\rn}^{\prime}_s \neq {\rn}_s}
  w^s_{{\rn}^{\prime}_s {\rn}_s} 
\nonumber \\ 
& = 
  \sum_{{\rn}_s}
  \sum_{\alpha}
  {\bar{\jmath}}^s_{\alpha , {\rn}_s}
  \sum_{{\rn}^{\prime}_s {\rn}_s \mid \alpha}
  \left( 
    \frac{{\rho}^s_{{\rn}_s}}{{\bar{\rho}}^s_{{\rn}_s}}
  \right) , 
\label{eq:HK_norm}
\end{align}
where $w^{\left( ji \right)}_{{\rn}^{\prime} \rn}$ is labeled by the
particular reaction connecting $\rn$ and
${\rn}^{\prime}$.\footnote{For notational simplicity, we suppose that
each ordered pair of complexes $\left( ji \right)$ is connected by at
most one reaction -- this still allows a common \emph{difference}
vector $Y_j - Y_i$ to be mediated by several distinct pairs $\left( ji
\right)$ -- the generalization to more complex cases is
straightforward.} 

We follow the construction of correlation functions for a counting
process from~\cite{Harris:fluct_thms:07}, for a quantity $r$ with
event-dependent values $r^{\left( ji \right)}$ defined as 
\begin{align}
  r^{\left( ji \right)}_{{\rn}^{\prime} \rn}
& \equiv 
  \log 
  \left(
    \frac{
      w^{\left( ji \right)}_{{\rn}^{\prime} \rn} 
      {\bar{\rho}}^s_{{\rn}_s}
      {\rho}^{e \mid s}_{{\rn}} 
    }{
      w^{\left( ji \right)}_{\rn {\rn}^{\prime}} 
      {\bar{\rho}}^s_{{\rn}^{\prime}_s}
      {\rho}^{e \mid s}_{{\rn}^{\prime}} 
    }
  \right) . 
\label{eq:r_npn_def}
\end{align}
A total housekeeping entropy rate denoted ${\dot{S}}^{\rm HK}$
is a sum over pairs of system states, of the quantity in
Eq.~(\ref{eq:sigdot_HK_def}).  Written as an expectation of $r$, it is
\begin{align}
  {\dot{S}}^{\rm HK}
& \equiv 
  \sum_{{\rn}_s}
  \sum_{{\rn}^{\prime}_s \neq {\rn}_s}
  {\sigma}^{\rm HK}_{{\rn}_s {\rn}^{\prime}_s} 
  w^s_{{\rn}_s {\rn}^{\prime}_s}
\nonumber \\ 
& = 
  \sum_{{\rn}_s}
  \sum_{\alpha}
  {\bar{\jmath}}^s_{\alpha , {\rn}_s}
  \sum_{{\rn}_s^{\prime} {\rn}_s \mid \alpha}
  \left( 
    \frac{{\rho}^s_{\rn}}{{\bar{\rho}}^s_{\rn}}
  \right) 
\nonumber \\ 
& \mbox{} \times 
  \frac{1}{w^s_{{\rn}^{\prime}_s {\rn}_s}}
  \sum_{\rn \mid {\rn}_s}
  \sum_{{\rn}^{\prime} \mid {\rn}^{\prime}_s}
  \sum_{
    ji \; 
  \mid \; 
    y_j - y_i  = 
    {\rn}^{\prime}  - \rn
  }
  w^{\left( ji \right)}_{{\rn}^{\prime} \rn} 
  {\rho}^{e \mid s}_{\rn}
  r^{\left( ji \right)}_{{\rn}^{\prime} \rn} . 
\label{eq:HK_intro}
\end{align}

The sign of Eq.~(\ref{eq:HK_intro}) can be deduced from convexity of
the observable $e^{-r}$, as in the usual fluctuation theorems. $r$ is
nonzero only on transitions, the terms in the sum in
Eq.~(\ref{eq:HK_norm}).  Let $dt$ be a short time interval.  Then the
expectation of any function of $r$ on the interval will be $\left( 1 -
dt \, J^s_{\rho} \right)$ times its value in an unchanging state, plus
a term $\propto dt$ from transitions.  Extracting the contribution
$\propto dt$ for expectations of $1$ and $e^{-r}$ over the interval,
denoted ${\left< \mbox{ } \right>}_{\rm dt}$, gives
\begin{align}
  {
    \left< 1 \right> 
  }_{\rm dt} - 
  \left( 1 - dt \, J^s_{\rho}  \right) 
& = 
  dt 
  \sum_{{\rn}_s}
  \sum_{\alpha}
  {\bar{\jmath}}_{\alpha , {\rn}_s}
  \sum_{{\rn}^{\prime}_s {\rn}_s \mid \alpha}
  \left( 
    \frac{{\rho}^s_{{\rn}_s}}{{\bar{\rho}}^s_{{\rn}_s}}
  \right) 
\nonumber \\ 
  {
    \left< e^{-r} \right> 
  }_{\rm dt} - 
  \left( 1 - dt \, J^s_{\rho}  \right) 
& = 
  dt 
  \sum_{{\rn}_s}
  \left( 
    \frac{{\rho}^s_{{\rn}_s}}{{\bar{\rho}}^s_{{\rn}_s}}
  \right) 
  \sum_{{\rn}^{\prime}_s \neq {\rn}_s}
  \sum_{\rn \mid {\rn}_s}
  \sum_{{\rn}^{\prime} \mid {\rn}^{\prime}_s}
\nonumber \\
& \mbox{} \times 
  \sum_{
    ji \; 
  \mid \; 
    y_j - y_i  = 
    {\rn}^{\prime}  - \rn
  }
  w^{\left( ji \right)}_{{\rn}^{\prime} \rn} 
  {\rho}^{e \mid s}_{\rn}
  {\bar{\rho}}^s_{{\rn}_s}
  e^{-r^{\left( ji \right)}_{{\rn}^{\prime} \rn}} 
\nonumber \\ 
& = 
  dt 
  \sum_{{\rn}_s}
  \left( 
    \frac{{\rho}^s_{{\rn}_s}}{{\bar{\rho}}^s_{{\rn}_s}}
  \right) 
  \sum_{{\rn}^{\prime}_s \neq {\rn}_s}
  \sum_{\rn \mid {\rn}_s}
  \sum_{{\rn}^{\prime} \mid {\rn}^{\prime}_s}
\nonumber \\
& \mbox{} \times 
  \sum_{
    ji \; 
  \mid \; 
    y_j - y_i  = 
    {\rn}^{\prime}  - \rn
  }
  w^{\left( ji \right)}_{\rn {\rn}^{\prime}} 
  {\rho}^{e \mid s}_{{\rn}^{\prime}}
  {\bar{\rho}}^s_{{\rn}^{\prime}_s}
\nonumber \\
& = 
  dt 
  \sum_{{\rn}_s}
  \left( 
    \frac{{\rho}^s_{{\rn}_s}}{{\bar{\rho}}^s_{{\rn}_s}}
  \right) 
  \sum_{{\rn}^{\prime}_s \neq {\rn}_s}
  w^s_{{\rn}_s {\rn}^{\prime}_s} 
  {\bar{\rho}}^s_{{\rn}^{\prime}_s}
\nonumber \\ 
& = 
  dt 
  \sum_{{\rn}_s}
  \sum_{\alpha}
  {\bar{\jmath}}_{\alpha , {\rn}_s}
  \sum_{{\rn}^{\prime}_s {\rn}_s \mid \alpha}
  \left( 
    \frac{{\rho}^s_{{\rn}^{\prime}_s}}{{\bar{\rho}}^s_{{\rn}^{\prime}_s}}
  \right) . 
\label{eq:exp_mr_ave}
\end{align}
Between the third and fourth lines in the evaluation of ${ \left<
e^{-r} \right> }_{\rm dt}$, index labels $\rn$ and ${\rn}^{\prime}$ are
switched.  Because the steady state currents decompose into cycles,
whether $\left( \rho / \bar{\rho} \right)$ is summed over the first
index or over the second index along a cycle, the sum is the same.
Hence the two expressions in Eq.~(\ref{eq:exp_mr_ave}) are the same.
By Jensen's inequality, $d { \left< r \right> }_{\rm dt} / dt =
{\dot{S}}^{\rm HK} \ge 0$, proving that Eq.~(\ref{eq:HK_intro}), which
is the second line of Eq.~(\ref{eq:Ddot_tot_decomp_HK}), is
non-decreasing. \textcolor{red}{$\blacksquare$}

\noindent \textbf{Remark:} 
The Hatano-Sasa generating function is constructed to replace the
transition matrix $\hat{\mathbb{T}}$ with its
adjoint~\cite{Harris:fluct_thms:07}.  In Eq.~(\ref{eq:exp_mr_ave}),
this relation is reflected in the way ${
\left< e^{-r} \right> }_{\rm dt}$ switches $\left( \rho / \bar{\rho}
\right)$ to the tail position of links, whereas in ${ \left< 1 \right>
}_{\rm dt}$ it is in the head position.  Exactly the same sum arises
if positivity of 
$\dot{D} \! \left( {\rho}^s \parallel {\bar{\rho}}^s \right)$
is proved by a similar convexity argument, from the expectation of
$\left( {\rho}^s_{{\rn}^{\prime}_s} /
{\bar{\rho}}^s_{{\rn}^{\prime}_s} \right) / 
\left( {\rho}^s_{{\rn}_s} / {\bar{\rho}}^s_{{\rn}_s} \right)$ 
over an interval $dt$, which is the inverse exponential of the
log-ratio of Hartley information appearing in 
Eq.~(\ref{eq:That_states_eval}).  Thus the two fluctuation theorems
are for generating functions spanning the same chord between two
probability measures, though the counting observables in the two
generating functions are distinct.

\section{Examples}
\label{sec:examples}

\subsection{Dualizing the housekeeping embedding thermodynamics}
\label{sec:Apol_hydr}

In the first example the system $s$ is elementary: a 2-state model of
polymerization and hydrolysis.  Our interest is in how the embedding
of such a model in a 1-parameter family of environments is captured in
the intensive and extensive thermodynamic parameters, when these
produce identical distributions ${\bar{\rho}}^s$, but with differing
housekeeping entropy rate.

The motivation for the model is a question from astrobiology: in how
far can two environments be considered analogues simply because they
produce similar distributions for a \emph{material} considered to be
of biological interest.  For instance, is Titan an analogue to early
Earth if both are believed to support significant polymerization of
small organic
molecules~\cite{Yung:org_Titan:84,Hebrard:Titan_atmos:12,Turse:org_Titan:13},
even if polymers on Titan are stable and near equilibrium at low water
activity, whereas Earth produced them (putatively) through a competition
between ligation driven by disequilibrium leaving groups such as
phosphates~\cite{Westheimer:phosphates:87,Pasek:why_P:11} or
thioesters~\cite{Goldford:P_free_metab:17,Goldford:bdry_expn:19} and
disequilibrium hydrolysis?

The polymerization/hydrolysis mechanism, though elementary, can be a
foundation for more complex heterogeneous polymerization
models~\cite{Esposito:copol_eff:10}, in which polymers once formed may
be sensitive to other consequences of water activity, such as
propensity to fold or to aggregate.  We ask, to what extent can the
rate balance that governs polymerization, and water activity
\textit{per se}, be decoupled as measures of environmental similarity.
The example will show that in the limit where one driving buffer goes
to zero concentration and a subset of reactions become strictly
irreversible, the two parameters can be made independent.

In this example \emph{all} influences on what we should call
thermodynamic order come from kinetics in buffered non-equilibrium
settings.  Total entropy production is uninformative because the
entropy associated with the polymer distribution sits atop a
housekeeping heat that can be varied freely.  Even knowing that
housekeeping heat depends on a difference of chemical potentials gives
no information when that difference is taken to infinity in an
irreversible limit.  Thus none of the conservation laws linking the
model to microscopically reversible mechanics provide information
beyond what is in the transition matrix.  Yet the
\emph{thermodynamics} of the process retains a regular representation:
dualization of the housekeeping heat introduces an intensive state
variable that is just the regular current through the
polymerization/hydrolysis cycle, even in the strictly irreversible
limit.

\subsubsection{One system, families of environments} 

\paragraph{Elementary reactions:} 

For simplicity we let environmental distributions ${\rho}^{e \mid s}$
be Poisson with large capacity as a model of chemostats, and omit them
from the explicit notation.  Half-reaction rate constants are denoted
${\hat{k}}_i$ and ${\hat{\bar{k}}}_i$ where $i$ is a label.

The first reaction type in the model is reversible dehydrating
polymerization and hydrolysis, with schema
\begin{align}
  A
& \overset{{\hat{k}}_1}{\underset{{\hat{\bar{k}}}_1}{\rightleftharpoons}}
  A^{\ast} + 
  {\rm H}_2 {\rm O} . 
\label{eq:Apol_hydr}
\end{align}
$A$ are monomers in solution, buffered at unit activity, and
$A^{\ast}$ are monomers attached to the end of a polymer by
dehydrating condensation.  $\rn$ denotes the number of $A^{\ast}$,
which we will interpret as polymer \emph{lengths}, and is the index
for the system state.  Water is also buffered, at activity $a_{{\rm
H}_2 {\rm O}}$, which is varied as a control parameter across a family
of environments.

A second process, involving only environment species to define a
reference scale for chemical potentials, is hydrolysis of a
phosphoanhydride bond, with schema
\begin{align}
  {\rm P}^{\ast} + 
  {\rm H}_2 {\rm O}
& \overset{{\hat{k}}_2}{\underset{{\hat{\bar{k}}}_2}{\rightleftharpoons}}
  {\rm P} . 
\label{eq:Pstar_hydr}
\end{align}
${\rm P}^{\ast}$ is a bound form with activity $a_{{\rm P}^{\ast}}$,
and $P$ a hydrolyzed form, with activity $a_{\rm
P}$.\footnote{Eq.~(\ref{eq:Pstar_hydr}) is a stand-in for a variety of
phosphoryl group-transfer processes, some linear as shown and some
higher-order in the hydrolyzed species~\cite{Pasek:why_P:11}.  For
higher-order reactions, appropriate powers of the activity would
replace $a_{\rm P}$ in the expressions below, without otherwise
changing the results.}

Monomer ligation driven by phosphate hydrolysis is the
stoichiometrically coupled reaction 
\begin{align}
  A + {\rm P}^{\ast}
& \overset{{\hat{k}}_3}{\underset{{\hat{\bar{k}}}_3}{\rightleftharpoons}}
  A^{\ast} + {\rm P} . 
\label{eq:Apol_Pstar_hydr}
\end{align}
Equilibrium constants for the three
reactions~(\ref{eq:Apol_hydr}--\ref{eq:Apol_Pstar_hydr}) are denoted
${\hat{K}}_i = {\hat{k}}_i / {\hat{\bar{k}}}_i$.  In actual chemistry,
detailed balance implies the relation ${\hat{K}}_3 = {\hat{K}}_1
{\hat{K}}_2$.  To simplify notation we will choose activity units to
set ${\bar{k}}_3 / {\bar{k}}_1 = 1$.  Reaction 2 contributes only to
dissipation internal to the environment (the third line in
Eq.~(\ref{eq:Ddot_tot_decomp_HK})), and we omit it, so the scale of
${\bar{k}}_2$ never enters.

\paragraph{Poisson stationary distribution over polymer lengths:}

The stationary distribution ${\bar{\rho}}^s$ for polymer length $\rn$
under the joint action of
reactions~(\ref{eq:Apol_hydr},\ref{eq:Apol_Pstar_hydr}) is Poisson
with parameter $K_{\rm eff}$, given by 
\begin{align}
  \frac{
    {\bar{\rho}}^s_{\rn + 1} \left( \rn + 1 \right)
  }{
    {\bar{\rho}}^s_{\rn}
  } 
& \equiv 
  K_{\rm eff} = 
  \frac{
    {\hat{K}}_1 + {\hat{K}}_3 a_{{\rm P}^{\ast}}
  }{
    a_{{\rm H}_2 {\rm O}} + a_{\rm P}
  } = 
  \underline{n} . 
\label{eq:Apol_K_eff_def}
\end{align}

\paragraph{The varieties of chemical disequilibrium:}

We are interested in families of environments that share the same
value of $K_{\rm eff}$ and so are indistinguishable from within the
system.  Chemical-potential coordinates within such a family are
\begin{align}
      \log 
      \left( 
        \frac{
          K_{\rm eff} a_{{\rm H}_2 {\rm O}}
        }{
          {\hat{K}}_1 
        }
      \right) 
& \equiv 
  {\mu}_H
\nonumber \\ 
      \log 
      \left( 
        \frac{
          {\hat{K}}_2 a_{{\rm P}^{\ast}} a_{{\rm H}_2 {\rm O}}
        }{
          a_{\rm P}
        }
      \right)
& \equiv 
  {\mu}_P . 
\label{eq:Apol_chem_pots}
\end{align}
${\mu}_H$ is the chemical potential for hydrolysis in the stationary
polymer distribution.  ${\mu}_P$ is the chemical potential of the
phosphate system for hydrolysis.

Within surfaces of constant $K_{\rm eff}$ we consider variation of
phosphate and water activities along contours of fixed $a_{\rm P} /
a_{{\rm H}_2 {\rm O}}$,\footnote{$a_{\rm P^{\ast}}$ is a nonlinear
function of the value of $a_{\rm P}$ along this contour.} and vary
this ratio later to define routes to irreversibility.  The
thermodynamic \textit{equilibrium} corresponds to ${\mu}_P = {\mu}_H =
0$, or 
\begin{align}
  {\hat{K}}_2 \underline{a_{{\rm P}^{\ast}}}
& = 
  \frac{a_{\rm P}}{a_{{\rm H}_2 {\rm O}}}
\nonumber \\
  \frac{\underline{a_{{\rm H}_2 {\rm O}}}}{{\hat{K}}_1}
& = 
  \frac{1}{K_{\rm eff}} . 
\label{eq:Apol_apast_eq}
\end{align}
At equilibrium the sub-system stationary distribution ${\bar{\rho}}^s$
is also the marginal ${\underline{\rho}}^s$ of the whole-system
equilibrium; to emphasize that it is fixed as ${\mu}_H$ and ${\mu}_P$
are varied across a family of environments, we reference distributions
to ${\underline{\rho}}^s$.

The activities governing reaction fluxes then depend on the
coordinates~(\ref{eq:Apol_chem_pots}) as 
\begin{align}
  a_{{\rm H}_2 {\rm O}}
& = 
  \frac{{\hat{K}}_1}{K_{\rm eff}}
  e^{{\mu}_H}
\nonumber \\ 
  a_{\rm P}
& = 
  \frac{{\hat{K}}_1}{K_{\rm eff}}
  \frac{
    e^{{\mu}_H}
    \left( e^{{\mu}_H} - 1 \right)
  }{
    \left( e^{{\mu}_P} - e^{{\mu}_H} \right)
  }
\nonumber \\ 
  a_{{\rm P}^{\ast}}
& = 
  \frac{1}{{\hat{K}}_2}
  \frac{
    e^{{\mu}_P}
    \left( e^{{\mu}_H} - 1 \right)
  }{
    \left( e^{{\mu}_P} - e^{{\mu}_H} \right)
  } . 
\label{eq:Apol_act_from_pot}
\end{align}
Fixed $a_{\rm P} / a_{{\rm H}_2 {\rm O}}$ contours near and far from
equilibrium satisfy 
\begin{align}
  \frac{{\mu}_P - {\mu}_H}{{\mu}_H}
& \underset{{\mu}_P \rightarrow 0}{\rightarrow}
  \frac{
    a_{{\rm H}_2 {\rm O}}
  }{
    a_{\rm P}
  }
\nonumber \\ 
  {\mu}_P - {\mu}_H
& \underset{{\mu}_P \rightarrow \infty}{\rightarrow}
  \log 
  \left( 
    \frac{
      a_{{\rm H}_2 {\rm O}}
    }{
      a_{\rm P}
    }
  \right) . 
\label{eq:Apol_lim_rat}
\end{align}

\paragraph{Cycle decomposition of steady-state currents:}

If environmental marginals ${\rho}^{e \mid s}$ are chemostats at the
indicated chemical potentials, then currents in the stationary system
distribution at a state $\rn$ are proportional to
${\underline{\rho}}^s_{\rn}$ by factors that do not depend on $\rn$,
and can be decomposed into three ``specific currents'' around cycles,
which are functions of the chemostat activities
\begin{align}
  {\hat{\jmath}}_{1}
& = 
  \frac{{\hat{K}}_1}{2}
  \left(
    e^{{\mu}_H} + 1 
  \right) 
\nonumber \\ 
  {\hat{\jmath}}_{3}
& = 
  \frac{{\hat{K}}_1}{2}
  \left( 
    \frac{
      e^{{\mu}_H} - 1 
    }{
      e^{{\mu}_P} - e^{{\mu}_H} 
    }
  \right) 
  \left(
    e^{{\mu}_P} + e^{{\mu}_H} 
  \right) 
\nonumber \\ 
  {\hat{\jmath}}_{\delta}
& = 
  {\hat{K}}_1
  \left(
    e^{{\mu}_H} - 1 
  \right) . 
\label{eq:Apol_3_jbars_red}
\end{align}
${\hat{\jmath}}_1$ is the average of forward and reverse currents in
reaction~(\ref{eq:Apol_hydr}), and ${\hat{\jmath}}_3$ the average of
forward and reverse currents in reaction~(\ref{eq:Apol_Pstar_hydr}),
divided by ${\underline{\rho}}^s_{\rn}$. ${\hat{\jmath}}_{\delta}$ is
the difference of forward and reverse currents, which must be equal
and opposite in reactions~(\ref{eq:Apol_hydr})
and~(\ref{eq:Apol_Pstar_hydr}) in stationary state, also divided by
${\underline{\rho}}^s_{\rn}$; it is the only current in Schnakenberg's
\textit{fundamental graph}~\cite{Schnakenberg:ME_graphs:76} for this
CRN.\footnote{${\hat{\jmath}}_{\delta}$ is so named because we have
called it a ``$\delta$-flow''
in~\cite{Krishnamurthy:CRN_moments:17,Smith:CRN_moments:17}.  The
number of possible independent net flows in the stationary state of a
CRN equals Feinberg's~\cite{Feinberg:notes:79} topological
characteristic termed \textit{deficiency} and denoted $\delta$.  The
CRN of this example has $\delta = 1$.}  The fundamental graph omits
currents ${\hat{\jmath}}_1$ and ${\hat{\jmath}}_3$ because they do not
lead to dissipation in the stationary state, but they do for more
general states.

\subsubsection{The housekeeping entropy rate}

The housekeeping entropy rate~(\ref{eq:HK_intro}), for an
arbitrary system distribution ${\rho}^s$, evaluates in terms of the
specific currents~(\ref{eq:Apol_3_jbars_red}) and the density
${\underline{\rho}}^s$, to
\begin{align}
\lefteqn{
  {\dot{S}}^{\rm HK}
\equiv 
  \sum_{\rn}
  \sum_{{\rn}^{\prime} \neq \rn}
  {\sigma}^{\rm HK}_{\rn {\rn}^{\prime}} \! 
  w^s_{\rn {\rn}^{\prime}} {\rho}^s_{{\rn}^{\prime}}   
} & 
\nonumber \\
& = 
  \sum_{\rn}
  \left\{
    {\left( \frac{\rho}{\underline{\rho}} \right)}^s_{\rn + 1} 
    \left[
      \left( 
        {\hat{\jmath}}_1 + \frac{{\hat{\jmath}}_{\delta}}{2} 
      \right) 
      {\mu}_H - 
      \left( 
        {\hat{\jmath}}_3 - \frac{{\hat{\jmath}}_{\delta}}{2} 
      \right)
      \left( {\mu}_P - {\mu}_H \right)
    \right]
  \right.
\nonumber \\ 
& \mbox{} + 
  \left. 
    {\left( \frac{\rho}{\underline{\rho}} \right)}^s_{\rn} 
    \left[ 
      \left( 
        {\hat{\jmath}}_3 + \frac{{\hat{\jmath}}_{\delta}}{2} 
      \right)
      \left( {\mu}_P - {\mu}_H \right) - 
      \left( 
        {\hat{\jmath}}_1 - \frac{{\hat{\jmath}}_{\delta}}{2} 
      \right)
      {\mu}_H 
    \right] 
  \right\}
  {\underline{\rho}}^s_{\rn}
\nonumber \\ 
& \sim
  {\hat{\jmath}}_{\delta}
  {\mu}_P + 
  \int dn \, 
  {\underline{\rho}}^s_n
  \frac{\partial}{\partial n}
  {\left( \frac{\rho}{\underline{\rho}} \right)}^s_n
  \left[
    {\hat{\jmath}}_1 
    {\mu}_H + 
    {\hat{\jmath}}_3 
    \left( {\mu}_H - {\mu}_P \right) 
  \right] . 
\label{eq:Apol_SHK_pos_form}
\end{align}
The second expression is a continuum approximation to first order in
derivatives, with $-\log {\rho}^s_n$ understood as usual to be approximated
by the appropriate large-deviation function $S_{\rm eff}$.

To check that the exact (discrete) form of
Eq.~(\ref{eq:Apol_SHK_pos_form}) is positive semidefinite for
arbitrary ${\rho}^s$, define measures  
\begin{align}
  p 
& \equiv 
  \frac{
    {\hat{\jmath}}_1 - {\hat{\jmath}}_{\delta} / 2
  }{
    {\hat{\jmath}}_1 + {\hat{\jmath}}_3
  }
& 
  1-p 
& \equiv 
  \frac{
    {\hat{\jmath}}_3 + {\hat{\jmath}}_{\delta} / 2
  }{
    {\hat{\jmath}}_1 + {\hat{\jmath}}_3
  }
\nonumber \\ 
  q 
& \equiv 
  \frac{
    {\hat{\jmath}}_1 + {\hat{\jmath}}_{\delta} / 2
  }{
    {\hat{\jmath}}_1 + {\hat{\jmath}}_3
  }
& 
  1-q 
& \equiv 
  \frac{
    {\hat{\jmath}}_3 - {\hat{\jmath}}_{\delta} / 2
  }{
    {\hat{\jmath}}_1 + {\hat{\jmath}}_3
  } . 
\label{eq:2pt_probs}
\end{align}
From the formulae~(\ref{eq:Apol_3_jbars_red}) it follows that 
\begin{align}
  \frac{p}{q}
& = 
  e^{- {\mu}_H}
& 
  \frac{1-p}{1-q}
& = 
  e^{{\mu}_P - {\mu}_H} , 
\label{eq:Apol_ratios_def}
\end{align}
and thus Eq.~(\ref{eq:Apol_SHK_pos_form}) can be written
\begin{align}
  {\dot{S}}^{\rm HK}
& = 
  \sum_{\rn}
  {\underline{\rho}}^s_{\rn}
  \left( {\hat{\jmath}}_1 + {\hat{\jmath}}_3 \right) 
  \left[
    {\left( \frac{\rho}{\underline{\rho}} \right)}^s_{\rn + 1} \! \! \!
    D \! \left( q \parallel p \right) + 
    {\left( \frac{\rho}{\underline{\rho}} \right)}^s_{\rn} 
    D \! \left( p \parallel q \right)
  \right] , 
\label{eq:Apol_SHK_pos_DKL}
\end{align}
with $D \! \left( q \parallel p \right)$ a Kullback-Leibler divergence
as elsewhere.

\paragraph{Housekeeping entropy rate as an embedding vector:} 

Eq.~(\ref{eq:Apol_SHK_pos_DKL}) can be put in the form 
\begin{align}
  {\dot{S}}^{\rm HK}
& = 
  \sum_{\rn}
  {\rho}^s_{\rn}
  {\dot{\sigma}}^{\rm HK}_{\rn} . 
\label{eq:Apol_SHK_pos_HKvec}
\end{align}
${\dot{\sigma}}^{\rm HK} \equiv \left[ {\dot{\sigma}}^{\rm HK}_{\rn}
\right]$ is a vector with positive-semidefinite components, which by
Eq.~(\ref{eq:Apol_ratios_def}) equals zero only at ${\mu}_P = {\mu}_H
= 0$.  

As noted, ${\bar{\rho}}^s$ is an extremal vector for the intensive
relative entropy $- D \! \left( {\rho}^s \parallel {\bar{\rho}}^s
\right)$ in the simplex of distributions ${\rho}^s$.  By
Eq.~(\ref{eq:Apol_SHK_pos_form}), at this extremal point of $s$,
$\sum_{\rn} {\bar{\rho}}^s_{\rn} {\dot{\sigma}}^{\rm HK}_{\rn} =
{\hat{\jmath}}_{\delta} {\mu}_P$.

\subsubsection{Legendre duality for housekeeping entropy rate}

The chemical potential ${\mu}_P \rightarrow \infty$ if the activity
$a_{\rm P} \rightarrow 0$ at fixed $a_{{\rm P}^{\ast}}$ and $a_{{\rm
H}_2 {\rm O}}$.  In this limit schema~(\ref{eq:Apol_Pstar_hydr})
becomes an irreversible reaction.  Phenomenologically, all
``thermodynamic'' characteristics of the system remain regular, only
the energetic accounting breaks down because it is referenced to the
equilibrium state variable $\log a_{\rm P}$, in a system that nowhere
couples to an equilibrium environment.

The divergence of ${\dot{S}}^{\rm HK}$ in this limit is like the
divergence of any extensive thermodynamic potential in the limit that
one of the extensive state variables diverges, except that
${\dot{S}}^{\rm HK}$ is an inherently \emph{non-equilibrium}
potential, and ${\mu}_P$ only behaves like an extensive variable with
respect to \emph{dissipation}.\footnote{Note that ${\dot{S}}^{\rm HK}$
has \emph{no status} with respect to equilibrium; while entropies are
extensive, ${\dot{S}}^{\rm HK}$ depends inherently on a \emph{rate}.} 

From the view that thermodynamics is about statistical organization
and not fundamentally about energy, it is natural to Legendre
transform ${\dot{S}}^{\rm HK}$ to expose the dynamically intensive
current that is dual to ${\mu}_P$.  For dualization, we work within
the tangent space to the family of constant $K_{\rm eff}$, but now
vary ${\mu}_P$ independently of ${\mu}_H$, rather than varying within
the non-linear contours~(\ref{eq:Apol_apast_eq}) at fixed $a_{\rm P} /
a_{{\rm H}_2 {\rm O}}$.

The gradient of ${\dot{S}}^{\rm HK}$ with respect to ${\mu}_P$ at
fixed ${\mu}_H$ is 
\begin{align}
\lefteqn{
  \frac{\partial {\dot{S}}^{\rm HK}}{\partial {\mu}_P} = 
} & 
\nonumber \\ 
& 
  {\hat{K}}_1
  \sum_{\rn}
  {\underline{\rho}}^s_{\rn}
  \left( 
    \frac{
        e^{{\mu}_H} - 1 
    }{
      e^{{\mu}_P} - e^{{\mu}_H} 
    }
  \right) 
  \left\{
    \left[
      e^{{\mu}_P}
      {\left( \frac{\rho}{\underline{\rho}} \right)}^s_{\rn} - 
      e^{{\mu}_H}
      {\left( \frac{\rho}{\underline{\rho}} \right)}^s_{\rn + 1} 
    \right] 
    \vphantom{
      \left( 
        \frac{
          e^{{\mu}_P} e^{{\mu}_H} 
        }{
          e^{{\mu}_P} + e^{{\mu}_H} 
        }
      \right) 
    }
  \right.
\nonumber \\ 
& \mbox{} 
  \phantom{\frac{{\hat{K}}_1}{K_{\rm eff}}} + 
  \left. 
    \left( 
      \frac{
        e^{{\mu}_P} e^{{\mu}_H} 
      }{
        e^{{\mu}_P} + e^{{\mu}_H} 
      }
    \right) 
    \left[
      {\left( \frac{\rho}{\underline{\rho}} \right)}^s_{\rn + 1} - 
      {\left( \frac{\rho}{\underline{\rho}} \right)}^s_{\rn + 1} 
    \right] 
    \left( 
      {\mu}_H - {\mu}_P
    \right) 
  \right\} 
\nonumber \\ 
& \sim 
  {\hat{\jmath}}_{\delta} - 
  \int dn \, 
  {\underline{\rho}}^s_n
  \frac{\partial}{\partial n}
  {\left( \frac{\rho}{\underline{\rho}} \right)}^s_n
  \left[ 
    {\hat{\jmath}}_3 + 
    \frac{\partial {\hat{\jmath}}_3}{\partial {\mu}_P}
    \left( {\mu}_P - {\mu}_H \right) 
  \right] . 
\label{eq:Sdot_HK_by_muP}
\end{align}
By Eq.~(\ref{eq:Apol_3_jbars_red}), ${\hat{\jmath}}_3 \left( {\mu}_P -
{\mu}_H \right)$ is convex in ${\mu}_P$, so its derivative, the term
in square brackets in the final line of Eq.~(\ref{eq:Sdot_HK_by_muP}),
is invertible to a value for ${\mu}_P$.  The Legendre dual potential
to ${\dot{S}}^{\rm HK}$ on ${\mu}_P$, which we denote ${\dot{F}}_P$,
is then given by
\begin{align}
\lefteqn{
  {\dot{F}}_P \! 
  \left( 
    \frac{\partial {\dot{S}}^{\rm HK}}{\partial {\mu}_P} , 
    {\mu}_H 
  \right) 
\equiv 
  {\mu}_P 
  \frac{\partial {\dot{S}}^{\rm HK}}{\partial {\mu}_P} - 
  {\dot{S}}^{\rm HK}
} &
\nonumber \\ 
& \approx 
  - \int dn \, 
  {\underline{\rho}}^s_n
  \frac{\partial}{\partial n}
  {\left( \frac{\rho}{\underline{\rho}} \right)}^s_n
  \left[
    \frac{\partial {\hat{\jmath}}_3}{\partial {\mu}_P}
    {\mu}_P
    \left( 
      {\mu}_P - {\mu}_H
    \right) + 
    \left( {\hat{\jmath}}_3 + {\hat{\jmath}}_1 \right) 
    {\mu}_H 
  \right] . 
\label{eq:Apol_Leg_muP}
\end{align}

\paragraph{Independent variation of control parameters in the
irreversible limit:}

We may now consider the effects of varying ${\mu}_H$, which remains
finite, across the one-parameter family of contours of different
$a_{\rm P} / a_{{\rm H}_2 {\rm O}}$ as $a_{\rm P} \rightarrow 0$.
$\left( {\hat{\jmath}}_3 + {\hat{\jmath}}_1 \right)$ approaches a
constant independent of ${\mu}_P$, and the only term in
Eq.~(\ref{eq:Apol_Leg_muP}) including multiples of diverging ${\mu}_P$
evaluates to
\begin{align}
\lefteqn{
  - \frac{\partial {\hat{\jmath}}_3}{\partial {\mu}_P}
  {\mu}_P
  \left( 
    {\mu}_P - {\mu}_H
  \right) 
  \rightarrow 
} & 
\nonumber \\ 
& 
  \rightarrow 
  {\hat{K}}_1
  \left(
    \frac{a_{\rm P} + a_{{\rm H}_2 {\rm O}}}{a_{{\rm H}_2 {\rm O}}} 
  \right)
  {\mu}_H 
  \left[ 
    1 - 
    \frac{1}{3!}
    {
      \left(
        \frac{a_{{\rm H}_2 {\rm O}}}{a_{\rm P}} 
      \right)
    }^2
    {\mu}_H^2
  \right] 
& : \; 
  {\mu}_H \rightarrow 0 
\nonumber \\ 
& 
  \rightarrow 
  2 {\hat{\jmath}}_3
  e^{- 
    \left( {\hat{K}}_1 / K_{\rm eff} a_{\rm P} \right) 
    {\mu}_H
  }
  {\mu}_P
  \left( {\mu}_P - {\mu}_H \right) 
& : \; 
  {\mu}_H \gtrsim 1 . 
\label{eq:Apol_curr_der_asymp}
\end{align}
It approaches ${\hat{K}}_1 {\mu}_H$ within a range around ${\mu}_H =
0$ that becomes vanishingly small as $a_{\rm P} \rightarrow 0$, and
converges exponentially to zero outside that range.

Thus the susceptibility that is the Legendre dual to ${\mu}_P$,
$\partial {\dot{S}}^{\rm HK} / \partial {\mu}_P \rightarrow
{\hat{\jmath}}_{\delta}$, a function only of ${\mu}_H$, almost
everywhere.  
The potential ${\dot{F}}_P$ retains only the ${\mu}_P$-independent
terms in Eq.~(\ref{eq:Apol_Leg_muP}).  If, for example, we choose
${\rho}^s$ a macrostate with mean $\bar{n}$, then $\int dn \,
{\underline{\rho}}^s_n \partial {\left( \rho / \underline{\rho}
\right)}^s_n / \partial n = \left( \bar{n} - \underline{n} \right)$
and 
${\dot{F}}_P \rightarrow - \left( \bar{n} - \underline{n} \right) 
\left( {\hat{\jmath}}_3 + {\hat{\jmath}}_1 \right) {\mu}_H$.

In the irreversible limit the problem is seen to separate.
${\bar{\rho}}^s$ dictates the intensive thermodynamics of the polymer
system, while along a contour that fixes $K_{\rm eff}$, the dual
potential ${\dot{F}}_P$ describes the non-trivial dependence of
entropy change in the environment on ${\rho}^s$.  ${\dot{S}}^{\rm HK}$
diverges harmlessly as an extensively-scaling potential independent of
${\rho}^s$ with a regular susceptibility ${\hat{\jmath}}_{\delta}$.

\subsection{Metastability and ${\bar{\rho}}^s$ as the reference
measure for relative entropy}
\label{sec:Schlogl_examp}

The second example uses the same buffered driving environment as the
first, but separates phosphate-driven ligation into a distinct,
self-catalyzed channel while leaving the reaction~(\ref{eq:Apol_hydr})
with water uncatalyzed.  The model for system catalysis is a variant
on the cubic Schl{\"{o}}gl model of
Fig.~\ref{fig:L0man_1spec}.  It is chosen to highlight
the difference between the entropy
rate~(\ref{eq:sigdot_env_def}) typically assigned as
``environmental'' and the housekeeping entropy
rates~(\ref{eq:sigdot_HK_def}).  A quantity ${\dot{S}}^{\rm
env}$ is a function only of the realized distributions ${\rho}^s$ and
${\rho}^{e \mid s}$.  Nowhere does it reflect the joint participation
of uncatalyzed and catalyzed reactions within the same \emph{system}.

The difference ${\dot{S}}^{\rm HK} - {\dot{S}}^{\rm env}$ may have
either sign.  We illustrate the case where phosphate activities are
chosen to put the system in its bistable regime, and ${\mu}_P$ is
chosen to make the unstable root $n_2$ of the rate
equation~(\ref{eq:cubic_n_EOM}) a fixed point.  For equivalent rate
constants a Poisson distribution ${\rho}^s$ at mean $n_2$ would be the
driven stationary distribution in the previous example, and the
${\dot{S}}^{\rm env}$ \emph{is the same} in the two models.  However,
${\dot{S}}^{\rm HK} - {\dot{S}}^{\rm env} < 0$ because the loss of
large-deviation accessibility in the system results not only from it
Shannon entropy change, but from the measure of that entropy
\emph{relative} to the true, bistable stationary distribution
${\bar{\rho}}^s$.

The autocatalytic model replaces the schema~(\ref{eq:Apol_Pstar_hydr})
with 
\begin{align}
  2 A^{\ast} + A + {\rm P}^{\ast}
& \overset{
    {\hat{k}}_4 
  }{
    \underset{
      {\hat{\bar{k}}}_4 
    }{
      \rightleftharpoons}
    }
  3 A^{\ast} + {\rm P} . 
\label{eq:Apol_PP_hydr}
\end{align}
Catalysis does not change energetics, so the rate constants must
satisfy ${\hat{K}}_4 = {\hat{K}}_1 {\hat{K}}_2$.  

In the mass-action rate law~(\ref{eq:cubic_n_EOM}), the roots $n_1 <
n_2 < n_3$ and $n_2$ is unstable.  We set a characteristic scale for
transitions between water and phosphate control by choosing relative
rate constants ${\bar{k}}_4 / {\bar{k}}_1 = 1 / n_2^2$.

The Schl{\"{o}}gl model is a birth-death process, so ${\bar{\rho}}^s$
is exactly solvable.  In place of $K_{\rm eff}$ from
Eq.~(\ref{eq:Apol_K_eff_def}) adjacent indices are related by a
position-dependent effective equilibrium constant 
\begin{align}
  \frac{
    {\bar{\rho}}^s_{\rn + 1} \left( \rn + 1 \right)
  }{
    {\bar{\rho}}^s_{\rn}
  }
& \equiv 
  K \! \left( \rn \right) = 
    \frac{
      {\hat{K}}_1 + {\hat{K}}_4 a_{\rm P^{\ast}} \, 
      \rn \left( \rn - 1 \right) / n_2^2 
    }{
      a_{{\rm H}_2 {\rm O}} + 
      a_{\rm P} \, 
      \rn \left( \rn - 1 \right) / n_2^2 
    } . 
\label{eq:K_rn_def}
\end{align}
Note that $K \! \left( n_2 \right) \rightarrow K_{\rm eff}$ to integer
rounding error, at all activity levels.

In terms of the same roots, choose water activity to set the lower
limit of Eq.~(\ref{eq:K_rn_def}) at $\rn \le 1$ to
\begin{align}
  \frac{{\hat{K}}_1}{a_{{\rm H}_2 {\rm O}}}
& = 
  \frac{1}{\sum_i 1 / n_i} \equiv 
  \underline{n} . 
\label{eq:Apolcat_H2O_act}
\end{align}
We will vary the activities in the phosphorus system so that
${\bar{\rho}}^s$ moves from a lower stable mode near equilibrium,
through a bistable phase, to an upper stable mode driven by
phosphorylation farther from equilibrium.  

Equilibrium is again defined as in Eq.~(\ref{eq:Apol_apast_eq}), and
there $K \! \left( \rn \right) \overset{\rm eq}{=} \underline{n}  \; ;
\;  \forall \rn$.  The marginal distribution for $s$ in the
whole-system equilibrium, ${\underline{\rho}}^s$, is then Poisson with
parameter $\underline{n}$.  Recall that, as a solution in detailed
balance, it does not and cannot reflect the fact that $s$ is a CRN
capable of bistability.  To match rate constants to the roots $\left\{
n_1, n_2 , n_3 \right\}$ of the Schl{\"{o}}gl model, we define a
reference equilibrium for the phosphorus system with activity levels
at
\begin{align}
  {\hat{K}}_2
  \underline{a_{\rm P^{\ast}}} = 
  \frac{\underline{a_{\rm P}}}{a_{{\rm H}_2 {\rm O}}}
& = 
  1 . 
\label{eq:Apolcat_Pacts_crit}
\end{align}

We now consider a family of driving environments in which the chemical
potential ${\mu}_P$ is partitioned between the activities $a_{\rm
P^{\ast}}$ and $a_{\rm P}$ in the proportions
\begin{align}
  \log 
  \left( 
    \frac{
      a_{\rm P^{\ast}}
    }{
      \underline{a_{\rm P^{\ast}}}
    } 
  \right)
& = 
  \frac{
    \log 
    \left[ 
      \left( \sum_i n_i \right) 
      n_2^2 / 
      \left( \prod_i n_i \right) 
    \right] 
  }{
    \log 
    \left[ 
      \left( \sum_i n_i \right) 
      \left( \sum_i 1 / n_i \right) 
    \right] 
  }
  {\mu}_P
\nonumber \\ 
  \log 
  \left( 
    \frac{
      a_{\rm P}
    }{
      \underline{a_{\rm P}}
    } 
  \right)
& = 
  - \frac{
    \log 
    \left[ 
      \left( \prod_i n_i \right) 
      \left( \sum_i 1 / n_i \right) / 
      n_2^2
    \right] 
  }{
    \log 
    \left[ 
      \left( \sum_i n_i \right) 
      \left( \sum_i 1 / n_i \right) 
    \right] 
  }
  {\mu}_P . 
\label{eq:Apolcat_expfam_rats}
\end{align}
The two potentials~(\ref{eq:Apolcat_expfam_rats}) are coordinates in
an exponential family for the phosphorus system.  Where ${\mu}_P =
\log \left[ \left( \sum_i n_i \right) \left( \sum_i 1 / n_i \right)
\right]$, the three roots of the mass action law pass through the
values $\left\{ n_1, n_2, n_3 \right\}$ from
Fig.~\ref{fig:L0man_1spec}.  

\begin{figure}[ht]
\begin{center} 
  \includegraphics[scale=0.425]{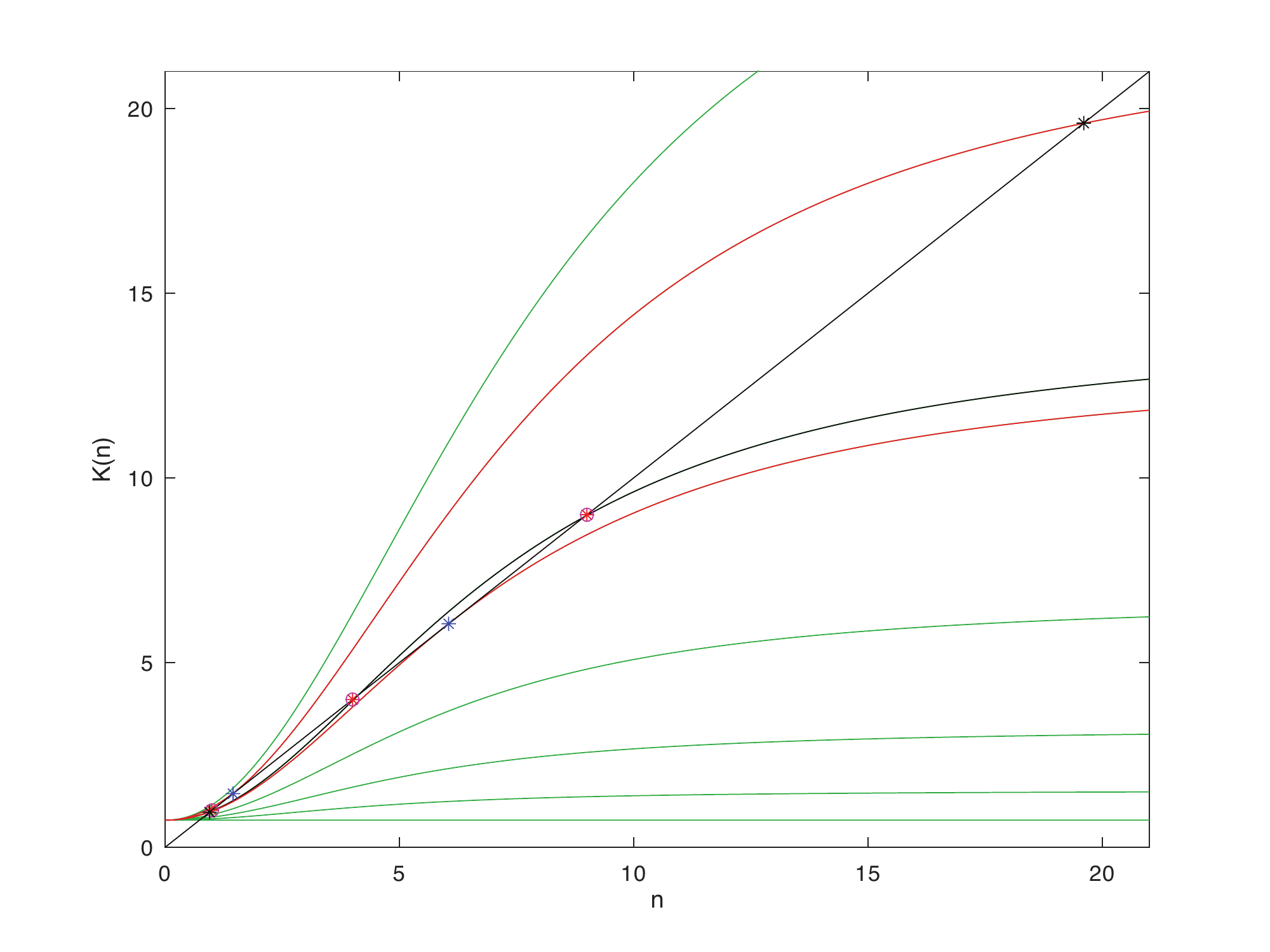} 
\caption{ 
  $K \! \left( \rn \right)$ on a sequence of increasing ${\mu}_P$
  values with proportions~(\ref{eq:Apolcat_expfam_rats}).  Curve at
  ${\mu}_P = \log \left[ \left( \sum_i n_i \right) \left( \sum_i 1 /
  n_i \right) \right]$ is black.  Boundaries of bistable range are
  red.  Fixed points  $K \! \left( \rn \right) = \rn$ shown with
  markers. 
  \label{fig:driven_Schlogl_Kn} 
}
\end{center}
\end{figure}

The difference in relative dissipations between the driven steady
state ${\bar{\rho}}^s$and ${\underline{\rho}}^s$, at adjacent $\rn$
indices, is given by the ratio of effective rate constants
\begin{align}
  \log
  \left(
    \frac{
      {\bar{\rho}}^s_{\rn + 1} / {\underline{\rho}}^s_{\rn + 1} 
    }{
      {\bar{\rho}}^s_{\rn} / {\underline{\rho}}^s_{\rn}
    }
  \right) 
& = 
  \frac{
    K \! \left( \rn \right) 
  }{
    \underline{n}
  } 
\nonumber \\ 
& = 
  \frac{
    1 + 
    \left( a_{\rm P^{\ast}} / \underline{a_{\rm P^{\ast}}} \right)
    \rn \left( \rn - 1 \right) / n_2^2 
  }{
    1 + 
    \left( a_{\rm P} / \underline{a_{\rm P}} \right)
    \rn \left( \rn - 1 \right) / n_2^2 
  }
\nonumber \\ 
& = 
  \frac{
    1 + 
    e^{{\mu}_P}
    \left( a_{\rm P} / \underline{a_{\rm P}} \right)
    \rn \left( \rn - 1 \right) / n_2^2 
  }{
    1 + 
    \left( a_{\rm P} / \underline{a_{\rm P}} \right)
    \rn \left( \rn - 1 \right) / n_2^2 
  } . 
\label{eq:K_rn_eval}
\end{align}
Eq.~(\ref{eq:K_rn_eval}) is a sigmoidal function of $\log \left( \rn /
n_2 \right)$ in the range  $K \! \left( \rn \right) / \underline{n}
\in \left[ 1 , e^{{\mu}_P} \right)$, graphed versus ${\mu}_P$ in
Fig.~\ref{fig:driven_Schlogl_Kn}. 
The hydrolysis potential ${\mu}_H$ from Eq.~(\ref{eq:Apol_chem_pots}
for the linear system is replaced in the autocatalytic case by a
position-dependent chemical potential
\begin{align}
  \mu \! \left( \rn \right)
& \equiv 
  \log 
  \left(
    \frac{
      a_{{\rm H}_2 {\rm O}}
      {\bar{\rho}}^s_{\rn + 1} \left( \rn + 1 \right)
    }{
      {\hat{K}}_1
      {\bar{\rho}}^s_{\rn}
    }
  \right) = 
  \log 
  \left( 
    \frac{K \! \left( \rn \right)}{\underline{n}}
  \right) . 
\label{eq:mu_rn_def}
\end{align}

In the cycle decomposition, the specific currents exiting and entering
a site $\rn$, now functions of $\rn$, depend on activities as 
\begin{align}
  {\hat{\jmath}}_{1}
& = 
  \frac{1}{2}
  \left(
    K \! \left( \rn \right) a_{{\rm H}_2 {\rm O}} + 
    {\hat{K}}_1
  \right) 
\nonumber \\ 
  {\hat{\jmath}}_{3}
& = 
  \frac{1}{2}
  \frac{
    \rn \left( \rn - 1 \right) 
  }{
    n_2^2
  }
  \left(
    {\hat{K}}_4 a_{\rm P^{\ast}} + 
    K \! \left( \rn \right) a_{\rm P}
  \right) 
\nonumber \\ 
  {\hat{\jmath}}_{\delta}
& = 
    K \! \left( \rn \right) a_{{\rm H}_2 {\rm O}} - 
    {\hat{K}}_1
\nonumber \\
& = 
  \rn \left( \rn - 1 \right) 
  \left( 
    {\hat{K}}_4 a_{\rm P^{\ast}} - 
    K \! \left( \rn \right) a_{\rm P}
  \right) . 
\label{eq:Apolcat_3_jbars_raw}
\end{align}
In the chemical-potential coordinates, these become 
\begin{align}
  {\hat{\jmath}}_{1}
& = 
  \frac{{\hat{K}}_1}{2}
  \left(
    e^{\mu \left( \rn \right)} + 1 
  \right) 
\nonumber \\ 
  {\hat{\jmath}}_{3}
& = 
  \frac{{\hat{K}}_1}{2}
  \frac{
    \rn \left( \rn - 1 \right)
  }{
    n_2^2
  }
  \left( 
    e^{{\mu}_P} + 
    e^{\mu \left( \rn \right)}
  \right) 
  \frac{
    a_{\rm P}
  }{
    \underline{a_{\rm P}}
  } 
\nonumber \\ 
  {\hat{\jmath}}_{\delta}
& = 
  {\hat{K}}_1
  \left(
    e^{\mu \left( \rn \right)} - 1 
  \right) 
\nonumber \\ 
& = 
  {\hat{K}}_1
  \frac{
    \rn \left( \rn - 1 \right)
  }{
    n_2^2
  }
  \left( 
    e^{{\mu}_P} - 
    e^{\mu \left( \rn \right)}
  \right) 
  \frac{
    a_{\rm P}
  }{
    \underline{a_{\rm P}}
  } . 
\label{eq:Apolcat_3_jbars_red}
\end{align}
Positive-semidefiniteness of ${\dot{S}}^{\rm HK}$ again follows by
using the ratios~(\ref{eq:2pt_probs}) to produce the
form~(\ref{eq:Apol_SHK_pos_DKL}), where now instead of
Eq.~(\ref{eq:Apol_ratios_def}), we have
\begin{align}
  \frac{p}{q}
& = 
  e^{- \mu \left( \rn \right)}
& 
  \frac{1-p}{1-q}
& = 
  e^{{\mu}_P - \mu \left( \rn \right)} . 
\label{eq:Apolcat_ratios_def}
\end{align}

For this example we separate out a term ${\dot{S}}^{\rm env}$, which
is the sum over transitions of Eq.~(\ref{eq:sigdot_env_def}), from
${\dot{S}}^{\rm HK}$, to give
\begin{align}
  {\dot{S}}^{\rm HK}
& = 
  \sum_{\rn}
  \left\{
    {\hat{K}}_1
    \frac{
      \rn \left( \rn - 1 \right)
    }{
      n_2^2
    }
    \left[
      \frac{
        a_{\rm P^{\ast}}
      }{
        \underline{a_{\rm P^{\ast}}}
      } 
      {\rho}^s_{\rn} - 
      \frac{
        a_{\rm P}
      }{
        \underline{a_{\rm P}}
      } 
      {\rho}^s_{\rn + 1}     
      \frac{
        \left( \rn + 1 \right)
      }{
        \underline{n}
      }
    \right] 
    {\mu}_P
    \vphantom{
      \left[ {\left( \frac{\rho}{\bar{\rho}} \right)}^s_{\rn} \right]
    }
  \right.
\nonumber \\
& \phantom{\sum_{\rn}} \quad + 
  \left.
    \left( {\hat{\jmath}}_1 + {\hat{\jmath}}_3 \right) 
    \left[
      {\rho}^s_{\rn + 1} 
      \frac{
        \left( \rn + 1 \right)
      }{
        K \! \left( \rn \right)
      } - 
      {\rho}^s_{\rn}
    \right] 
    \mu \! \left( \rn \right) 
  \right\}
\nonumber \\ 
& = 
  \sum_{\rn}
  \left\{
    {\hat{K}}_1
    \frac{
      \rn \left( \rn - 1 \right)
    }{
      n_2^2
    }
    {\underline{\rho}}^s_{\rn}
    \left[
      \frac{
        a_{\rm P^{\ast}}
      }{
        \underline{a_{\rm P^{\ast}}}
      } 
      {\left( \frac{\rho}{\underline{\rho}} \right)}^s_{\rn} - 
      \frac{
        a_{\rm P}
      }{
        \underline{a_{\rm P}}
      } 
      {\left( \frac{\rho}{\underline{\rho}} \right)}^s_{\rn + 1} 
    \right] 
    {\mu}_P
    \vphantom{
      \left[ {\left( \frac{\rho}{\bar{\rho}} \right)}^s_{\rn} \right]
    }
  \right.
\nonumber \\
& \phantom{\sum_{\rn}} \quad + 
  \left.
    \left( {\hat{\jmath}}_1 + {\hat{\jmath}}_3 \right) 
    {\bar{\rho}}^s_{\rn}
    \left[
      {\left( \frac{\rho}{\bar{\rho}} \right)}^s_{\rn + 1} - 
      {\left( \frac{\rho}{\bar{\rho}} \right)}^s_{\rn} 
    \right] 
    \mu \! \left( \rn \right) 
  \right\}
\nonumber \\ 
& = 
  {\dot{S}}^{\rm env} + 
  \sum_{\rn}
    \left( {\hat{\jmath}}_1 + {\hat{\jmath}}_3 \right) 
    {\bar{\rho}}^s_{\rn}
    \left[
      {\left( \frac{\rho}{\bar{\rho}} \right)}^s_{\rn + 1} - 
      {\left( \frac{\rho}{\bar{\rho}} \right)}^s_{\rn} 
    \right] 
    \mu \! \left( \rn \right) . 
\label{eq:Apolcat_SHK_split}
\end{align}
${\dot{S}}^{\rm env}$ does not depend on ${\bar{\rho}}^s$, and differs
from the sum of terms multiplying ${\mu}_P$ in
Eq.~(\ref{eq:Apol_SHK_pos_form}) only by the $\rn$-dependence now
multiplying ${\hat{K}}_1$ in Eq.~(\ref{eq:Apolcat_SHK_split}).

\begin{figure}[ht]
\begin{center} 
  \includegraphics[scale=0.425]{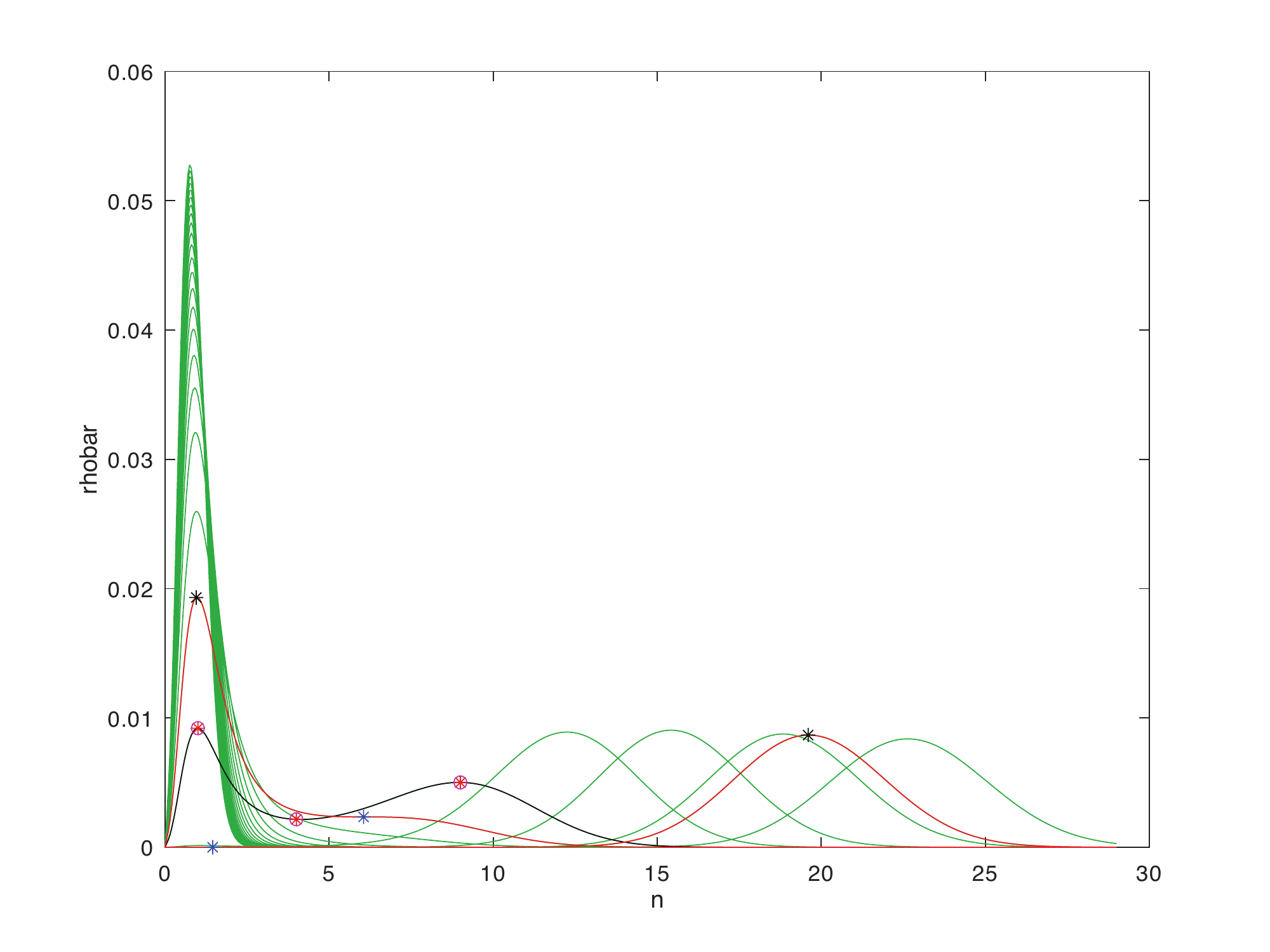} 
\caption{ 
  densities ${\bar{\rho}}^s$ for a scale factor $4 \times \left\{ 1, 4,
  9 \right\}$ for roots
  \label{fig:driven_Schlogl_rhobar} 
}
\end{center}
\end{figure}

\begin{figure}[ht]
\begin{center} 
  \includegraphics[scale=0.425]{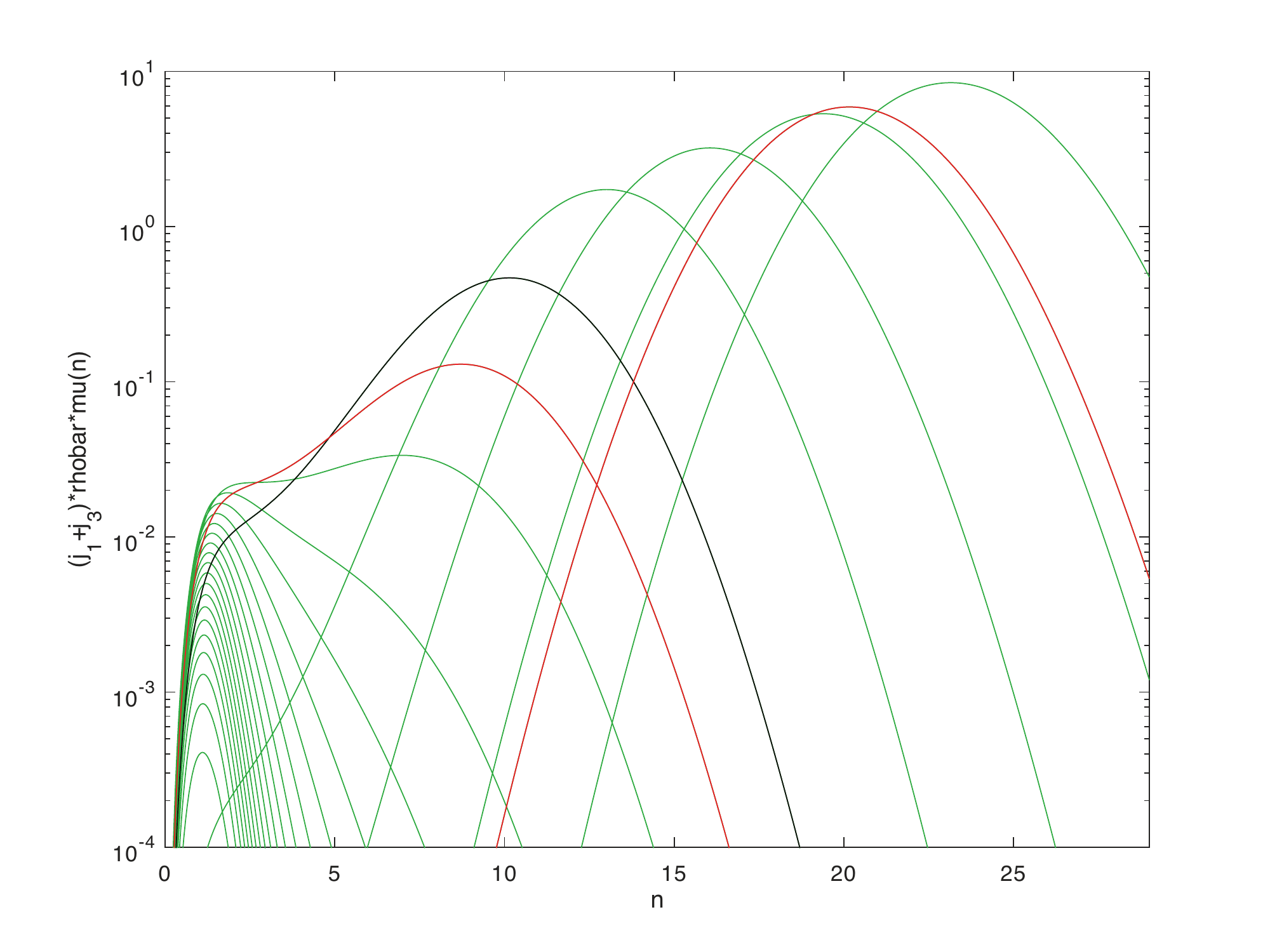} 
\caption{ 
  The combination $\left( {\hat{\jmath}}_1 + {\hat{\jmath}}_3 \right)
  {\bar{\rho}}^s_{\rn} \mu \! \left( \rn \right)$
  \label{fig:driven_Schlogl_HK_grad_coeff} 
}
\end{center}
\end{figure}

All dependence of ${\dot{S}}^{\rm HK}$ on ${\bar{\rho}}^s$ comes from
the second term involving $\mu \! \left( \rn \right)$.
The shape of ${\bar{\rho}}^s_{\rn}$ is shown as a function of
${\mu}_P$ in Fig.~\ref{fig:driven_Schlogl_rhobar}, and the lumped
coefficient $\left( {\hat{\jmath}}_1 + {\hat{\jmath}}_3 \right)
{\bar{\rho}}^s_{\rn} \, \mu \! \left( \rn \right)$ multiplying the
difference term in $\left( {\rho}^s / {\bar{\rho}}^s \right)$ in
Eq.~(\ref{eq:Apolcat_SHK_split}) is shown in 
Fig.~\ref{fig:driven_Schlogl_HK_grad_coeff}.

As an example, consider the value ${\mu}_P = \log \left[ \left( \sum_i
n_i \right) \left( \sum_i 1 / n_i \right) \right]$ that makes $\left\{
n_1, n_2 , n_3 \right\}$ the three fixed points, and evaluate the
embedding entropies for a Poisson distribution ${\rho}^s$ with mean
$n_2$, the unstable fixed point.  The ``environmental'' component 
\begin{align}
  {\dot{S}}^{\rm env}
& = 
  {\hat{K}}_1
  \sum_{\rn}
  {\rho}^s_{\rn}
  \frac{
  \rn \left( \rn - 1 \right)
  }{
    n_2^2
  }
  \left( 
    e^{{\mu}_P} - 
    \frac{n_2}{\underline{n}}
  \right)
  \frac{
    a_{\rm P}
  }{
    \underline{a_{\rm P}}
  } 
  {\mu}_P
\nonumber \\ 
& = 
  {\hat{K}}_1
  \left( 
    e^{{\mu}_P} - 
    e^{\mu \left( n_2 \right)}
  \right)
  \frac{
    a_{\rm P}
  }{
    \underline{a_{\rm P}}
  } 
  {\mu}_P
\nonumber \\ 
& = 
  {\hat{\jmath}}_{\delta} \! \left( n_2 \right) 
  {\mu}_P . 
\label{eq:Apolcat_Sdot_env_unst_ACK}
\end{align}
is the same as the value ${\hat{\jmath}}_{\delta} {\mu}_P$ following
from Eq.~(\ref{eq:Apol_SHK_pos_form}) for the non-catalytic CRN, if we
set ${\mu}_H = \log \left( n_2 / \underline{n} \right)$ in
Eq.~(\ref{eq:Apol_chem_pots}), where $K_{\rm eff} \rightarrow n_2$ and
this ${\rho}^s$ is a driven steady state.  It is also the value of
${\dot{S}}^{\rm HK}$ in Eq.~(\ref{eq:Apol_SHK_pos_HKvec}) at the
extremal distribution ${\rho}^s = {\bar{\rho}}^s$ for the linear model.

For the same ${\rho}^s$, recognizing that 
$\left( \mu \! \left( \rn \right) - {\mu}_H \right) = \log \left( K \!
\left( \rn \right) / n_2 \right)$, 
the gradient relative to the bistable
reference distribution ${\bar{\rho}}^s$ in
Eq.~(\ref{eq:Apolcat_SHK_split}) expands to ${\bar{\rho}}^s_{\rn}
\left[ {\left( \rho / \bar{\rho} \right)}^s_{\rn + 1} - {\left( \rho /
\bar{\rho} \right)}^s_{\rn} \right] \approx - {\rho}^s_{\rn} \left[
\mu \! \left( \rn \right) - {\mu}_H \right]$.  The difference of
environment from housekeeping entropy changes then evaluates to 
\begin{align}
  {\dot{S}}^{\rm HK} - 
  {\dot{S}}^{\rm env}
& \approx 
  - \sum_{\rn}
  \left( {\hat{\jmath}}_1 + {\hat{\jmath}}_3 \right) 
  {\rho}^s_{\rn} 
  \mu \! \left( \rn \right)
  \left[ \mu \! \left( \rn \right) - {\mu}_H \right] . 
\label{eq:Apolcat_diff_SHK_Senv_cohn2}
\end{align}
The combination $ \left( {\hat{\jmath}}_1 + {\hat{\jmath}}_3 \right)
{\rho}^s_{\rn} \mu \! \left( \rn \right)$ is increasing through $n_2$
(See Fig.~\ref{fig:driven_Schlogl_HK_grad_coeff}), and $\left[ \mu \!
\left( \rn \right) - {\mu}_H \right]$ is antisymmetric, so
Eq.~(\ref{eq:Apolcat_diff_SHK_Senv_cohn2}) is negative.   

Hence not all of the entropy change by ${\dot{S}}^{\rm env}$ 
reflects a loss of large-deviation accessibility attributable to the
environment.  A quantity equal to ${\dot{S}}^{\rm env} -
{\dot{S}}^{\rm HK}$ is assigned in the natural
decomposition~(\ref{eq:Ddot_tot_decomp_HK}) to the system, because a
Poisson distribution around $n_2$ is unstable and has excess net 
probability to relax toward the bistable distribution
${\bar{\rho}}^s$.

\section{Discussion}
\label{sec:discussion}

\subsubsection*{The problem of macroworlds as an alternative central
contribution} 

A suggestion that thermodynamics should not be mainly about the
relation between work and heat flows would have been non-sequitur at
the turn of the 20th century.  Not only were applications centered
around interactions with a thermal system through mechanical
boundaries~\cite{Carnot:essays:60}; the nature of irreversibility and
dissipation as scientific problems was framed in their relation to
mechanical principles of time-reversibility~\cite{Laplace:probs:51}
and conservation of
energy~\cite{Clausius:equivalence:62,Joule:sci_papers:44}.

Innovations in scaling methods, however, in context of the unification
of statistical and quantum theories in the second half of the 20th
century, have exposed \emph{understanding the existence of
macroworlds} as one of the central problems in science.
Sec.~\ref{sec:Ham_Jacob} explained how large-deviations scaling
formalizes a concept of macrostate in terms of selected distributions
over microstates.  When methods derived for phase transitions and
critical phenomena in condensed matter~\cite{Wilson:RG:74} were
merged with the renormalization-group approach~\cite{GellMann:RG:54}
to vacuum quantum field theory~\cite{Polchinski:RGEL:84}, it became
clear that renormalization-group flow, a closely related
operation\footnote{Through the correspondence of Keldysh to 2-field
statistical field methods, the algebras of the two can often be
interconverted~\cite{Kamenev:DP:02,Smith:DP:08}.} of dimensional
reduction to large-deviations scaling, provides a concept of 
\textit{effective fields} describing the robust properties of
macroworlds through phenomenological
Lagrangians~\cite{Weinberg:phenom_Lagr:79}.  Effective fields are the
counterparts under renormalization to classical states under
large-deviations scaling. 

In effective field theories nested through phase transitions (as in
the models of this paper) microstates are assimilated back to
macrostates.  The result has been a theory of the vacuum and the
hierarchy of matter in which all earlier, partly phenomenological
concepts of elementary and composite objects and interactions has been
subsumed within a semantics of distributions.  While these physics
examples continue to be organized by symmetries and conservation laws,
similar scaling methods developed independently for reliable
coding~\cite{Shannon:MTC:49,Cover:EIT:91} offer non-mechanical
counterparts (see~\cite{Smith:geo13:16}, Ch.~7).

The alternative characterization offered, then, is that 
\emph{
thermodynamics is not fundamentally the theory of the movements of
heat, but rather the theory of the emergence of macroworlds from
microworlds.}

\subsubsection*{The end of entropy flow, the natural partition, and
Hartley information}

In the absence of an energy-mediated constraint on the states jointly
realizable by a system and its environment, the notion of ``entropy
flow'' between subsystems may never arise.\footnote{Even in classical
thermodynamics, it is only well-defined for adiabatic transformations,
in which both subsystems remain in classical states constrained by
energy at every moment.}  The default description of entropy change
ceases to be the metaphor of a fluid and becomes the literal one: it
is the loss of large-deviation accessibility of classical states due
to relaxation of the whole-system distribution.  The concept of a
tangent surface delimiting the amount of entropy gain in one component
that can be influenced by the state of another, instead of the
adiabatic transformation, is filled by the the vector of Hartley
informations $- \log {\bar{\rho}}^s$ of the stationary state for the
${\mathbb{T}}^s$ at that moment.  The natural system-environment
partition~(\ref{eq:Ddot_tot_decomp_HK}) encodes the conditional
independence of components of the loss of states accessible by
fluctuations.

Hartley information appears as a random variable along trajectories in
the construction of the generating functional for housekeeping
heat~\cite{Speck:HS_heat_FT:05,Seifert:stoch_thermo_rev:12}.  In
standard treatments, its interpretation as an entropy is taken to
depend on its association with a dissipated heat.\footnote{The exact
statement, referring to Eq.~(123)
in~\cite{Seifert:stoch_thermo_rev:12}, is: ``In the absence of a first
law for the master equation dynamics, which would require further
physical input not available at this general stage, this
identification is by analogy only.''}  Here, $- \log {\bar{\rho}}^s$
gains its interpretation as an entropy through its role as a carrier
of information about the interaction of processes within a system on
its own limiting large-deviation function and on its capacity to limit
large-deviations in the environment.

\subsubsection*{Making trajectories first-class citizens}

The persistence over 90 years of Onsager's
program~\cite{Onsager:RRIP1:31,Onsager:RRIP2:31} of making equilibrium
Gibbs entropies the foundation of non-equilibrium thermodynamics has
built an asymmetry into this concept even within physics, at the same
time as growth in statistical methods for path ensembles has reduced
the reason for an asymmetry to exist, by defining the same tools for
paths as for states.  The equilibrium Gibbs entropy is defined from
the large-deviation function on ensembles of states.  The stochastic
effective action~(\ref{eq:S_oflogzn_from_HJ})~\cite{Smith:LDP_SEA:11}
is the corresponding function\emph{al} for trajectories.  However, the
role that energy conservation plays in stochastic thermodynamics, as a
constraint on which states can be jointly occupied by components
within a system and within its environment, will not generally have a
counterpart for the jointly realizable trajectories involving these
components.\footnote{Consider, as examples, non-Markovian noise
sources and error-correcting codes exploiting time correlations to
optimize against those environments~\cite{Verdu:Shannon:98}.}

Borrowing a term from computer science, the privileged roles of
equilibrium entropy and energy conservation in current
non-equilibrium thermodynamics makes spaces of states
``first-class citizens''~\cite{Abelson:SICP:96}, and admits
thermodynamic interpretations for ensembles only in so far as those
derive from equilibrium heat.  Jaynes
anticipated~\cite{Jaynes:caliber:80,Jaynes:LOS:03} a self-contained
thermodynamic interpretation for path ensembles, though still only in
a limited form referenced to the \textit{calibers} of cross-sections
of states.  If there is to be a thermodynamics in which trajectories
become first-class citizens on par with states, it will need a
foundation in more general primitives such as large-deviation
accessibility and separation of scales, as in the domain-agnostic
framework presented here.

\subsubsection*{Rule-based systems and life}

This paper's formula for a self-contained and substrate-agnostic
thermodynamics is meant to support concept discovery in two domains
that should clearly have such a thermodynamics, and for which energy
conservation should play a limited role or no role at all for many
interesting questions.\footnote{A well-known examples of a constraint
that is proved by a fluctuation theorem, but not binding because a
more proximal constraint exists, is that of total entropy production
versus the ``excess heat'' in a driven
system.~\cite{Speck:FDT_NESS:06} The former bounds the non-equilibrium
entropy change within a distribution, but is uninformative because it
diverges on long timescales.  The tight bound, under some conditions,
is the finite excess heat.}

The first domain is rule-based
modeling~\cite{Danos:rule_based_modeling:08} which recognizes the
finitary-to-infinitary mappings exhibited here between CRN generators
and state spaces as a widely generalizable organizing principle.  The
algebra and combinatorics of non-commutative rule systems is a very
active area of
study~\cite{Harmer:info_carriers:10,Andersen:generic_strat:14,Behr:alg_graph_rewrt:16}
which connects to chemistry, systems biology, theories of algorithms,
process calculus, and much more.  A thermodynamics of rule-based
systems that expands the conceptual scope of science will not be a
reduction to their consequences for heat generation. 

The second domain is the thermodynamic nature of
life~\cite{Smith:geo13:16}, encompassing rule-based order in its
chemical substrate, the multi-level control systems and error
correction required to maintain complex dynamical phases, and the
nested population dynamics and historical contingency of evolution.
There will be order in such systems that derives, sometimes
cryptically, from microscopic time-reversibility.  But it is
inconceivable that there will not be much \emph{more} order, of novel
kinds, that originates from the abundance of architectures -- most of
them irreversible -- standing between the microscopic substrate and
robust macrophenomena.

\section{Concluding remarks}

The presentation above is not a review of recent technical
innovations, but an attempt to select core concepts that depend on
formalization to be adequately expressed, and to exhibit them in
relation to each other and in a context that does not presume
mechanics.  The Hamilton-Jacobi and CRN formalisms are of course not
new, but needed to be reviewed within a coherent framework to make
points about them that are not made elsewhere.  Likewise most
monotonicity results are known from fluctuation theorems, though
alternative proofs for a few are given here.  It may therefore be
useful to summarize the novel contributions of the paper that are
interleaved with standard results above:

\begin{trivlist}

\item Monotonic change of Kullback-Leibler divergence is obvious and
positivity of housekeeping heat was
known~\cite{Speck:HS_heat_FT:05,Harris:fluct_thms:07}.  An emphasis
here on entropy increase as loss of accessibility by large deviations,
together with the observation that thermalization in a multi-scale
system produces separation in the conditional dependence between
system and environment degrees of freedom, leads to the decomposition
of $\dot{D} \! \left( \rho \parallel \underline{\rho} \right)$ in
Eq.~(\ref{eq:Ddot_tot_decomp_HK}) as the \emph{natural} partition of
irreversibility between system and environment changes, irrespective
within each of the degree of irreversibility in the other.

\item The ``non-equilibrium entropy production'' of stochastic
thermodynamics is a complicated sum, of an entropy functional of
arbitrary distributions (over stochastic microstates) and a Gibbs
entropy state-function (for bath macrostates).  The designation
``violations of the 2nd law'' by stochastic events uses that summation
to conflate these two entropy summands which have different relations
to constraints by boundary conditions.  The decomposition in
Eq.~(\ref{eq:DKL_ray_Seff_rays_chain}) of $D \! \left( {\rho}^{\left(
\bar{n} \right)} \parallel \underline{\rho} \right)$ in terms of the
Lyapunov entropy state function $S_{\rm eff}$ and a remainder  $D \!
\left( 1_{\underline{n} \left( \bar{n} \right)} \parallel
\underline{\rho} \right)$  that must be relaxed by instantons, retains
an un-violated 2nd law for relative entropy, and also the Lyapunov
and large-deviation roles of the classical state-function entropy, in
the proper relation.

\item The dualization by Baish~\cite{Baish:DP_duality:15} of
observable and response fields in Doi-Peliti theory stands in obvious
analogy to the construction of the adjoint transition matrix in the
fluctuation theorem for housekeeping
heat~\cite{Speck:HS_heat_FT:05,Harris:fluct_thms:07}.  The
standardization of the CRN generator~(\ref{eq:gen_L_defform_Doi})
makes the analogy explicit in the descaling~(\ref{eq:state_space_TM})
of $\mathbb{T}$ and the descaling~(\ref{eq:bbAhat_def}) of
$\mathbb{A}$.  Moreover, it shows the analogy as an expression of the
general relation of microstates to macrostates, and relates the
preservation or loss of information in the classical limit to CRN
complexity classes.

\item The distinction between relaxation and escape trajectories in
terms of formal momenta is widely developed in \textit{momentum-space
WKB theory}~\cite{Assaf:mom_WKB:17}, as Hamilton-Jacobi
theory~\cite{Bertini:macro_NEThermo:09}, and in relation to
information geometry~\cite{Smith:IGDP:19}.  The convexity proof in
Eq.~(\ref{eq:DKL_dot_ctm_mimicdisc_class}) relating the Lyapunov to the
large-deviation role of the classical-state entropy, and to the
structure of the $\mathcal{L} = 0$ manifold, is novel. 

\item The use of cycle decompositions in CRNs to compute heat
dissipation in the stationary state is standard since
Schnakenberg~\cite{Schnakenberg:ME_graphs:76}.  Here a more complete
cycle decomposition (keeping detailed-balanced currents) unifies all
proofs of monotonicity of entropies for non-stationary distributions,
relating the Lyapunov and large-deviation changes in macrostate
entropy on complex-balanced CRNs in
Eq.~(\ref{eq:cycle_disses_nonorm}), and its
counterpart~(\ref{eq:That_states_eval}) in the state space for all
CRNs.  It also provides an interesting view of the (known) equivalence
of the adjoints constructed in the generating
function~(\ref{eq:HK_intro}) for housekeeping heat in the environment,
and an equivalent function for the relative
entropy~(\ref{eq:That_states_eval}) in the system.

\item The Legendre duality of ${\dot{S}}^{\rm HK}$ to the potential
${\dot{F}}_P$ in Eq.~(\ref{eq:Apol_Leg_muP}) defines extensivity and
intensivity with respect to a different scaling than that for entropy, 
but a natural one if time-reversal and energy conservation are not
the only basis for taking macroscopic limits. 

\end{trivlist}

\vfill 
\eject 

% \bibliographystyle{unsrt} 
% \bibliography{DES}

\begin{thebibliography}{100}

\bibitem{Szilard:MD:29}
Leo Szilard.
\newblock Uber die entropieverminderung in einem thermodynamischen system bei
  eingriffen intelligenter wesen.
\newblock {\em Zeitschrift f{\"{u}}r Physik}, 53:840--856, 1929.

\bibitem{Landauer:IHGCP:61}
R.~Landauer.
\newblock Irreversibility and heat generation in the computing process.
\newblock {\em IBM J.~Res.~Development}, 3:183--191, 1961.

\bibitem{Bennett:TC:82}
Charles~H. Bennett.
\newblock The thermodynamics of computation -- a review.
\newblock {\em Int.~J.~Theor.~Phys.}, 21:905--940, 1982.

\bibitem{Wolpert:stoch_thermo_comp:19}
David~H. Wolpert.
\newblock The stochastic thermodynamics of computation.
\newblock {\em J.~Phys.~A: Math.~Theor.}, 52:193001, 2019.

\bibitem{Seifert:stoch_thermo_rev:12}
Udo Seifert.
\newblock Stochastic thermodynamics, fluctuation theorems, and molecular
  machines.
\newblock {\em Rep.~Prog.~Phys.}, 75:126001, 2012.
\newblock arXiv:1205.4176v1.

\bibitem{Onsager:RRIP1:31}
Lars Onsager.
\newblock {\protect Reciprocal Relations in Irreversible Processes. I.}
\newblock {\em Phys.~Rev.}, 37:405--426, 1931.

\bibitem{Onsager:RRIP2:31}
Lars Onsager.
\newblock {\protect Reciprocal Relations in Irreversible Processes. II.}
\newblock {\em Phys.~Rev.}, 38:2265--2279, 1931.

\bibitem{Glansdorff:structure:71}
P.~Glansdorff and I.~Prigogine.
\newblock {\em Thermodynamic Theory of Structure, Stability, and Fluctuations}.
\newblock Wiley, New York, 1971.

\bibitem{Prigogine:MT:98}
Dilip Kondepudi and Ilya Prigogine.
\newblock {\em Modern Thermodynamics: From Heat Engines to Dissipative
  Structures}.
\newblock Wiley, New York, 1998.

\bibitem{Searles:fluct_thm:99}
Denis~J. Evans and Debra~J. Searles.
\newblock Fluctuation theorem for stochastic systems.
\newblock {\em Phys.~Rev.~E}, 60:159--164, 1999.

\bibitem{Evans:fluct_thm:02}
Denis~J. Evans and Debra~J. Searles.
\newblock The fluctuation theorem.
\newblock {\em Adv.~Phys.}, 51:1529--1585, 2002.

\bibitem{Gallavotti:dyn_ens_NESM:95}
G.~Gallavotti and E.~D.~G. Cohen.
\newblock Dynamical ensembles in non-equilibrium statistical mechanics.
\newblock {\em Phys.~Rev.~Lett.}, 74:2694--2697, 1995.

\bibitem{Gallavotti:dyn_ens_SS:95}
G.~Gallavotti and E.~D.~G. Cohen.
\newblock Dynamical ensembles in stationary states.
\newblock {\em J.~Stat.~Phys.}, 80:931--970, 1995.

\bibitem{Jarzynski:fluctuations:08}
C.~Jarzynski.
\newblock Nonequilibrium work relations: foundations and applications.
\newblock {\em Eur.~Phys.~J.~B}, 64:331--340, 2008.

\bibitem{Chetrite:fluct_diff:08}
Rapha{\"{e}}l Chetrite and Krzysztof Gawedzki.
\newblock Fluctuation relations for diffusion processes.
\newblock {\em Commun.~Math.~Phys.}, 282:469--518, 2008.

\bibitem{Esposito:fluct_theorems:10}
Massimiliano Esposito and Christian Van~den Broeck.
\newblock Three detailed fluctuation theorems.
\newblock {\em Phys.~Rev.~Lett.}, 104:090601, 2010.

\bibitem{Crooks:NE_work_relns:99}
Gavin~E. Crooks.
\newblock Entropy production fluctuation theorem and the nonequilibrium work
  relation for free energy differences.
\newblock {\em Phys.~Rev.~E}, 6:2721--2726, 1999.

\bibitem{Crooks:path_ens_aves:00}
Gavin~E. Crooks.
\newblock Path-ensemble averages in systems driven far from equilibrium.
\newblock {\em Phys.~Rev.~E}, 61:2361--2366, 2000.

\bibitem{Kurchan:NEWRs:07}
Jorge Kurchan.
\newblock Non-equilibrium work relations.
\newblock {\em J.~Stat.~Mech.}, 2007:P07005, 2007.

\bibitem{England:statphys_selfrepl:13}
Jeremy~L. England.
\newblock Statistical physics of self-replication.
\newblock {\em J.~Chem.~Phys.}, 139:121923, 2013.

\bibitem{Perunov:adaptation:15}
Nikolai Perunov, Robert Marsland, and Jeremy England.
\newblock Statistical physics of adaptation.
\newblock 2015.
\newblock arXiv:1412.1875v1 [physics.bio-ph].

\bibitem{Luce:choice:59}
R.~Duncan Luce.
\newblock {\em Individual Choice Behavior}.
\newblock Wiley, New York, 1959.

\bibitem{McFadden:quantal_choice:76}
Daniel McFadden.
\newblock Quantal choice analysis: a survey.
\newblock {\em Ann.~Econ.~Soc.~Measurement}, 5:363--390, 1976.

\bibitem{Lenton:climate_tip_ew:12}
T.~M. Lenton, V.~N. Livina, V.~Dakos, E.~H. van Nes, and M.~Scheffer.
\newblock Early warning of climate tipping points from critical slowing down:
  comparing methods to improve robustness.
\newblock {\em Phil.~Trans.~R.~Soc.~A}, 370:1185--1204, 2011.

\bibitem{Joule:sci_papers:44}
J.~P. Joule.
\newblock {\em The Scientific Papers of James Prescott Joule}.
\newblock London: Physical Society, Open Library, 1844.
\newblock
  https://openlibrary.org/books/OL239730M/The\_scientific\_papers\_of\_James\_Prescott\_Joule.

\bibitem{Clausius:mech_the_heat:65}
T.~Archer Hirst, editor.
\newblock {\em The Mechanical Theory of Heat}, London, 1865. John van Voorst.

\bibitem{Boltzmann:second_law:86}
Ludwig Boltzmann.
\newblock The second law of thermodynamics.
\newblock In {\em Popul{\"{a}}re Schriften}, pages 25--50. J.~A.~Barth,
  Leipzig, 1905.
\newblock re-issued Braunschweig: F. Vieweg, 1979.

\bibitem{Hald:prob_pre_1750:90}
Anders Hald.
\newblock {\em A history of probability and statistics and their applications
  before 1750}.
\newblock Wiley, New York, 1990.

\bibitem{Jaynes:ITSM_I:57}
E.~T. Jaynes.
\newblock Information theory and statistical mechanics.
\newblock {\em Phys.~Rev.}, 106:620--630, 1957.
\newblock reprinted in~\cite{Jaynes:Papers:83}.

\bibitem{Jaynes:ITSM_II:57}
E.~T. Jaynes.
\newblock {\protect Information theory and statistical mechanics. II}.
\newblock {\em Phys.~Rev.}, 108:171--190, 1957.
\newblock reprinted in~\cite{Jaynes:Papers:83}.

\bibitem{Rota:lect_notes:08}
Gian-Carlo Rota.
\newblock {Lectures on \textit{Being and Time} (1998)}.
\newblock {\em New Yearbook for Phenomenology and Phenomenological Philosophy},
  8:225--319, 2008.

\bibitem{GellMann:RG:54}
Murray Gell-Mann and Francis Low.
\newblock Quantum electrodynamics at small distances.
\newblock {\em Phys.~Rev.}, 95:1300--1312, 1954.

\bibitem{Wilson:RG:74}
K.~G. Wilson and J.~Kogut.
\newblock The renormalization group and the $\varepsilon$ expansion.
\newblock {\em Phys.~Rep., Phys.~Lett.}, 12C:75--200, 1974.

\bibitem{Weinberg:phenom_Lagr:79}
Steven Weinberg.
\newblock {Phenomenological Lagrangians}.
\newblock {\em Physica A}, 96:327--340, 1979.

\bibitem{Polchinski:RGEL:84}
Joseph~G. Polchinski.
\newblock Renormalization group and effective lagrangians.
\newblock {\em Nuclear Physics B}, 231:269--295, 1984.

\bibitem{Fermi:TD:56}
Enrico Fermi.
\newblock {\em Thermodynamics}.
\newblock Dover, New York, 1956.

\bibitem{Bertini:macro_NEThermo:09}
L.~Bertini, A.~De~Sole, D.~Gabrielli, G.~Jona-Lasinio, and C.~Landim.
\newblock Towards a nonequilibrium thermodynamics: a self-contained macroscopic
  description of driven diffusive systems.
\newblock {\em J.~Stat.~Phys.}, 135:857--872, 2009.

\bibitem{Amari:inf_geom:01}
Shun-Ichi Amari.
\newblock {\em Methods of Information Geometry}.
\newblock Amer.~Math.~Soc., 2001.

\bibitem{Ay:info_geom:17}
Nihat Ay, J{\"{u}}rgen Jost, H{\^{o}}ng~V{\^{a}}n L{\^{e}}, and Lorentz
  Schwachh{\"{o}}fer.
\newblock {\em Information Geometry}.
\newblock Schwinger International, Cham, Switzerland, 2017.

\bibitem{Ellis:ELDSM:85}
Richard~S. Ellis.
\newblock {\em Entropy, Large Deviations, and Statistical Mechanics}.
\newblock Springer-Verlag, New York, 1985.

\bibitem{Touchette:large_dev:09}
Hugo Touchette.
\newblock The large deviation approach to statistical mechanics.
\newblock {\em Phys.~Rep.}, 478:1--69, 2009.
\newblock arxiv:0804.0327.

\bibitem{Shannon:MTC:49}
Claude~Elwood Shannon and Warren Weaver.
\newblock {\em The Mathematical Theory of Communication}.
\newblock U.~Illinois Press, Urbana, Ill., 1949.

\bibitem{Hartley:information:28}
R.~V.~L. Hartley.
\newblock Transmission of information.
\newblock {\em Bell system technical journal}, July:535--563, 1928.

\bibitem{Speck:HS_heat_FT:05}
T.~Speck and U.~Seifert.
\newblock Integral fluctuation theorem for the housekeeping heat.
\newblock {\em J.~Phys.~A Math.~Gen.}, 38:L581--L588, 2005.

\bibitem{Harris:fluct_thms:07}
R.~J. Harris and G.~M. Sch{\"{u}}tz.
\newblock Fluctuation theorems for stochastic dynamics.
\newblock {\em J.~Stat.~Mech.}, page P07020, 2007.
\newblock doi:10.1088/1742-5468/2007/07/P07020.

\bibitem{Cover:EIT:91}
Thomas~M. Cover and Joy~A. Thomas.
\newblock {\em Elements of Information Theory}.
\newblock Wiley, New York, 1991.

\bibitem{Hatano:NESS_Langevin:01}
Takahiro Hatano and Shin-ichi Sasa.
\newblock {Steady state thermodynamics of Langevin systems}.
\newblock {\em Phys.~Rev.~Lett.}, 86:3463--3466, 2001.

\bibitem{Smith:LDP_SEA:11}
Eric Smith.
\newblock Large-deviation principles, stochastic effective actions, path
  entropies, and the structure and meaning of thermodynamic descriptions.
\newblock {\em Rep.~Prog.~Phys.}, 74:046601, 2011.
\newblock http://arxiv.org/submit/199903.

\bibitem{GellMann:EC:96}
Murray Gell-Mann and Seth Lloyd.
\newblock Information measures, effective complexity, and total information.
\newblock {\em Complexity}, 2:44--52, 1996.

\bibitem{GellMann:eff_complx:04}
Murray Gell-Mann and Seth Lloyd.
\newblock Effective complexity.
\newblock In Murray Gell-Mann and Constantino Tsallis, editors, {\em
  Nonextensive Entropy -- Interdisciplinary Applications}, pages 387--398.
  Oxford U. Press, New York, 2004.

\bibitem{Polettini:open_CNs_I:14}
Matteo Polettini and Massimiliano Esposito.
\newblock {Irreversible thermodynamics of open chemical networks. I. Emergent
  cycles and broken conservation laws}.
\newblock {\em J.~Chem.~Phys.}, 141:024117, 2014.

\bibitem{Polettini:stoch_macro_thermo:16}
Matteo Polettini, Gregory Bulnes-Cuetara, and Massimiliano Esposito.
\newblock Bridging stochastic and macroscopic thermodynamics.
\newblock 2016.
\newblock arXiv:1602.06555v1.

\bibitem{Horn:mass_action:72}
F.~Horn and R.~Jackson.
\newblock General mass action kinetics.
\newblock {\em Arch.~Rat.~Mech.~Anal}, 47:81--116, 1972.

\bibitem{Feinberg:notes:79}
Martin Feinberg.
\newblock Lectures on chemical reaction networks.
\newblock lecture notes, 1979.
\newblock https://crnt.osu.edu/LecturesOnReactionNetworks.

\bibitem{Krishnamurthy:CRN_moments:17}
Supriya Krishnamurthy and Eric Smith.
\newblock Solving moment hierarchies for chemical reaction networks.
\newblock {\em J.~Phys.~A: Math.~Theor.}, 50:425002, 2017.

\bibitem{Smith:CRN_moments:17}
Eric Smith and Supriya Krishnamurthy.
\newblock Flows, scaling, and the control of moment hierarchies for stochastic
  chemical reaction networks.
\newblock {\em Phys.~Rev.~E}, 96:062102, 2017.

\bibitem{Seifert:FDT:10}
Udo Seifert and Thomas Speck.
\newblock Fluctuation-dissipation theorem in nonequilibrium steady states.
\newblock {\em Europhysics Lett.}, 89:10007, 2010.

\bibitem{Schnakenberg:ME_graphs:76}
J.~Schnakenberg.
\newblock Network theory of microscopic and macroscopic behavior of master
  equation systems.
\newblock {\em Rev.~Mod.~Phys}, 48:571--585, 1976.

\bibitem{Andersen:generic_strat:14}
Jakob~L. Andersen, Christoph Flamm, Daniel Merkle, and Peter~F. Stadler.
\newblock Generic strategies for chemical space exploration.
\newblock {\em Int.~J.~Comput.~Biol.~Drug Des.}, 7:225--258, 2014.

\bibitem{Assaf:mom_WKB:17}
Michael Assaf and Baruch Meerson.
\newblock Wkb theory of large deviations in stochastic populations.
\newblock {\em J.~Phys.~A}, 50:263001, 2017.

\bibitem{Freidlin:RPDS:98}
M.~I. Freidlin and A.~D. Wentzell.
\newblock {\em Random Perturbations in Dynamical Systems}.
\newblock Springer, New York, second edition, 1998.

\bibitem{Martin:MSR:73}
P.~C. Martin, E.~D. Siggia, and H.~A. Rose.
\newblock Statistical dynamics of classical systems.
\newblock {\em Phys.~Rev.~A}, 8:423--437, 1973.

\bibitem{Kamenev:DP:02}
Alex Kamenev.
\newblock Keldysh and doi-peliti techniques for out-of-equilibrium systems.
\newblock In I.~V. Lerner, B.~L. Althsuler, V.~I. Fal${}^{\prime}$ko, and
  T.~Giamarchi, editors, {\em Strongly Correlated Fermions and Bosons in
  Low-Dimensional Disordered Systems}, pages 313--340, Heidelberg, 2002.
  Springer-Verlag.

\bibitem{Smith:NS_thermo_II:08}
Eric Smith.
\newblock {\protect Thermodynamics of natural selection II: Chemical Carnot
  cycles}.
\newblock {\em J.~Theor.~Biol.}, 252:198--212, 2008.
\newblock PMID: 18367209.

\bibitem{Smith:NS_thermo_I:08}
Eric Smith.
\newblock {\protect Thermodynamics of natural selection I: Energy and entropy
  flows through non-equilibrium ensembles}.
\newblock {\em J.~Theor.~Biol.}, 252:185--197, 2008.
\newblock PMID: 18367210.

\bibitem{Esposito:copol_eff:10}
Massimiliano Esposito, Katja Lindenberg, and Christian Van~den Broeck.
\newblock {Extracting chemical energy by growing disorder: efficiency at
  maximum power}.
\newblock {\em J.~Stat.~Mech.}, doi:10.1088/1742-5468/2010/01/P01008:1--11,
  2010.

\bibitem{Kittel:TP:80}
Charles Kittel and Herbert Kroemer.
\newblock {\em Thermal Physics}.
\newblock Freeman, New York, second edition, 1980.

\bibitem{Siegmund:IS_seq_tests:76}
D.~Siegmund.
\newblock {Importance sampling in the Monte Carlo study of sequential tests}.
\newblock {\em Ann.~Statist.}, 4:673--684, 1976.

\bibitem{Goldstein:ClassMech:01}
Herbert Goldstein, Charles~P. Poole, and John~L. Safko.
\newblock {\em Classical Mechanics}.
\newblock Addison Wesley, New York, third edition, 2001.

\bibitem{Anderson:product_dist:10}
David~F. Anderson, George Craciun, and Thomas~G. Kurtz.
\newblock Product-form stationary distributions for deficiency zero chemical
  reaction networks.
\newblock {\em Bull.~Math.~Bio.}, 72:1947--1970, 2010.

\bibitem{Doi:SecQuant:76}
M.~Doi.
\newblock Second quantization representation for classical many-particle
  system.
\newblock {\em J.~Phys.~A}, 9:1465--1478, 1976.

\bibitem{Doi:RDQFT:76}
M.~Doi.
\newblock Stochastic theory of diffusion-controlled reaction.
\newblock {\em J.~Phys.~A}, 9:1479--, 1976.

\bibitem{Peliti:PIBD:85}
L.~Peliti.
\newblock Path-integral approach to birth-death processes on a lattice.
\newblock {\em J.~Physique}, 46:1469, 1985.

\bibitem{Peliti:AAZero:86}
L.~Peliti.
\newblock Renormalization of fluctuation effects in $a + a \rightarrow a$
  reaction.
\newblock {\em J.~Phys.~A}, 19:L365, 1986.

\bibitem{Cardy:Instantons:78}
J.~L. Cardy.
\newblock Electron localisation in disordered systems and classical solutions
  in ginzburg-landau field theory.
\newblock {\em J.~Phys.~C}, 11:L321 -- L328, 1987.

\bibitem{Coleman:AoS:85}
Sidney Coleman.
\newblock {\em Aspects of Symmetry}.
\newblock Cambridge, New York, 1985.

\bibitem{Smith:CRN_CTM:20}
Eric Smith and Supriya Krishnamurthy.
\newblock Eikonal solutions for moment hierarchies of chemical reaction
  networks in the limits of large particle number.
\newblock {\em J.~Phys.~A Math.~Theor.}, submitted, 2020.

\bibitem{Gunawardena:CRN_for_bio:03}
Jeremy Gunawardena.
\newblock Chemical reaction network theory for \textit{in-silico} biologists.
\newblock lecture notes, June 2003.
\newblock vcp.med.harvard.edu/papers/crnt.pdf.

\bibitem{Smith:evo_games:15}
Eric Smith and Supriya Krishnamurthy.
\newblock {\em Symmetry and Collective Fluctuations in Evolutionary Games}.
\newblock IOP Press, Bristol, 2015.

\bibitem{Andersen:NP_autocat:12}
Jakob~L. Andersen, Christoph Flamm, Daniel Merkle, and Peter~F. Stadler.
\newblock {Maximizing output and recognizing autocatalysis in chemical reaction
  networks is NP-complete}.
\newblock {\em J.~Sys.~Chem.}, 3:1, 2012.

\bibitem{Berge:hypergraphs:73}
Claude Berge.
\newblock {\em Graphs and Hypergraphs}.
\newblock North-Holland, Amsterdam, rev.~ed. edition, 1973.

\bibitem{Baez:QTRN_eq:14}
John~C. Baez and Brendan Fong.
\newblock Quantum techniques for studying equilibrium in reaction networks.
\newblock {\em J.~Compl.~Netw.}, 3:22--34, 2014.
\newblock
  https://academic.oup.com/comnet/article-abstract/3/1/22/490572/Quantum-techniques-for-studying-equilibrium-in?redirectedFrom=fulltext.

\bibitem{Baish:DP_duality:15}
Andrew~James Baish.
\newblock Deriving the jarzynski relation from doi-peliti field theory.
\newblock Bucknell University Honors Thesis, 2015.

\bibitem{Feinberg:def_01:87}
Martin Feinberg.
\newblock {Chemical reaction network structure and the stability of complex
  isothermal reactors -- I. The deficiency zero and deficiency one theorems}.
\newblock {\em Chem.~Enc.~Sci.}, 42:2229--2268, 1987.

\bibitem{Schlogl:near_SS_thermo:71}
F.~Schl{\"{o}}gl.
\newblock On thermodynamics near a steady state.
\newblock {\em Z.~Phys.}, 248:446--458, 1971.

\bibitem{Dykman:chem_paths:94}
M.~I. Dykman, Eugenia Mori, John Ross, and P.~M. Hunt.
\newblock Large fluctuations and optimal paths in chemical kinetics.
\newblock {\em J.~Chem.~Phys.}, 100:5735--5750, 1994.

\bibitem{vanKampen:Stoch_Proc:07}
N.~G. van Kampen.
\newblock {\em Stochastic Processes in Physics and Chemistry}.
\newblock Elsevier, Amsterdam, third edition, 2007.

\bibitem{Danos:rule_based_modeling:08}
Vincent Danos, J{\'{e}}r{\^{o}}me Feret, Walter Fontana, Russell Harmer, and
  Jean Krivine.
\newblock Rule-based modelling, symmetries, refinements.
\newblock {\em Formal methods in systems biology: lecture notes in computer
  science}, 5054:103--122, 2008.

\bibitem{Baez:QTRN:13}
John~C. Baez.
\newblock Quantum techniques for reaction networks.
\newblock 2013.
\newblock https://arxiv.org/abs/1306.3451.

\bibitem{Yung:org_Titan:84}
Yuk Yung, M.~Allen, and J.~P. Pinto.
\newblock {Photochemistry of the atmosphere of Titan: comparison between model
  and observations}.
\newblock {\em Astrophys.~J.~Suppl.~Ser.}, 55:465--506, 1984.

\bibitem{Hebrard:Titan_atmos:12}
E.~H{\'{e}}brard, M.~Dobrijevic, J.~C. Loison, A.~Bergeat, and K.~M. Hickson.
\newblock {Neutral production of hydrogen isocyanide (HNC) and hydrogen cyanide
  (HCN) in Titan's upper atmosphere}.
\newblock {\em Astron.~Astrophys.}, 541:A21, 2012.

\bibitem{Turse:org_Titan:13}
Carol Turse, Johannes Leitner, Maria Firneis, and Dirk Schulze-Makuch.
\newblock {Simulations of prebiotic chemistry under post-imact conditions on
  Titan}.
\newblock {\em Life (Basel)}, 3:538--549, 2013.

\bibitem{Westheimer:phosphates:87}
Frank~H. Westheimer.
\newblock Why nature chose phosphates.
\newblock {\em Science}, 235:1173--1178, 1987.

\bibitem{Pasek:why_P:11}
Matthew~A. Pasek and Terence~P. Kee.
\newblock On the origin of phosphorylated biomolecules.
\newblock In Richard Egel, Dirk-Henner Lankenau, and Armen~Y. Mulkidjanian,
  editors, {\em Origins of Life: The Primal Self-Organization}, pages 57--84.
  Springer-Verlag, Berlin, 2011.

\bibitem{Goldford:P_free_metab:17}
Joshua~E. Goldford, Hyman Hartman, Temple~F. Smith, and Daniel Segr{\`{e}}.
\newblock Remnants of an ancient metabolism without phosphate.
\newblock {\em Cell}, 168:1126--1134, 2017.

\bibitem{Goldford:bdry_expn:19}
Joshua~E. Goldford, Hyman Hartman, Robert Marsland~III, and Daniel Segr{\`{e}}.
\newblock Environmental boundary conditions for early life converge to an
  organo-sulfur metabolism.
\newblock {\em Nat.~Ecol.~Evol.}, 2019.
\newblock doi:10.1038/s41559-019-1018-8.

\bibitem{Carnot:essays:60}
Sadi Carnot.
\newblock {\em Reflections on the Motive Power of Fire}.
\newblock Dover, New York, e.~mendoza (ed.) edition, 1960.

\bibitem{Laplace:probs:51}
Pierre~Simon Laplace.
\newblock {\em A Philosophical Essay on Probabilities}.
\newblock Dover, New York, original french 6th ed.~by truscott, f.~w. and
  emory, f.~l. edition, 1951.

\bibitem{Clausius:equivalence:62}
R.~Clausius.
\newblock On the application of the theorem of the equivalence of
  transformations to interior work.
\newblock In Hirst \cite{Clausius:mech_the_heat:65}, pages 215--250.
\newblock Fourth Memoir.

\bibitem{Smith:DP:08}
Eric Smith.
\newblock Quantum-classical correspondence principles for locally
  non-equilibrium driven systems.
\newblock {\em Phys.~Rev.~E}, 77:021109, 2008.
\newblock originally as SFI preprint \# 06-11-040.

\bibitem{Smith:geo13:16}
Eric Smith and Harold~J. Morowitz.
\newblock {\em The origin and nature of life on Earth: the emergence of the
  fourth geosphere}.
\newblock Cambridge U.~Press, London, 2016.
\newblock ISBN: 9781316348772.

\bibitem{Verdu:Shannon:98}
Sergio Verd{\'{u}}.
\newblock \protect{Fifty years of Shannon theory}.
\newblock {\em IEEE Trans.~Information Theory}, 44:2057--2078, 1998.

\bibitem{Abelson:SICP:96}
Harold Abelson, Gerald~Jay Sussman, and Julie Sussman.
\newblock {\em Structure and Interpretation of Computer Programs}.
\newblock MIT Press, Cambridge, MA, second edition, 1996.

\bibitem{Jaynes:caliber:80}
E.~T. Jaynes.
\newblock The minimum entropy production principle.
\newblock {\em Annu.~Rev.~Phys.~Chem.}, 31:579--601, 1980.

\bibitem{Jaynes:LOS:03}
E.~T. Jaynes.
\newblock {\em Probability Theory: The Logic of Science}.
\newblock Cambridge U.~Press, New York, 2003.

\bibitem{Speck:FDT_NESS:06}
T.~Speck and U.~Seifert.
\newblock Restoring a fluctuation-dissipation theorem in a nonequilibrium
  steady state.
\newblock {\em Europhys.~Lett.}, 74:391--396, 2006.

\bibitem{Harmer:info_carriers:10}
Russ Harmer, Vincent Danos, J{\'{e}}r{\^{o}}me Feret, Jean Krivine, and Walter
  Fontana.
\newblock Intrinsic information carriers in combinatorial dynamical systems.
\newblock {\em Chaos}, 20:037108, 2010.

\bibitem{Behr:alg_graph_rewrt:16}
Nicolas Behr, Vincent Danos, Ilias Garnier, and Tobias Heindel.
\newblock The algebras of graph rewriting.
\newblock 2016.
\newblock arXiv:1612.06240v1 [math-ph] 19 Dec 2016.

\bibitem{Smith:IGDP:19}
Eric Smith.
\newblock The information geometry of 2-field functional integrals.
\newblock submitted, 2019.
\newblock http://arxiv.org/abs/1906.09312.

\bibitem{Jaynes:Papers:83}
R.~D. Rosenkrantz, editor.
\newblock {\em Jaynes, E.~T.: Papers on Probability, Statistics and Statistical
  Physics}.
\newblock D.~Reidel, Dordrecht, Holland, 1983.

\end{thebibliography}

\end{document}